\documentclass[preprint,12pt,sort&compress]{elsarticle}

\usepackage{amsmath,amssymb}
\usepackage[]{graphicx}
\usepackage[latin1]{inputenc}
\usepackage{color}
\usepackage{enumerate}
\usepackage{hyperref}
\hypersetup{colorlinks=true,citecolor=blue}

\journal{Chemical Engineering Science}

\usepackage[top=3.0cm, bottom=3.0cm, left=3.0cm, right=3cm]{geometry}

\begin{document}

\begin{frontmatter}

\title{Integration of in situ Imaging and Chord Length Distribution Measurements for Estimation of Particle Size and Shape}

\author[a1]{Okpeafoh S. Agimelen\corref{cor1}}
\ead{okpeafoh.agimelen@strath.ac.uk}

\author[a1]{Anna Jawor-Baczynska}

\author[a1]{John McGinty}

\author[a2]{Christos Tachtatzis}

\author[a3]{Jerzy Dziewierz}

\author[a4]{Ian Haley} 

\author[a1]{Jan Sefcik}

\author[a6]{Anthony J. Mulholland\corref{cor1}}
\ead{anthony.mulholland@strath.ac.uk} 

\cortext[cor1]{Corresponding authors}
\address[a1]{EPSRC Centre for Innovative Manufacturing in Continuous Manufacturing and Crystallisation, Department of Chemical and Process Engineering, University of Strathclyde, James Weir Building, 75 Montrose Street, Glasgow, G1 1XJ, United Kingdom.}

\address[a2]{Centre for Intelligent Dynamic Communications, Department of Electronic and Electrical Engineering, University Of Strathclyde, Royal College Building, 204 George Street, Glasgow, G1 1XW, 
United Kingdom.
}

\address[a3]{The Centre for Ultrasonic Engineering, Department of Electronic and Electrical Engineering, University Of Strathclyde, Royal College Building, 204 George Street, Glasgow, G1 1XW, 
United Kingdom.
}

\address[a4]{Mettler-Toledo Ltd., 64 Boston Road, Beaumont Leys Leicester, LE4 1AW, United Kingdom}

\address[a6]{Department of Mathematics and Statistics, University of Strathclyde, Livingstone Tower, 26 Richmond Street, Glasgow G1 1XH, United Kingdom
}

\begin{abstract}

Efficient processing of particulate products across various manufacturing steps requires that particles possess desired attributes such as size and shape. Controlling the particle production process to obtain required attributes will be greatly facilitated using robust algorithms providing the size and shape information of the particles from in situ measurements. However, obtaining particle size and shape information in situ during manufacturing has been a big challenge. This is because the problem of estimating particle size and shape (aspect ratio) from signals provided by in-line measuring tools is often ill posed, and therefore it calls for appropriate constraints to be imposed on the problem. One way to constrain uncertainty in estimation of particle size and shape from in-line measurements is to combine data from different measurements such as chord length distribution (CLD) and imaging. This paper presents two different methods for combining imaging and CLD data obtained with in-line tools in order to get reliable estimates of particle size distribution and aspect ratio, where the imaging data is used to constrain the search space for an aspect ratio from the CLD data. 

\end{abstract}


\end{frontmatter}

\section{Introduction}
\label{sec1}

One of key steps in the manufacture of particulate products in the pharmaceuticals and fine chemicals industry is crystallisation, which is widely used for separation and purification of intermediates, fine chemicals and active pharmaceutical ingredients. The crystals come in different sizes and shapes. Subsequent steps in the manufacturing process, such as filtration, drying, blending and formulation of final products, require that the particle sizes and shapes lie within some desirable range. In order to provide monitoring and control of crystallisation processes it is necessary to develop techniques for estimating the shape and size distribution of particles in situ. There are a number of off line tools \cite{Washington1992} that can be used to estimate the particle size distribution\footnote{The term particle size distribution is broadly used here to refer both to continuous analytical probability density functions for particle sizes and discretised probability histograms of the particle sizes.} (PSD) of crystals produced in a crystallisation process. However, of particular importance to the control of a crystallisation process are tools that can be used in situ. These tools should be suitable for estimating size and shape information of particles dispersed in a slurry without the need for sampling and/or dilution. Examples of such instruments are the focused beam reflectance measurement (FBRM), the three dimensional optical reflectance (3D-ORM) \cite{Heinrich2012} and the particle vision and measurement (PVM) \cite{Barrett2002} sensors.

In-line sensors such as FBRM and 3D-ORM measure a chord length distribution (CLD)\footnote{Similar to the case of PSD, the term chord length distribution is used to cover both continuous analytical functions and discretised probability histograms.} which is related to the size and shape of the particles in a slurry. It has been a long standing challenge to be able to deduce the actual PSD and particle shape from experimental CLD data. In order to do this, an inverse problem needs to be solved. This is usually achieved by suitably discretising both CLD (which is already measured as a discrete distribution) and PSD and then constructing an appropriate transformation matrix relating these two distributions \cite{Wynn2003,Li2005n1,Agimelen2015}. The transformation matrix depends on the choice of size bins used to discretise the two distributions and the corresponding size ranges as well as the shape of particles. The transformation matrix is usually not known in advance and needs to be estimated along with the corresponding PSD (discretised) so that the convolution of the transformation matrix with the PSD yields a CLD which agrees with the experimentally measured CLD. However, this problem is ill posed. There are a number of different transformation matrices and PSDs whose convolutions give rise to the same CLD. Hence the challenge is how to estimate a combination of transformation matrix and PSD whose convolution will agree with an experimentally measured CLD for a given slurry as well as the PSD estimated being physically reasonable and representative of the particles in the slurry. 
 
The approach which was used in previous works \cite{Ruf2000,Worlitschek2005,Li2014,Kail2009} when estimating the transformation matrix was to assume the same shape (quantified by a metric referred to as aspect ratio) for all the particles in the slurry, and then use a previously estimated\footnote{The approach was to estimate the range of particle sizes in a sample by techniques such as sieving, laser diffraction, microscopy or use information supplied by the manufacturer before suspending the particles.} range of particle sizes in the slurry to construct the transformation matrix. This approach is not suitable for monitoring a crystallisation process where nucleation and/or growth of particles was present as neither the range of particle sizes nor their aspect ratio would be known in advance. A technique which was suitable for estimating the range of particle sizes in a slurry in situ was presented in our previous work \cite{Agimelen2015}. However, like in other previous works, the transformation matrix was constructed with a single aspect ratio for all particles. This leaves open a possibility that the transformation matrix is constructed with inappropriate aspect ratio or that there is a wider range of aspect ratios present for particles of same or different sizes. It was demonstrated in our previous work \cite{Agimelen2015} that it was still possible to calculate different CLDs that all had a very good agreement with an experimentally measured CLD even though some of the transformation matrices were constructed at aspect ratios that were far from the shape of the particles described. However, it was also shown that as the aspect ratio deviated further from the true shape of the particles, then the corresponding PSD became increasingly noisy. This situation led to the introduction of a penalty function in order to eliminate unrealistic aspect ratios.
However, when there is a wide variation of aspect ratios of the particles in the slurry, there is a need to introduce further constraints on the aspect ratio to reduce the search space and regularize the inverse problem. One way to do this is to get estimates of aspect ratio (within some reasonable bounds) using imaging, and then use this information to constrain the search for a representative aspect ratio. However, the imaging needs to be done in situ in order to develop techniques for estimation of PSD and particle shape which could be used for real time monitoring and control of particle production processes. 

While it would be desirable to get good estimates of both PSD and particle aspect ratio using in situ imaging alone, this is currently not the case. The currently available in-line imaging tools (for example, the PVM used in this work) produce 2D projection images. Furthermore, the objects in the images may be partially or completely out of focus, parts of imaged object may cross the image frame or objects may overlap each other\footnote{The issue of objects overlapping each other would not be a problem if an appropriate image processing algorithm which can resolve the objects is used.}.
Although advanced measurement equipment have been developed which can be used to capture 3D images of particles in a slurry and make good estimates of PSD and shape of particles, it requires sampling and dilution flow loops\footnote{The dilution is necessary to avoid instances of overlapping particles in images.} to allow capturing 3D images of individual particles in a flow-through cell \cite{Eggers2008,Kempkes2010,Schorsch2012,Schorsch2014}. Therefore this approach may not be generally applicable for in-line monitoring of particle manufacturing processes. Hence the current situation is that PSD cannot be estimated to a good degree of accuracy using routinely available in-line imaging tools. To overcome this challenge, we propose to combine in-line CLD measurements with imaging data to provide more reliable estimate of quantitative particle attributes.

In this paper, we present two different methods for combining imaging data with CLD data for particle size and shape estimation. The first method presented here calculates an estimate of the mean aspect ratio of all the particles in the slurry and then uses this information to constrain the search space for size and shape estimation from the CLD data. In the second method, a distribution of aspect ratios for each particle size is used for the PSD estimation. The distribution of aspect ratios is based on the data from the captured images.

\section{Experiment}
\label{sec2}

To demonstrate the technique for estimating particle size and shape information using a combination of the CLD and imaging data, experiments were performed in slurries containing particles of different shapes. 
The materials and procedure are described below.

\subsection{Materials}
\label{ssec2_1}

Three different samples were used for the measurements. Sample 1 consisted of polystyrene (PS) microspheres purchased from EPRUI nanoparticles and Microspheres Co. Ltd. with batch number 2012-5-7, and 0.2g of the PS microspheres were dispersed in 100g of isopropanol (IPA) purchased from VWR (20842.323) giving a concentration of 0.2\% by weight. Sample 2 consisted of cellobiose octaacetate (COA) particles obtained from GSK. The particles were dispersed in methanol (purchased from VWR (20847.307)) with the same concentration as in sample 1. Sample 3 consisted of glycine (Glycine) crystals obtained by cooling crystallisation from an aqueous solution. The solution with glycine concentration of 340mg/ml was prepared using glycine (purchased from Sigma-Aldrich (G8898, $\geq 99\%$ TLC)) and deionized water (from an in-house Millipore Water System (18M$\Omega$/cm)). The solution was cooled from a temperature of 90°C to a temperature of 43$^{\circ}$C at a rate of 3$^{\circ}$C/min. During this process the glycine crystallised out of solution until an equilibrium particle size distribution was reached. The crystallisation of glycine was monitored with the FBRM probe which showed an initial increase (in time) of chord lengths before eventually reaching a steady state.

\subsection{Experimental Setup}
\label{ssec2_2}

The suspension of particles for all samples was made in the Mettler Toledo EasyMax 102 system. The EasyMax system consists of a cylindrical jacketed vessel of volume 100ml with different stirrer and blade options. An anchored overhead stirrer with pitched (45$^{\circ}$ pitch angle) blades was used in all the experiments in this work. The stirrer shaft and probes were inserted into the slurries through ports located at the top of the set up. The stirring speed was set at 400\,rpm in all experiments.

 \begin{figure}[tbh]
\centerline{\includegraphics[width=0.8\textwidth]{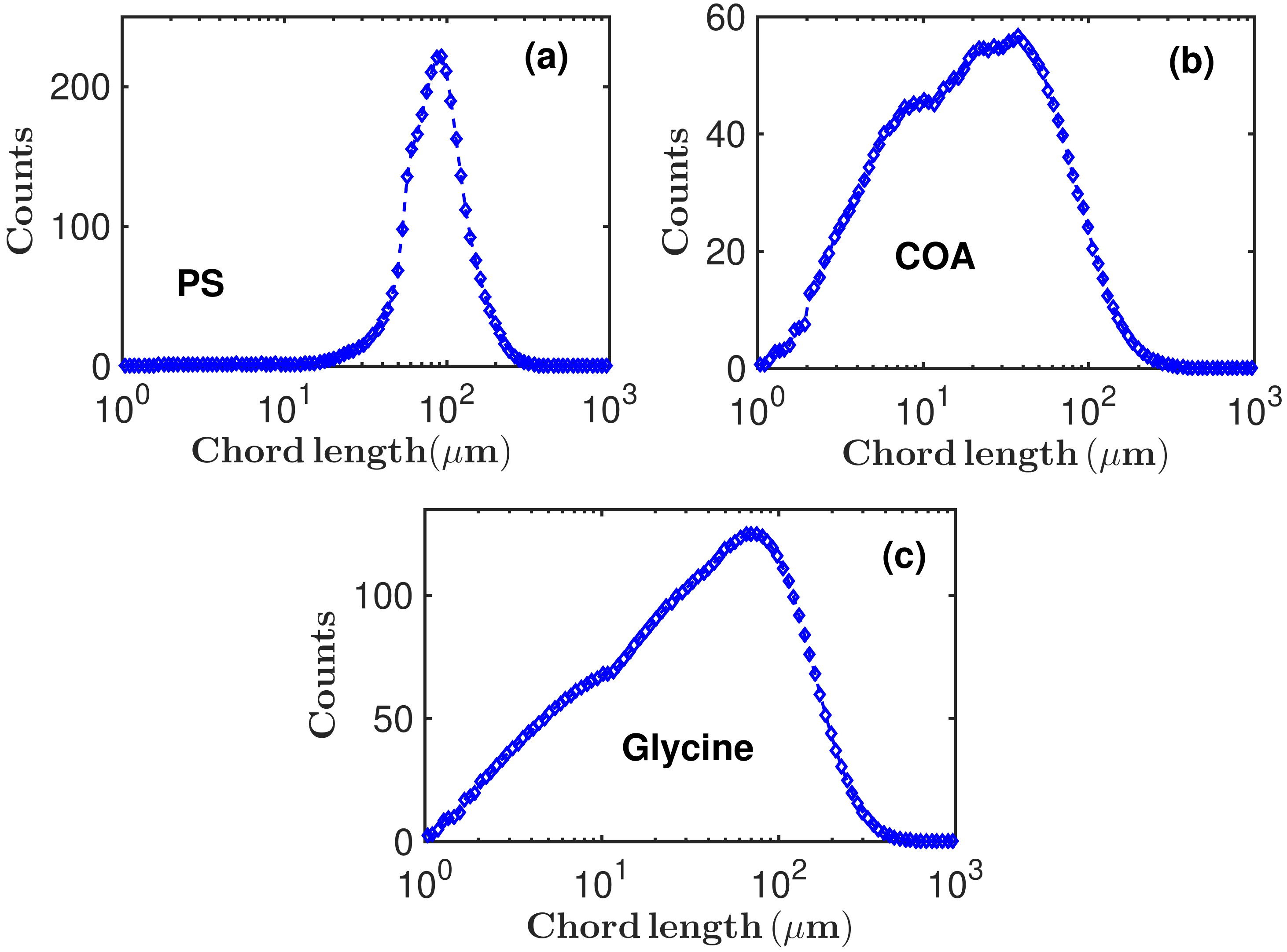}}
\caption{The CLD (measured with the FBRM G400 sensor) for the (a)\,PS, (b)\,COA and (C)\,Glycine samples.}
\label{fig1}
 \end{figure}

The CLD measurements were made with a Mettler Toledo FBRM G400 probe. The images of the particles were  captured with a Mettler Toledo PVM V819 probe during the period of CLD measurement. The FBRM sensor consists of a system of lenses which focus a laser beam onto a spot near the probe window in the slurry. The laser spot moves in a circular trajectory and the back scattered light is detected. The chord length is then calculated as the speed of the laser spot multiplied by the duration of the back scattered light as the laser traverses a particle. The FBRM  sensor records the lengths of the chords for a pre-set duration after which the CLD is reported \cite{Heinrich2012,Kail2007,Kail2008,Worlitschek2003,Agimelen2015}.

The PVM is an in situ microscope which consists of eight laser sources enclosed in a cylindrical tube. The six forward lasers and two back lasers (achieved by reflecting two lasers off a Teflon cap at the probe window) illuminate the particles in the slurry. The back scattered light is detected by a CCD element from which grayscale images are constructed. The image frame of the CCD array consists of 1360$\times$1024 pixels with a pixel size of 0.8$\mu$m. The PVM V819 sensor has a maximum acquisition rate of 5 images per second, although lower rates of acquisition could be set depending on requirements. The depth of the focal zone is restricted to about 50$\mu$m so that all objects that are in focus result in images that have identical magnification levels. Each of the lasers can be switched on or off so that different degrees of illumination can be achieved. 

    \begin{figure}[tbh]
   \centerline{\includegraphics[width=0.8\textwidth]{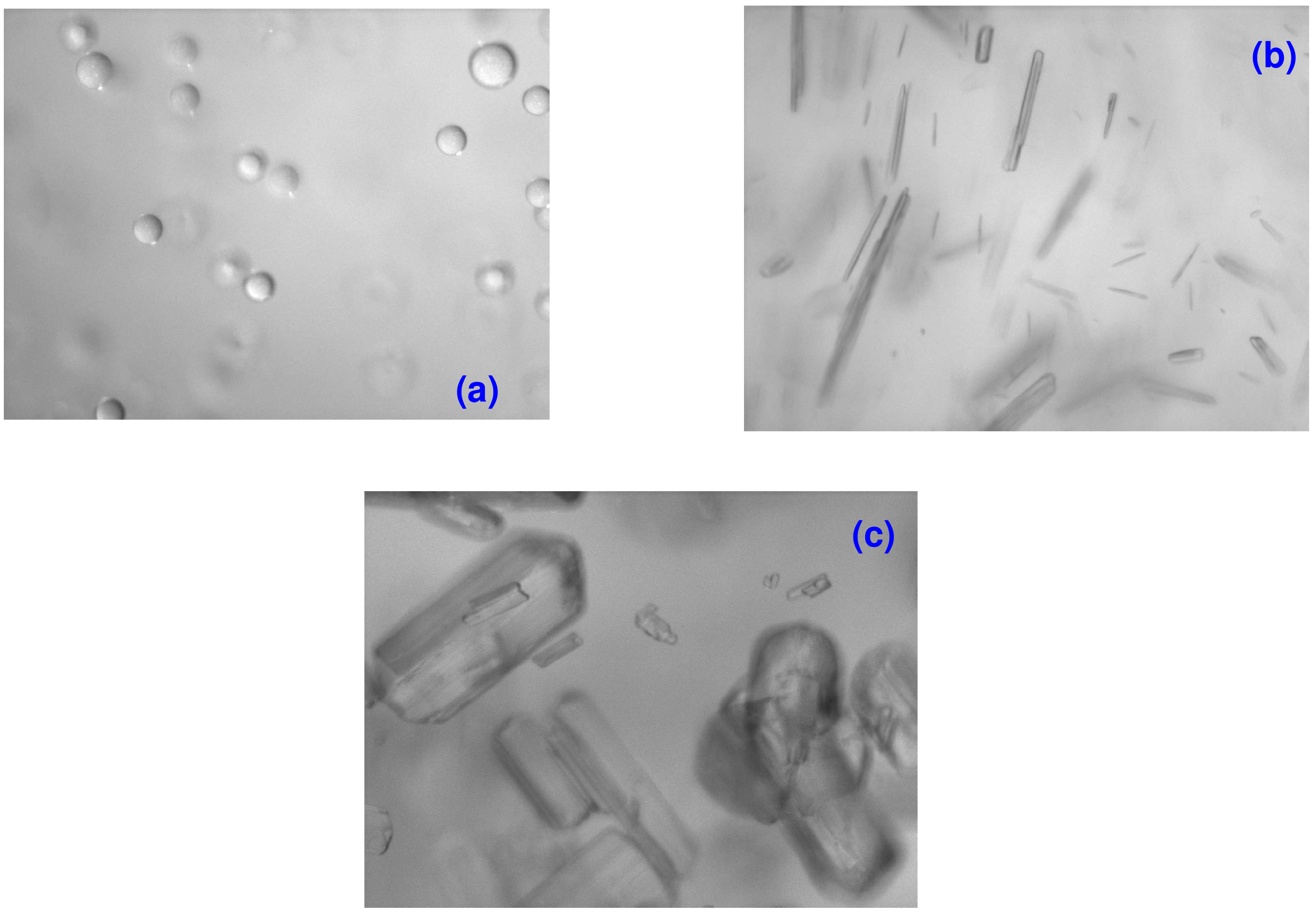}}
   \caption{Representative images (obtained with the PVM V819 sensor) for the (a)\,PS, (b)\,COA and (c)\,Glycine samples. The images have the same width of 1088$\mu$m and height of 819$\mu$m.}
   \label{fig2}
    \end{figure}

\section{Experimental Data}
\label{sec3}

The CLD data measured with the FBRM sensor for PS is shown in Fig. \ref{fig1}(a). Images were captured with the PVM sensor during the CLD measurement. A representative image from the PS sample is shown in Fig. \ref{fig2}(a). 

Similarly, the CLD data recorded for COA is shown in Fig. \ref{fig1}(b) while that of Glycine is shown in Fig. \ref{fig1}(c). Representative images for COA and Glycine are shown in Figs. \ref{fig2}(b) and \ref{fig2}(c) respectively. A total of 600 images were acquired each for PS and COA, and 121 images for Glycine. 

\section{Image Analysis}
 \label{sec4}
 
 As mentioned in the introductory section, an estimate of the PSD and particle shape can be made from images alone without the need to include CLD data. However, due to the reasons discussed in the introductory section, this approach is not always convenient. The techniques presented here utilise images obtained with an in-line measuring tool. However, due to the limitations in the images (as discussed in the introductory section), it is necessary to combine the imaging data with CLD to obtain reasonable aspect ratio and/or size estimates.
  
 The images captured with the PVM are processed in order to detect the objects contained in them and hence obtain information about the shape of the particles in the slurry. However, the image processing algorithm used in this work does not have features to resolve an object completely when it does not lie entirely in focal plane of the PVM sensor. Also, it does not have functionalities to resolve overlapping objects in images. For this reason the samples used in this work were deliberately prepared 
 dilute\footnote{Low slurry densities have been used here for the purpose of methods development. Future work will involve the investigation of the applicability of the methods developed here at higher slurry densities (with a more advanced image processing algorithm) using suspensions of particles of known PSD, then the degree of deviation of the results from the known PSD can be quantified at different slurry densities.}
    to reduce instances of overlapping objects. The parameters\footnote{The parameters of the algorithm need to be tuned for different samples due to variation of contrast.} of the image processing algorithm were tuned to reject most of the particles that were not entirely in the focal plane of the PVM sensor. However, the imaging data still has some degree of inaccuracy as seen in the error bars of the data (see subsection 4.6 of the supplementary information). This limitation not withstanding, the data from the images was sufficiently accurate to demonstrate the techniques developed in this work. Furthermore, images which do not contain  objects that are contained completely within the image frame were also discarded. This situation of having to discard some images reduces the number of data sets that can be gathered from the images. 
 However, it can be shown (see section 2 of the supplementary information) that with a sample size (number of objects) of about 500 the error incurred in estimating the aspect ratio is reduced to a reasonable extent. However, for a more robust estimate of the PSD using imaging data, a larger number of objects will be required. The number\footnote{A total of 1393 objects were detected for PS, 1810 for COA and 526 for Glycine.} of detected objects used in this work is just sufficient to demonstrate the methods developed here.
   The issue of objects not completely in focus can be dealt with if additional functionalities are added to the image processing algorithm, but this is beyond the scope of this work. The key steps for detecting objects in the images captured by the PVM sensor are summarised in subsection \ref{ssec4_1}.

 \subsection{Object Detection}
 \label{ssec4_1}
 
 The raw grayscale images from the PVM sensor are passed through a median filter to remove speck noise from the image background which is homogeneous. At this stage objects on the boundary of the image frame are removed. Any object with surface area below 900 pixels\footnote{The surface area of 900 pixels represents length dimensions of approximately 30 $\times$ 30 pixels (assuming a square geometry). This implies that objects that are smaller than approximately 24$\mu$m are rejected by the image processing algorithm. The consequence is that there is no estimate of aspect ratio for these small objects. However, the particles used in this work have sizes mostly in the range of 100$\mu$m so that the effects of this are minimal.} is considered noise and excluded from processing. Finally, a closing operation with a disk structural element is used to join broken edges. The resulting blob properties such as area, centroid, eccentricity, convex area, and major and minor axes can be obtained. These steps are summarised in Fig. \ref{fig3}. The tunable parameters of the image processing algorithm are summarised in subsection 4.5 of the supplementary information. 
 
        \begin{figure}[tbh]
       \centerline{\includegraphics[width=0.8\textwidth]{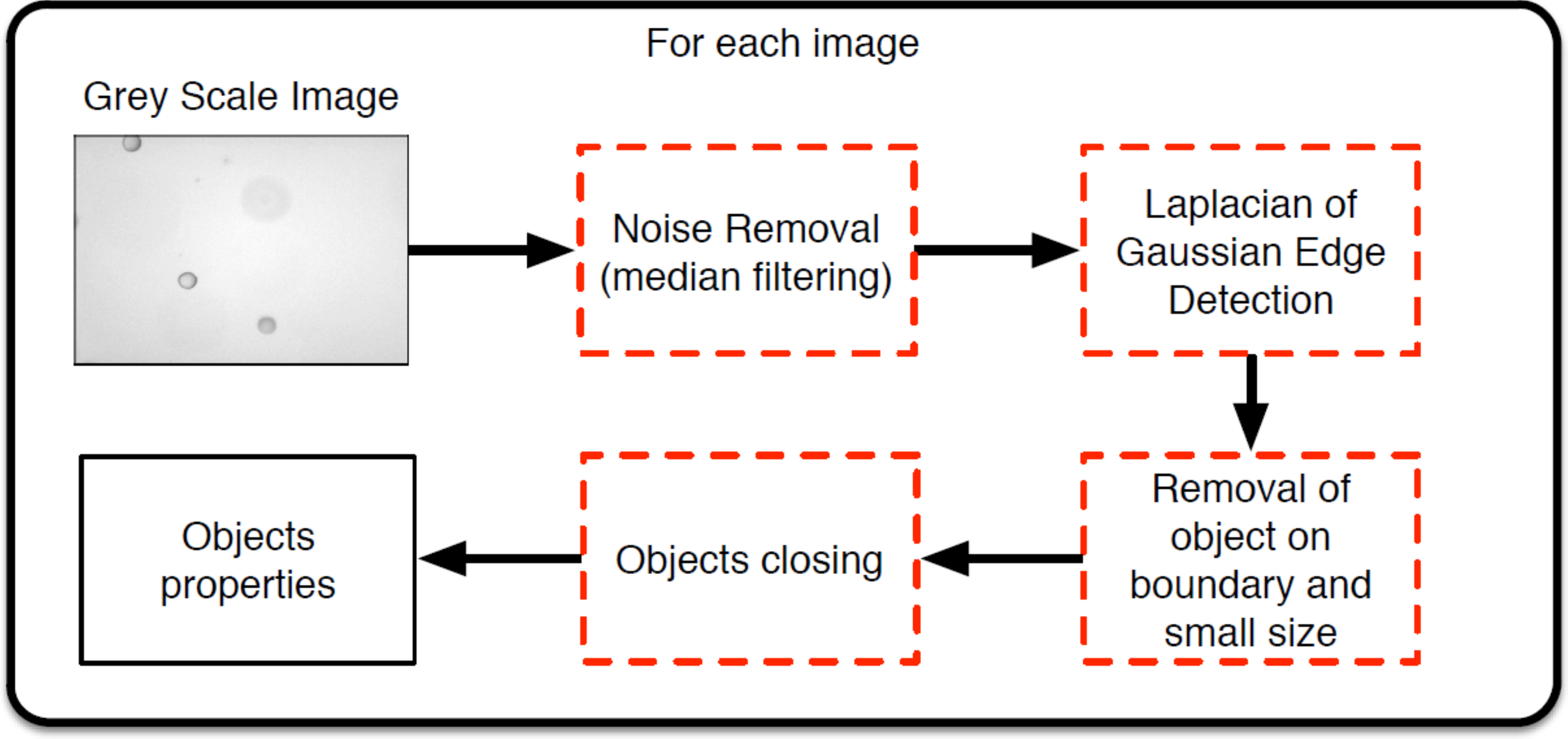}}
       \caption{A schematic for the key steps of the image processing algorithm used in this work.}
       \label{fig3}
        \end{figure}
       
         \begin{figure}[tbh]
        \centerline{\includegraphics[width=0.8\textwidth]{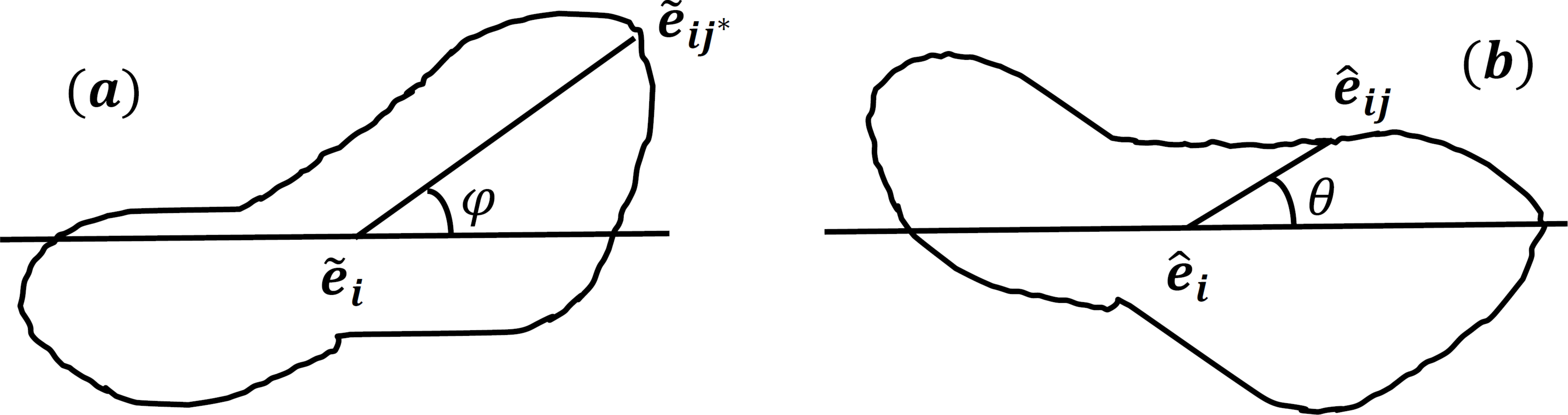}}
        \caption{A sketch of the particle characterisation procedure. The boundary pixels before rotation are shown in (a). The line joining the farthest boundary pixel $\mathbf{\tilde{e}}_{ij^{\ast}}$ to the centroid $\mathbf{\tilde{e}}_i$ makes an angle $\varphi$ with the horizontal axis. (b)\,A rotation is performed so that the farthest pixel from the centroid lies on the horizontal axis. The indices of the remaining pixels are assigned relative to this pixel.}
        \label{fig4}
         \end{figure}
        
          \begin{figure}[tbh]
         \centerline{\includegraphics[width=0.8\textwidth]{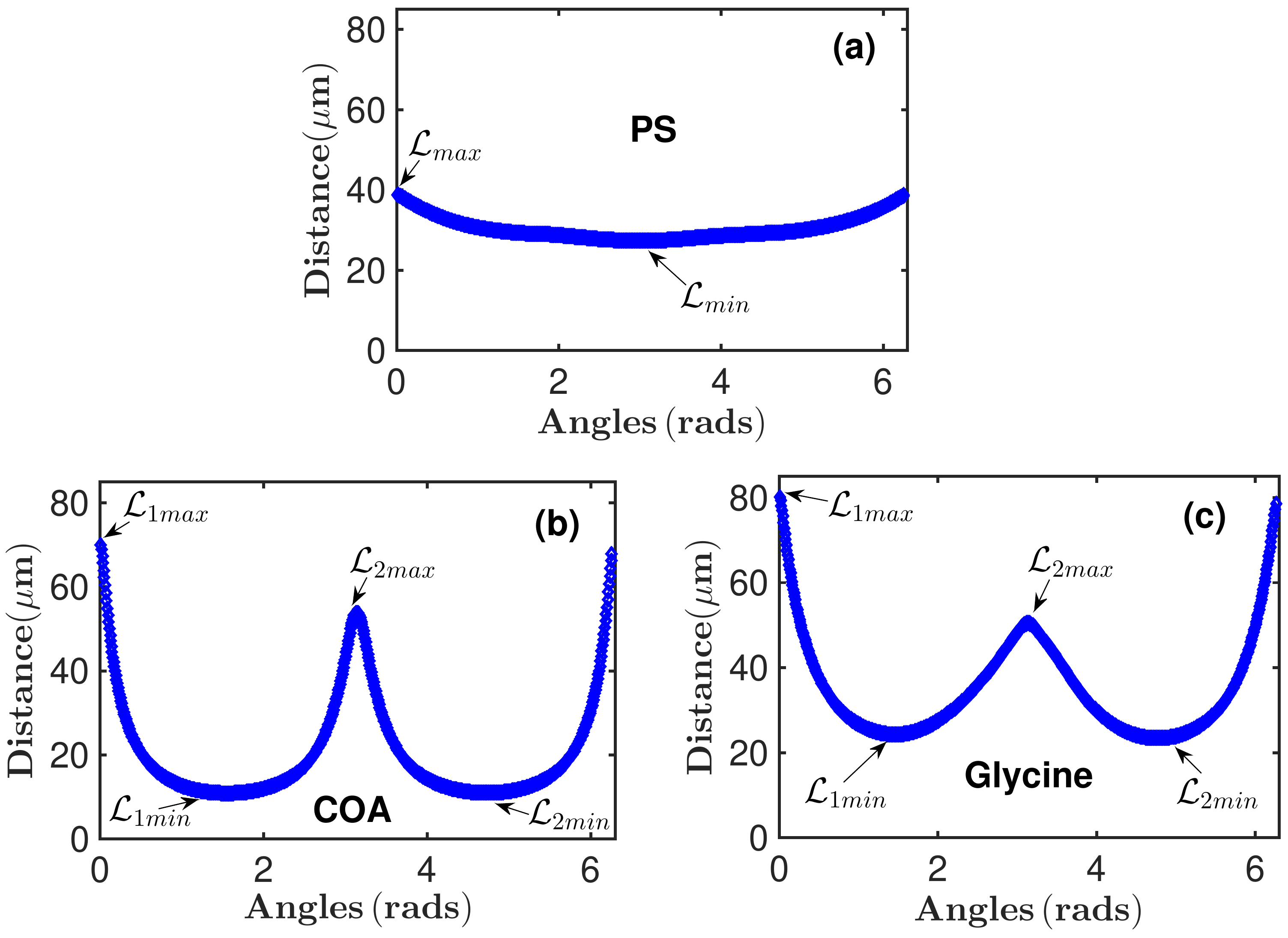}}
         \caption{The shape descriptor ($\mathbf{\overline{d}}$ in Eq. \eqref{eq4}) as a function of the angle ($\theta$) for the (a)\,PS, (b)\,COA and (c)\,Glycine samples.}
         \label{fig5}
          \end{figure}
   
 \subsection{Characterising Particle Shape}
 \label{ssec4_2}
 
The following procedure is used to obtain a shape descriptor, which is then used to characterise the shape of each particle. Each boundary pixel $j$ of object $i$ has coordinates $\mathbf{\tilde{e}}_{ij}=[\tilde{e}_{ij,x},\tilde{e}_{ij,y}]$, where the Cartesian coordinates system has been used. The centroid of object $i$ has coordinates $\mathbf{\tilde{e}}_i$. The distance $\tilde{d}_{ij}$ of the centroid of object $i$ to the pixel $j$ is given as
   \begin{equation}
   \tilde{d}_{ij} = \left\|\mathbf{\tilde{e}}_i - \mathbf{\tilde{e}}_{ij}\right\|.
   \label{eq1}
   \end{equation}
  Then for each object $i$, locate the pixel (with index $j^{\ast}$) whose distance from the centroid $\tilde{d}_{ij^{\ast}}$ is largest as
     \begin{equation}
     \tilde{d}_{ij^{\ast}} = \max_j{\{\tilde{d}_{ij}\}}. 
     \label{eq2}
     \end{equation}
  The angle between the line joining this farthest pixel to the centroid and the horizontal axis is $\varphi$ as shown in Fig. \ref{fig4}(a). Then, for each object $i$, transform the coordinates of each pixel by performing a rotation through the angle $\varphi$ so that the line joining the centroid to the farthest pixel becomes parallel to the horizontal axis. This operation transforms the coordinates of each pixel $j$ to $\mathbf{\hat{e}}_{ij}$ and the centroid to $\mathbf{\hat{e}}_i$. The line joining the pixel $j$ to the centroid now makes an angle $\theta$ with the horizontal axis as shown in Fig. \ref{fig4}(b). Due to the rotation, the farthest pixel from the centroid corresponds to $\theta=0$. The distance of each pixel $j$ from the centroid is the same of course.
  
  Since the sample rate with respect to $\theta$ is not uniform, then the pixel distances were resampled with $N_p$ uniformly spaced $\theta$ values constructed as $\theta_p = p\Delta\theta, \quad \Delta\theta = 2\pi/N_p,\quad  p=1,2,\ldots,N_p$. This allows the vector of all pixels distances (from the centroid) for object $i$ to be written as 
       \begin{equation}
       \mathbf{d}_i = [\tilde{d}_{i1}, \tilde{d}_{i2},\ldots,\tilde{d}_{iN_p}].
       \label{eq3}
       \end{equation}
Finally, the average vector $\mathbf{\overline{d}}$ of all pixel distances for all objects detected by the image processing algorithm can be calculated as
        \begin{equation}
       \mathbf{\overline{d}} = \frac{1}{N_{obj}}\sum_{i=1}^{N_{obj}}{\mathbf{d}_i},
        \label{eq4}
        \end{equation}
 where $N_{obj}$ is the number of objects detected from all the images analysed.
 
 A plot of $\mathbf{\overline{d}}$ versus angle ($\theta$) can be made as shown in Figs. \ref{fig5}(a) to \ref{fig5}(c). For near spherical particles, the shape descriptor $\mathbf{\overline{d}}$ is nearly constant as in the case of PS in Fig. \ref{fig5}(a). However, for elongated particles, the shape descriptor $\mathbf{\overline{d}}$ has two minima and maxima as in the cases of CoA and Glycine in Figs. \ref{fig5}(b) and \ref{fig5}(c). The shape descriptor shown in \ref{fig5}(a) to \ref{fig5}(c) is similar to the type described in \cite{Schorsch2012}.
 
 In an ideal situation, the shape descriptor for spherical particles will be constant at a value representing the radius of the spherical particles. However, since the PS particles are not perfectly spherical there is slight variation in the dimensions so that an average aspect ratio $r$ (the ratio of the minor to the major dimension) can be estimated. Similarly, the maxima in the shape descriptors for the CoA and Glycine particles (in Figs. \ref{fig5}(b) and \ref{fig5}(c)) represent the major dimension of the particles while the minima represent the minor dimension of the particles. 
 
 For the case of the PS particles the mean aspect ratio $r$ is estimated from the minimum dimension $\mathcal{L}_{min}$ and maximum dimension $\mathcal{L}_{max}$ (see Fig. \ref{fig5}(a)) as 
   \begin{equation}
   r = \frac{\mathcal{L}_{min}}{\mathcal{L}_{max}}.
   \label{eq5}
   \end{equation}
 In the case of elongated particles, the maximum dimension (average length of particles) is given as  
 $\mathcal{L}_{max}=\mathcal{L}_{1max} + \mathcal{L}_{2max}$ (shown in Figs. \ref{fig5}(b) and \ref{fig5}(c)) and the minimum dimension (average width of particles) is given as $\mathcal{L}_{min}=\mathcal{L}_{1min} + \mathcal{L}_{2min}$. So that the average aspect ratio can then be estimated using Eq. \eqref{eq5}.
 
\begin{figure}[tbh]
                  \centerline{\includegraphics[width=\textwidth]{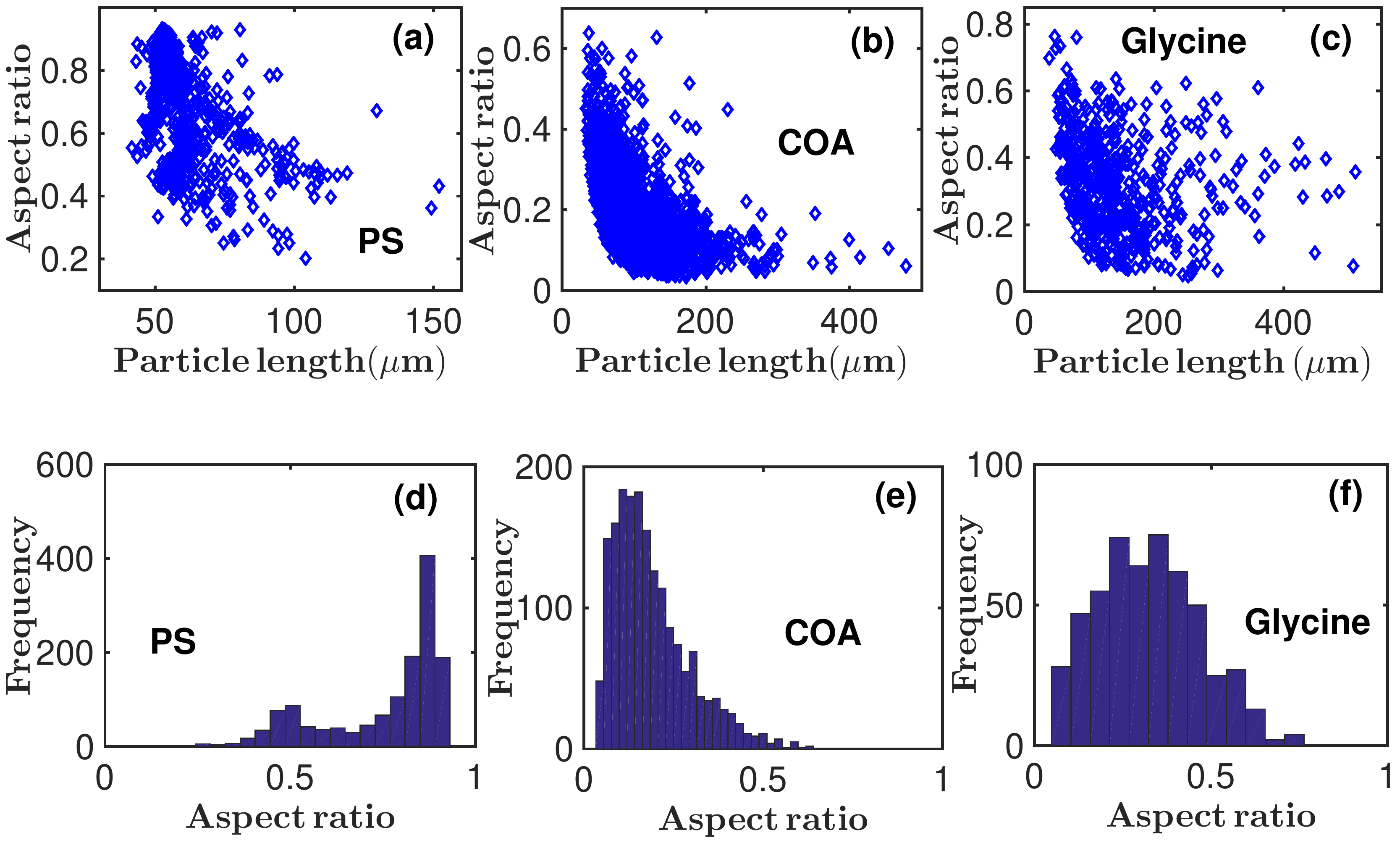}}
                  \caption{Scatter plots of the aspect ratio versus particle length for individual objects for (a)\,PS, (b)\,COA and (c)\,Glycine. The histograms of the aspect ratios are shown in (d)\,PS, (e)\,COA and (f)\,Glycine.}
                  \label{fig6}
                   \end{figure}

  However, the aspect ratios for individual particles will be different from $r$. The aspect ratio for each particle is estimated from the shape descriptor corresponding to that particle. Once the aspect ratios of individual particles are estimated, then a scatter plot of aspect ratio versus particle length can be made as shown in Figs. \ref{fig6}(a) to \ref{fig6}(c). The shape descriptors for individual particles are not always smooth as in the cases shown in Figs. \ref{fig5}(a) to \ref{fig5}(c). They contain different degrees of variation due to imperfections in the particles and images (see subsection 4.6 of the supplementary information for details).
  
Furthermore, the aspect ratios estimated from the shape descriptor for individual particles contain some artefacts (as can be seen in Fig. \ref{fig6}) due to a number of factors. These factors include deformations in the particles (that is, particles whose shapes deviate from the majority of particles in the slurry), impurity objects in the monitored slurry and particles not completely in focus\footnote{The image processing algorithm parameters are tuned to remove particles that are out of focus. However, when the particle is only partially in focus, the image processing algorithm detects only part of the particle and this leads to an error in the estimated aspect ratio for that particle.}.
 For example, aspect ratios  as low as about 0.2 in the case of the spherical particles in Fig. \ref{fig6}(a) or the minor peak at aspect ratio $r\approx0.5$ in Fig. \ref{fig6}(d) are artefacts.
 
 Figures \ref{fig6}(b) and \ref{fig6}(c) suggest the presence of particles of sizes up to about 500$\mu$m in the COA and Glycine samples respectively. However, the number of data points (in Figs. \ref{fig6}(b) and \ref{fig6}(c)) corresponding to these large particles may not be representative of the actual number of these large particles in the slurry. This is because the image processing algorithm has been designed to remove objects making contact with the image frame, and larger particles have a higher probability of making contact with the image frame. In the situation where the PSD is to be estimated from image data alone, then this probability will need to be taken into account \cite{ISO13322}. However, since the objective here is just to estimate the aspect ratio of particles of different sizes, then this is not a crucial issue.
 
 The histogram (in Fig. \ref{fig6}(d))\footnote{The uniform bin widths of the histograms in Figs. \ref{fig6}(d) to \ref{fig6}(f) were estimated using the Freedman-Diaconis rule \cite{Izenman1991}. The number of bins were then estimated as 18 bins for PS, 28 bins for COA and 13 bins for Glycine.} for the spherical particles has a dominant mode close to the aspect ratio of 0.85 which is close to the average aspect ratio $r\approx 0.8$ estimated from the average shape descriptor (in Fig. \ref{fig5}(a)) for this slurry. Similarly the modes of the histograms (in Figs. \ref{fig6}(e) and \ref{fig6}(f)) for COA and Glycine occur close to aspect ratios of 0.2 and 0.4 respectively. These values are close to the average aspect ratios $r\approx 0.2$
(for COA) and $r\approx 0.4$ (for Glycine) obtained from their respective shape descriptors in Figs. \ref{fig5}(b) and \ref{fig5}(c).
 
\section{Modelling Chord Length Distribution}
\label{sec5}

The sizes of particles in a slurry can be represented by the equivalent spherical diameter as was done in \cite{Agimelen2015}. However, a characteristic length $L$ could also be used, which can be chosen as the distance between the two extreme points in the particles' geometry.  Since the estimated sizes from the images is $L$, then this metric is used here for consistency with the image data. Once the metric for particle sizes has been chosen, then the PSD can then be expressed in terms of the chosen particle size metric. The PSD $\mathbf{X}$ is related to the CLD $\mathbf{C}$ by means of a convolution function \cite{Hobbel1991,Heinrich2012,Agimelen2015} and the relationship can be written in matrix form as \cite{Agimelen2015} \footnote{Note that the symbol $L$ was used to represent the length of a chord in \cite{Agimelen2015}. However, the symbol $L$ is used to represent the characteristic  length and $s$ the length of a chord in this work. The CLD $\mathbf{C}$ and PSD $\mathbf{\tilde{X}}$ in Eq. \eqref{eq2} have been discretised. As such they are not continuous probability density functions and the term distribution is used in this work for simplicity.}
\begin{equation}
\mathbf{C}(s) = \mathbf{A}(s,L)\mathbf{\tilde{X}}(L),
\label{eq6}
\end{equation}
where $s$ is the chord length, and $\mathbf{\tilde{X}}$ is the length weighted PSD \cite{Agimelen2015} given as 
 \begin{equation}
 \tilde{X}_i = \overline{L}_iX_i, \quad i = 1,2,3,\ldots ,N.
 \label{eq7}
 \end{equation}
The PSD $X_i$ (which is actually a histogram) consists of $N$ bins. The characteristic size $\overline{L}_i$ of particle size bin $i$ is the geometric mean of sizes $L_i$ and $L_{i+1}$ as $\overline{L}_i=\sqrt{L_iL_{i+1}}$. The bin boundaries are calculated as
   \begin{equation}
 L_i = L_{min}\omega^{i-1}, \quad i=1,2, \ldots , N+1
   \label{eq8}
   \end{equation}
   where
      \begin{equation}
     \omega = \left(\frac{L_{max}}{L_{min}}\right)^{\frac{1}{N}},
      \label{eq9}
      \end{equation}
where $L_{min}$ is the left boundary of the first particle size bin and $L_{max}$ is the right boundary of the last particle size bin. 

In previous works \cite{Ruf2000,Worlitschek2005,Li2005n2,Li2013,Li2014,Yu2008} the values of $L_{min}$ and $L_{max}$ were estimated from suitable measurements. However, the technique of estimating $L_{min}$ and $L_{max}$ directly from the bin boundaries of the CLD histogram using a moving window technique  has been demonstrated to yield more accurate results \cite{Agimelen2015}. This window technique is more suitable for estimating the sizes of particles in-line in a process where particle size information is obtained from the  CLD \cite{Agimelen2015}.

The length weighting applied to the PSD $\mathbf{X}$ in Eq. \eqref{eq7} is necessary because the CLD for a population of particles is biased towards particles of larger sizes \cite{Agimelen2015,Hobbel1991,Simmons1999,Vaccaro2006}. The forward problem in Eq. \eqref{eq6} is implemented by considering a chord length histogram $C_j$ of $M$ bins where the characteristic chord length $\overline{s}_j$ of bin $j$ is the geometric mean of the chord lengths of $s_j$ and $s_{j+1}$ as outlined in \cite{Li2005n1,Agimelen2015}. 
 
If the PSD for a population of particles is known, then the CLD can be calculated using Eq. \eqref{eq6}. However, in practical situations, the particle size histogram $X_i$ is not known in advance resulting in the inverse problem of calculating an unknown PSD $X_i$ from a known CLD $C_j$. For this reason, the forward problem in Eq. \eqref{eq6} is reformulated as 
 \begin{equation}
 \mathbf{C}(s) = \mathbf{\tilde{A}}(s,L)\mathbf{X}(L),
 \label{eq10}
 \end{equation}
 where the matrix $\mathbf{\tilde{A}}$ is obtained from matrix $\mathbf{A}$ by multiplying each column of $\mathbf{A}$ by the corresponding particle length as described in \cite{Agimelen2015}.
 
 Each column $i$ of matrix $A_{ji}$ is calculated from the CLD of a single particle (single particle CLD) of length $\overline{L}_i$ and given aspect ratio $r_i$. In the current work, the single particle CLD used in constructing the columns of matrix $\mathbf{A}$ are obtained from the analytical Li and Wilkinson (LW) model \cite{Li2005n1}\footnote{Even though the model by Vaccaro et al. \cite{Vaccaro2006} gave estimates of particle aspect ratios that were closer to the estimates from images in \cite{Agimelen2015}, the LW model is used in this work for the CLD calculation. The reason is that the Vaccaro model is restricted to small values of aspect ratios $r\lesssim 0.4$, whereas the image data in Fig. \ref{fig3} cover aspect ratios of $r\approx 0.1$ to $r\approx 0.9$. The LW model covers the entire range from $r=0$ to $r=1$.}.
  
 The process of calculating the single particle CLD involves computing the relative likelihood of obtaining a chord of length $s$ from a particle of a given length and aspect ratio \cite{Li2005n1}. The LW model gives a probability density function (PDF) which can be used in making this calculation for ellipsoidal shaped 
 particles\footnote{The shapes of the COA and Glycine particles in this work have been approximated as ellipsoids. However, this is only an approximation as Fig. \ref{fig2}(c) clearly shows that the Glycine particles are faceted. The use of ellipsoids to represent faceted objects introduces some discrepancy between the single particle CLDs of both objects (see section 8 of the supplementary information for details). However, the ellipsoid approximation used here is sufficient to illustrate the methods presented here.}. 
 The PDF is derived from an ellipsoid of semi major axis length $a$, semi minor axis length $b$ and aspect ratio $r=b/a$ \cite{Li2005n1}. The LW model gives the probability $p_{\overline{L}_i}(s_j,s_{j+1})$ of obtaining a chord whose length lies between $s_j$ and $s_{j+1}$ from an ellipsoid of characteristic length $\overline{L}_i = 2a_i$ (see section 1 of the supplementary information and \cite{Agimelen2015,Li2005n1} for the mathematical expression for $p_{\overline{L}_i}(s_j,s_{j+1})$). Once the probabilities are calculated, then for each row $j$ of matrix $\mathbf{A}$ the columns are constructed as
        \begin{equation}
          A_j = \left[p_{\overline{L}_1}(s_j,s_{j+1}), p_{\overline{L}_2}(s_j,s_{j+1}), \ldots ,  p_{\overline{L}_i}(s_j,s_{j+1}), \ldots , p_{\overline{L}_N}(s_j,s_{j+1}) \right].
         \label{eq11}
        \end{equation}
 
\section{Incorporating aspect ratio from images}
\label{sec6}

As stated in section \ref{sec5} the calculation of a column of matrix $\mathbf{A}$ requires the characteristic size of the particle size bin corresponding to that column, as well as the aspect ratio $r$ of the particle of that characteristic size. In the previous work \cite{Agimelen2015}, all particles were assumed to have the same aspect ratio, and its value was estimated using an algorithm based solely on CLD data. The single aspect ratio approach will be used in subsection \ref{ssec6_1}, where the single aspect ratio value is estimated from imaging data. This approach is most suitable for the case of spherical particles where the aspect ratios of the individual particles are tightly packed around some mean value. However, for the case of particles where there is a wider spread of aspect ratios, a variation of aspect ratios for different particle sizes can also be used. The corresponding technique is outlined in subsection \ref{ssec6_2}.
  
  \subsection{Method 1: Population CLD with a single aspect ratio}
   \label{ssec6_1}
   
   When the aspect ratios are tightly packed around some mean value as in the case of spherical (or near spherical) particles in Figs. \ref{fig6}(a) and \ref{fig6}(d), it may be desirable to use a mean value for all particles. Although, the mean aspect ratio estimated from the images provides the best estimate based on available data, there is an uncertainty due to sampling limitations and various artefacts discussed above. This necessitates a search (within a suitable confidence interval) around the mean aspect ratio estimated from the images for an aspect ratio which best matches the experimentally measured CLD. This results in the search space being narrowed down leading to significantly lower computation time and less uncertainty in subsequent calculations.
     
    In this section, the transformation matrix $\tilde{\mathbf{A}}$ in Eq. \eqref{eq10} is constructed with a single aspect ratio for all its columns. The aspect ratio $r$ is chosen from a range given as
      \begin{equation}
    r\in[\overline{r} - \mathcal{N}\sigma_{r}, \overline{r} + \mathcal{N}\sigma_{r}],
      \label{eq12}
      \end{equation}
   where $\sigma_{r}$ is the standard deviation of all aspect ratios estimated from images and $\mathcal{N}$ is the number of standard deviations chosen. The purpose of Eq. \eqref{eq12} is to constrain the search space for the inverse problem. The details of the inverse problem will be given in section \ref{sec7}. Hence, Method 1 corresponds to the case described in \cite{Agimelen2015} where each column of matrix $\mathbf{\tilde{A}}$ consists of the single particle CLD of a particle of length 
  $\overline{L}$
   and aspect ratio 
  $r$, 
   where the search space for 
  $r$
   is reduced by means of Eq. \eqref{eq12}.
    
\subsection{Method 2: Population CLD with multiple aspect ratios}
\label{ssec6_2}

\begin{figure}[tbh]
               \centerline{\includegraphics[width=0.38\textwidth]{fig7a.pdf}
               \includegraphics[width=0.62\textwidth]{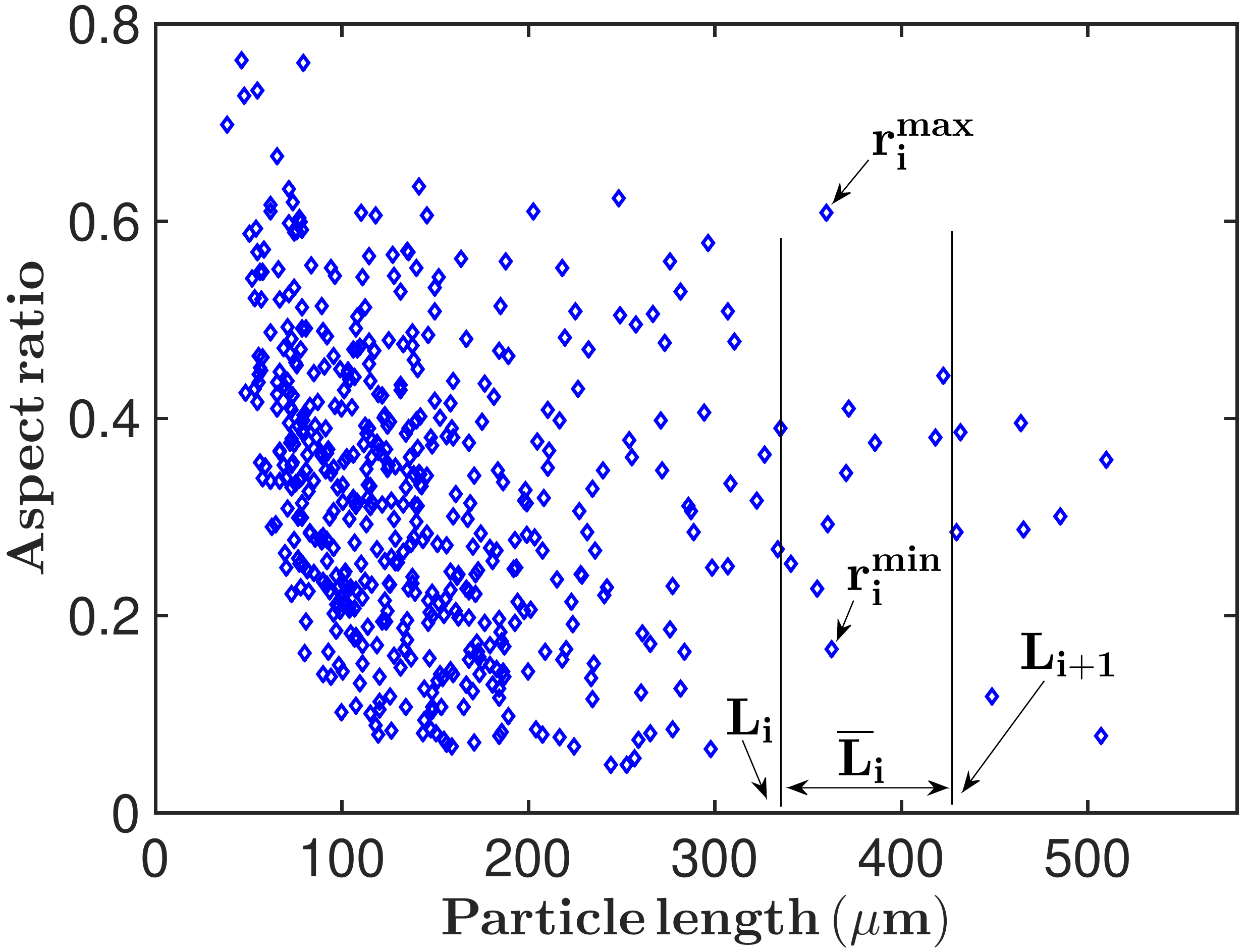}}
               \caption{Left: A schematic to illustrate the assignment of different aspect ratios to particles of the same length. Right: A scatter plot to illustrate the maximum aspect ratio $r_i^{max}$ and minimum aspect ratio $r_i^{min}$ for a particular bin $i$ for calculating the columns (of slice $i$) of the 3 D matrix in Eq. \eqref{eq15}.}
               \label{fig7}
                \end{figure}

The Method 1 presented in the previous subsection assigns the same aspect ratio to all the particles in the slurry. This method is capable of getting reasonable estimates of the PSD. However, to take the variation of particle shape into account, a second method is presented here in which multiple aspect ratios are assigned to particles of the same size. This method is particularly relevant for particles whose shape is needle-like or near rectangular as illustrated on the left of Fig. \ref{fig7}. The second method is outlined below.

In Method 2, the particles in bin $i$ are assigned the same characteristic size $\overline{L}_i$ but different aspect ratios. The aspect ratios assigned to the particles run from $r_i^{min}$ to $r_i^{max}$ as illustrated in Fig. \ref{fig7}. The procedure is to divide the particles in bin $i$ into $N_{r}$ subgroups ($N_r=50$ in this case, see subsection 4.1 of the supplementary information for details), and the particles in subgroup $k$ are assigned the same aspect ratio $r_i^k$. The aspect ratios of the subgroups are uniformly spaced so that the aspect ratios are given as
      \begin{equation}
    r\in \left\{r_i^1=r_i^{min}, r_i^1+\Delta r, \ldots, r_i^k, \ldots, r_i^{N_{r}}=r_i^{max}\right\},
      \label{eq13}
      \end{equation}
where
      \begin{equation}
    \Delta r = \frac{r_i^{max} - r_i^{min}}{N_{r}-1}.
      \label{eq14}
      \end{equation}

However, the aspect ratios $r_i^k$ of the different subgroups need to be weighted by different amounts as the
 numbers of particles with given aspect ratios in the images may not necessarily be uniform from $r_i^{min}$ to $r_i^{max}$.  The data on the right of Fig. \ref{fig7} suggests an approximately uniform distribution of the aspect ratios from $r_i^{min}$ to $r_i^{max}$, hence the aspect ratios were assigned to each of the subgroup $k$ with equal weight in this case. 

Once aspect ratios have been assigned to different particle size bins\footnote{Details of the technique for combining the aspect ratios with the windowing technique developed in \cite{Agimelen2015} can be found in section 5 of the supplementary information.}, then a 3D probability array $\mathcal{A}_{jki}$ is constructed. Each slice $i$ of the array corresponds to a particle of size $\overline{L}_i$, and each column $k$ of slice $i$ contains the probabilities $p_{\overline{L}_i^k}(s_j,s_{j+1})$ of obtaining chords whose lengths lie between $s_j$ and $s_{j+1}$ from a particle of size $\overline{L}_i$ and aspect ratio $r_i^k$. Hence the 3D array consists of $M$ rows, $N_{r}$ columns and $N$ slices. The transformation matrix $A_{ji}$ in Eq. \eqref{eq10} is then obtained from the 3D array by averaging over the slices as
   \begin{equation}
 A_{ji} = \frac{1}{N_{r}}\sum_{k=1}^{N_{r}}{\mathcal{A}_{jki}}.
   \label{eq15}
   \end{equation}
This simple averaging is carried out since the aspect ratios $r_i^k$ are assigned to the subgroups $k$ with equal weights. This is the simplest way to construct the matrix $A_{ji}$ from the 3D array $\mathcal{A}_{jki}$, and this approach is supported by the data from the images as shown on the right of Fig. \ref{fig7}. It is possible to introduce a probability distribution for the aspect ratios assigned to the subgroups, but this simple approach has been used here for the purpose of illustrating the technique. Once the matrix $A_{ji}$ has been constructed, then the forward problem in Eq. \eqref{eq10} can be solved for a given PSD $X_i$.
 
 \section{PSD Estimation}
\label{sec7}

As mentioned in section \ref{sec5}, the problem encountered in practical situations is the estimation of the PSD $X_i$ corresponding to an experimentally measured CLD $C_j^{\ast}$. This is the inverse problem to the forward problem given in Eq. \eqref{eq10}. One of the key steps in the process of the PSD estimation is to determine the transformation matrix $\mathbf{\tilde{A}}$ in Eq. \eqref{eq10} as accurately as possible. The level of accuracy of the matrix $\mathbf{\tilde{A}}$ depends on the values of $L_{min}$ and $L_{max}$ as well as the aspect ratio(s) used in calculating its columns.

To determine the best possible values of $L_{min}$ and $L_{max}$, the forward problem in Eq. \eqref{eq10} is rewritten as
   \begin{equation}
  \mathbf{C} = \mathbf{\tilde{A}X}+\mbox{\boldmath{$\epsilon$}},
   \label{eq16}
   \end{equation}
  where $\mbox{\boldmath{$\epsilon$}}$ is an additive error between the calculated and experimentally measured CLD. Then for given values of $L_{min}$, $L_{max}$, values of the fitting parameter $\gamma$ are found\footnote{The Levenberg-Marquardt (LM) algorithm as implemented in Matlab was used in this work to solve the optimisation problem here. The PSD $X_i$ is estimated by means of the parameter $\gamma_i$. An initial value of $\gamma_i$ is passed on to the LM algorithm which then searches for the optimum value of $\gamma_i$ to fit the given CLD. Since the PSD $X_i$ is defined as an exponential function in Eq. 18, then the parameter $\gamma_i$ can take values in the interval $(-\infty, +\infty)$ and still give $X_i \geq 0$. This implies that the non negativity requirement on the PSD is maintained by the formulation of $X_i$ in Eq. 18. Therefore the LM algorithm was run without the use of lower or upper bounds as the parameter $\gamma_i$ is defined in $(-\infty, +\infty)$.} which minimises the objective function $f_1$ given as
    \begin{equation}
   f_1 = \sum_{j=1}^{M}{\left[C_j^{\ast} - \sum_{i=1}^N{\tilde{A}_{ji}X_i}\right]^2},
    \label{eq17}
    \end{equation}
  where
          \begin{equation}
         X_i = e^{\gamma_i}, i=1,2,3, \ldots, N.
          \label{eq18}
          \end{equation}

A trial solution of $\gamma_i=0$ was used in the calculation of the vector $X_i$ from Eq. \eqref{eq17}. Once the solution vector $X_i$ is obtained, then it is used to solve the forward problem in Eq. \eqref{eq10} to obtain a calculated CLD $C_j,\, j=1,2,\ldots,M$, where $M$ is the number of bins in the CLD histogram\footnote{The value of $M=100$ was used in all the calculations here to mimic the number of bins set in the FBRM G400 sensor. The values of $N=70$ and $N=50$ were used in Methods 1 and 2 respectively. See subsection 4.2 of the supplementary information  for more details on the choice of the values of $N$ for the two methods.}. The procedure is repeated until the optimum values of $L_{min}$ and $L_{max}$ are found for which there is the best match between the calculated and experimentally measured CLD.

The objective function $f_1$ given in Eq. \eqref{eq17} is suitable for estimating the optimum values of $L_{min}$ and $L_{max}$ whether the same aspect ratio is assigned to all particles or a distribution of aspect ratios is assigned to particles of the same size. In the case where a single aspect ratio is assigned to all particles, the objective function $f_1$ is not suitable for picking out the best aspect ratio within the confidence interval of aspect ratios. This task is accomplished with another objective function $f_2$ (to be introduced in subsection \ref{ssec7_1}). The calculation for estimating the best aspect ratio (within the confidence interval of aspect ratios) using the objective function $f_2$ is carried out with the values of $L_{min}$ and $L_{max}$ estimated with the objective function $f_1$. However, in the other case where a distribution of aspect ratios is assigned to particles of the same size, the objective function $f_2$ is not used as the problem of determining the best aspect ratio has been removed.

Once the optimum values of $L_{min}$ and $L_{max}$ and (in the case of Method 1) the best aspect ratio have been estimated, then the PSDs (both number and volume based) can be calculated. However, these PSDs may not be reasonably smooth, showing non-physical oscillations as is often the case when solving ill-posed problems. In such cases, a third objective function $f_3$ (to be introduced in subsection \ref{ssec7_3}) is used to calculate smooth PSDs. The calculation is carried out using the optimum values of $L_{min}$ and $L_{max}$ obtained with the objective function $f_1$ and in the case of Method 1, the best aspect ratio obtained with the objective function $f_2$. The calculation of the smooth PSDs with the objective function $f_3$ is done using suitable criteria described in subsection 4.4 of the supplementary information.

As stated above, the objective function $f_1$ is used to obtain the optimum values of $L_{min}$ and $L_{max}$ for both Methods 1 and 2. A given pair of $L_{min}$ and $L_{max}$ are said to be optimum when the corresponding calculated CLD $\mathbf{C}$ has the best match with the experimentally measured CLD $\mathbf{C}^{\ast}$. The level of agreement between the calculated and experimentally measured CLD is assessed by computing the $L_2$ norm
          \begin{equation}
         \|\mathbf{C}^{\ast} - \mathbf{C}\| = \sqrt{f_1}. 
          \label{eq19}
          \end{equation}
 The values of $L_{min}$ and $L_{max}$ for which the $L_2$ norm in Eq. \eqref{eq19} reaches a minimum are chosen as the optimum values.   

\subsection{PSD estimation for Method 1}
\label{ssec7_1}

In Method 1 the search for the optimum values of $L_{min}$ and $L_{max}$ using the objective function $f_1$ is done at each aspect ratio within the confidence interval in Eq. \eqref{eq12}. However, for particles of a given shape, the $L_2$ norm in Eq. \eqref{eq19} initially decreases with increasing aspect ratio and then becomes level (see subsection 4.3 of the supplementary information for details). This leads to non-uniqueness in determining the optimum value of $r$ \cite{Agimelen2015}. This problem of non-uniqueness is removed by using a modified objective function $f_2$ which contains a penalty term to control the size of the calculated PSD vector as 
  \begin{equation}
 f_2 = \sum_{j=1}^{M}{\left[C_j^{\ast} - \sum_{i=1}^N{\tilde{A}_{ji}X_i}\right]^2} + \lambda_1\sum_{i=1}^N{X_i^2},
  \label{eq20}
  \end{equation}
where the parameter $\lambda_1$ sets the level of imposed penalty. The value of $\lambda_1$ is chosen by comparing the magnitudes of the terms in Eq. \eqref{eq20} (see subsection 4.3 of the supplementary information). The aspect ratio at which the objective function $f_2$ reaches a minimum is then chosen as the optimum. 

The solution vector (which is a number based PSD) obtained from Eq. \eqref{eq20} is not necessarily smooth as the penalty imposed on the solution vector only restricts its magnitude. With this penalty function, the LM algorithm could settle on a solution vector that contains some local fluctuations but whose value of $f_2$ is slightly less than a nearby solution that is smooth. For this reason, a new objective function $f_3$ (see subsection \ref{ssec7_3} for details) which contains a penalty term to control the second derivative (to improve the smoothness of the solution vector) of the solution vector is used to estimate a number based PSD whose corresponding CLD is compared with the experimental data.

\subsection{PSD estimation for Method 2}
\label{ssec7_2}

In Method 2 (described in subsection \ref{ssec6_2}) particles of different characteristic sizes $\overline{L}_i$ are assigned a range of aspect ratios as outlined in subsection \ref{ssec6_2}. This eliminates the need to search for the best global aspect ratio which is the situation in Method 1. Hence in Method 2, it is only necessary to search for the best values of $L_{min}$ and $L_{max}$ using the objective function $f_1$. Once the optimum transformation matrix is obtained using the objective function $f_1$, then the corresponding smoothed solution is obtained with the objective function $f_3$ (given in Eq. \eqref{eq21}).

\subsection{Volume based PSD}
\label{ssec7_3}

It is often necessary to recast the PSD $X_i$ (which is number based) in Eq. \eqref{eq17} as a volume based PSD since most instruments for measuring PSD give the data in terms of a volume based PSD. A new technique which allows suitable penalties to be imposed on the calculated volume based PSD $\mathbf{X}^v$ was introduced in \cite{Agimelen2015}. In the current work, a smoothing penalty (referred to in subsection \ref{ssec7_1}, see also \cite{Roths2001}) is imposed. This is because the estimated volume based PSD $\mathbf{X}^v$ may contain significant non-physical oscillations even though the corresponding number based PSD only contains none or minor fluctuations. The objective function $f_3$ (see Eq. \eqref{eq21}) used to impose smoothness on the volume based PSD can also be used to obtain a smooth number based PSD. Hence the function $f_3$ is given in Eq. \eqref{eq21} in terms of generic quantities depending on whether the number based or volume based PSD is being computed. 

The function $f_3$ is given as 
   \begin{equation}
  f_3 = \sum_{j=1}^{M}{\left[C_j^+ - \sum_{i=1}^N{A^+_{ji}X^+_i}\right]^2} + \lambda_2\sum_{i=1}^{N}{\left[\nabla_h^2\left[X^+_i\right]\right]^2}.
   \label{eq21}
   \end{equation}
 In the case of a number based PSD, the CLD $C^+_j=C_j^{\ast}$ (the experimentally measured CLD), the matrix $A^+_{ji}=\tilde{A}_{ji}$\footnote{The matrix $\tilde{A}_{ij}$ is the transformation matrix in the forward problem in Eq. \eqref{eq10} initially estimated with the objective function $f_1$ in Eq. \eqref{eq17}. The smoothed solution is then calculated using the function $f_3$ in Eq. \eqref{eq21} at a unique aspect ratio determined using the function $f_2$ in Eq. \eqref{eq20}. The solution vector $X_i$ from Eq. \eqref{eq17} is used to construct a trial solution as $\gamma_i = \ln(X_i)$ in the case of the number based PSD. For the volume based PSD, the corresponding number based PSD from Eq. \eqref{eq21} is used to construct a trial solution as described in section 3 of the supplementary information.}, and the solution vector $X_i^+=X_i$. However, in the case of the volume based PSD, the vector $X_i^+=X^v_i$ (in the case where smoothing is not required\footnote{The vector $X_i^+$ is defined as an exponential function of a fitting parameter similar to Eq. \eqref{eq18} for $X_i$. The optimisation is then performed to obtain the optimum value of the fitting parameter using the LM algorithm similar to the case of $X_i$ in Eq. \eqref{eq18}.}, then the volume based PSD $X^v_i$ is obtained from an objective function similar to $f_1$ in Eq. \eqref{eq17} as described in section 3 of the supplementary information, otherwise Eq. \eqref{eq21} is used), the matrix $A^+_{ji}$ will be scaled accordingly (see section 3 of the supplementary information for details), and the vector $C_j^+=\hat{C}^{\ast}$ where the transformed CLD $\hat{C}^{\ast}$ is calculated as 
  \begin{equation}
\hat{C}^{\ast}_j = \sum_{i=1}^{N}{\tilde{A}_{ji}}\hat{X}_i,
  \label{eq22}
  \end{equation}
  where
    \begin{equation}
   \hat{X}_i=\frac{X_i}{\sum_{i=1}^{N}{X_i}}.
    \label{eq23}
    \end{equation}
 The operator $\nabla_h^2$ is a finite difference approximation to the second derivative of the vector $X^+_i$ given as \cite{Veldman1992}\footnote{The form of the central difference approximation to the second derivative of the vector $X^+_i$ given in Eq. \eqref{eq18} is necessary since the grid for $\overline{L}$ is non uniform as seen in Eq. \eqref{eq8}.}
  \begin{equation}
  \nabla_h^2 = \frac{h_{-}X^+_{i+1} - (h_{+} + h_{-})X^+_i 
  + h_{+}X^+_{i-1}}{\frac{1}{2}h_{+}h_{-}(h_{+} + h_{-})},
   \label{eq24}
   \end{equation}
      where
             \begin{align}
            	h_{-} & = \overline{L}_i - \overline{L}_{i-1} \nonumber \\
            	h_{+} & = \overline{L}_{i+1} - \overline{L}_i
            	\label{eq25}
            		\end{align}
   and $X^+_i$ has been treated as a function of the characteristic particle size $\overline{L}$. The parameter $\lambda_2$ sets the level of penalty imposed on the second derivative of $X^+_i$. If the value of $\lambda_2$ is sufficiently large, then the penalty on the second derivative causes the LM algorithm to search for a solution vector which is smooth thereby avoiding solutions with localised oscillations (see subsection 4.4 of the supplementary information).
    
  The volume based PSD obtained from Eq. \eqref{eq21} is normalised and converted to a probability density distribution as
      \begin{equation}
	\tilde{X}^v_i = \frac{\overline{X}_i^v}{(L_{i+1} - L_i)\sum_{i=1}^N{\overline{X}_i^v}}.
      \label{eq26}
      \end{equation}

\section{Results and Discussion}
\label{sec8}

The results obtained with the two methods outlined in sections \ref{sec6}  and \ref{sec7} are presented in this section. More details of the choice of parameter values are presented in the supplementary information.

     \begin{figure}[tbh]
    \centerline{\includegraphics[width=0.8\textwidth]{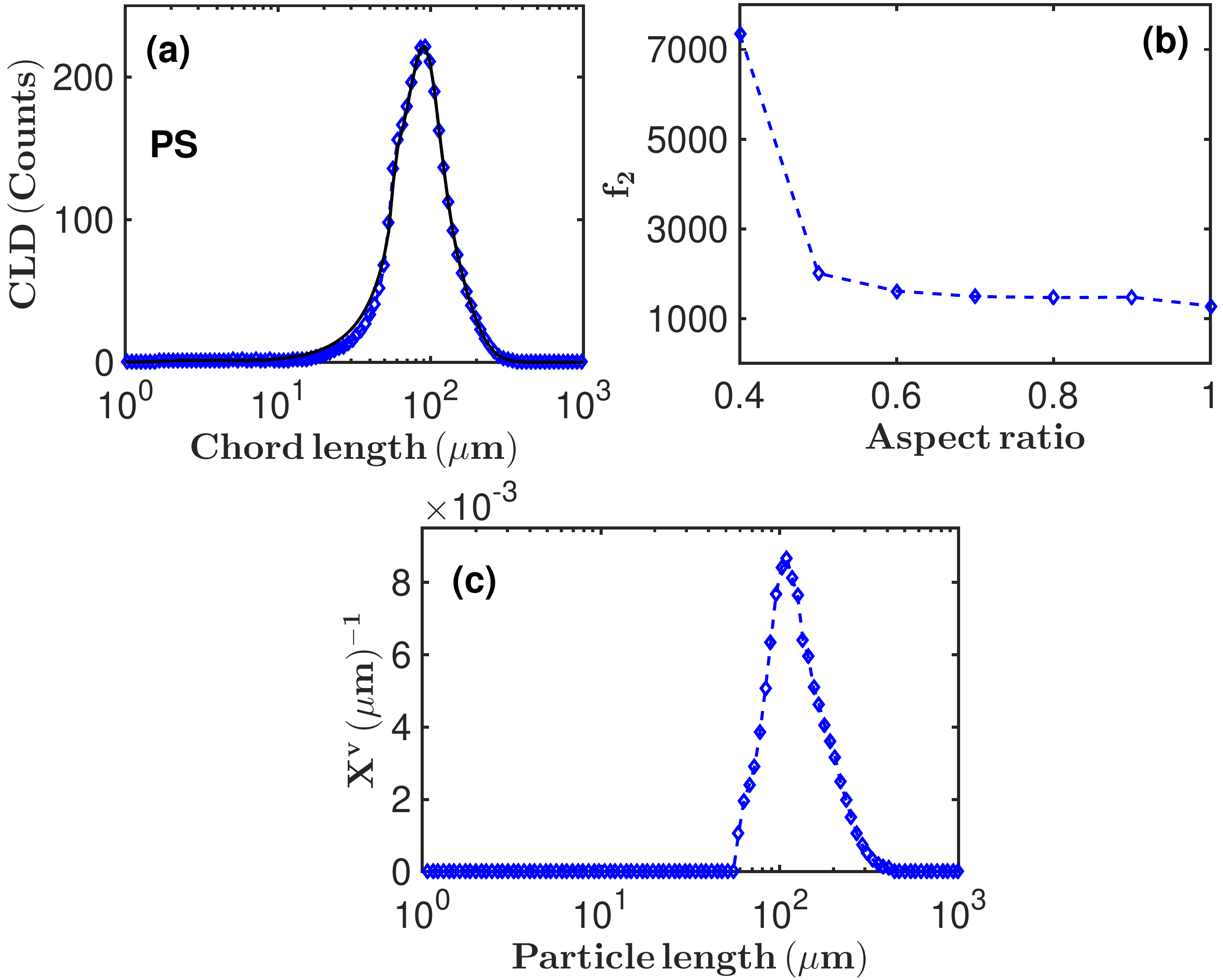}}
    \caption{(a)\,Experimentally measured (symbols) and calculated (solid line) CLD for PS. The calculated CLD was obtained by solving the forward problem in Eq. \eqref{eq10} using the number based PSD calculated with the objective function $f_3$ (using $\lambda_2 = 10^{-2}$) in Eq. \eqref{eq21}. The calculation was done at the aspect ratio $r=1$ where the objective function $f_2$ (in Eq. \eqref{eq20} using $\lambda_1\approx0.95$) reaches a minimum. The matrix $\mathbf{\tilde{A}}$ in Eq. \eqref{eq10} was calculated by Method 1 (described in subsection \ref{ssec6_1}). (b)\,The objective function $f_2$ (using $\lambda_1\approx0.95$) in Eq. \eqref{eq20} for different aspect ratios for PS. (c)Calculated (by Method 1) volume based PSD for the PS sample. The volume based PSD was calculated using the objective function $f_2$ (at $\lambda_1\approx 0.95$). The objective function $f_3$ was not used in the calculation of the volume based PSD in this case as smoothing was not required.}
    \label{fig8}
     \end{figure}
 
\subsection{Results from Method 1}
\label{ssec8-1}

Figure \ref{fig8}(b) shows the objective function $f_2$ as a function of the aspect ratio $r$ for PS. The function reaches a minimum at $r=1$ suggesting spherical particles. This is consistent with the shape of the particles in Fig. \ref{fig2}(a) and the mean aspect ratio of $\overline{r}\approx 0.8$ obtained from the shape descriptor in Fig. \ref{fig5}(d) for this sample. This is also in agreement with the histogram in Fig. \ref{fig6}(d) which suggests that the majority of the particles in the sample are near spherical. Hence the aspect ratio predicted with Method 1 gives a reasonable description of the shape of the particles in the population as previously established \cite{Agimelen2015}.

The calculation in Fig. \ref{fig8}(b) was done with $\lambda_1\approx0.95$ in Eq. \eqref{eq20} (see subsections 4.3 and 4.4 of the supplementary information for details on how the values of $\lambda_1$ and $\lambda_2$ are chosen in Eqs. \eqref{eq20} and \eqref{eq21}) using the optimum transformation matrix from Eq. \eqref{eq17}. Using this optimum transformation matrix and $r=1$, a number based PSD is calculated from Eq. \eqref{eq21} with the smoothness penalty set by $\lambda_2 = 10^{-2}$. The CLD corresponding to this number based PSD is shown by the solid line in Fig. \ref{fig8}(a). Furthermore, the volume based PSD (calculated at $r=1$) is shown in Fig. \ref{fig8}(c). In this case, the volume based PSD was calculated from the objective function $f_2$ (with the CLD $C^{\ast}_j$ replaced with transformed CLD $\hat{C}^{\ast}_j$ and the matrix $\tilde{A}_{ji}$ rescaled as described in section 3 of the supplementary information) at $\lambda_1\approx0.95$. The objective function $f_3$ was not used in calculating the volume based PSD in this case as smoothing was not required. 

The calculated CLD in Fig. \ref{fig8}(a) has a near perfect match with the measured CLD for PS which is shown by the symbols in Fig. \ref{fig8}(a). The calculations in Fig \ref{fig8} were done at a value of $\mathcal{N}=2$ (where $\mathcal{N}$ is defined in Eq. \eqref{eq12}). This value was sufficient to give a wide enough range of aspect ratios to find a good match to the experimentally measured CLD. If the value of $\mathcal{N}$ is not large enough, then the calculated CLD may not match the experimentally measured CLD as
the particles do not have exactly the same shape and Method 1 only uses a single aspect ratio to describe the shape of all the particles in the population. The single aspect ratio chosen will then not be representative of all the particles in the population. However, the imaging data narrows down the search space for a representative aspect ratio, and hence reduce the risk of predicting an unreliable aspect ratio.

\begin{figure}[tbh]
        \centerline{\includegraphics[width=0.8\textwidth]{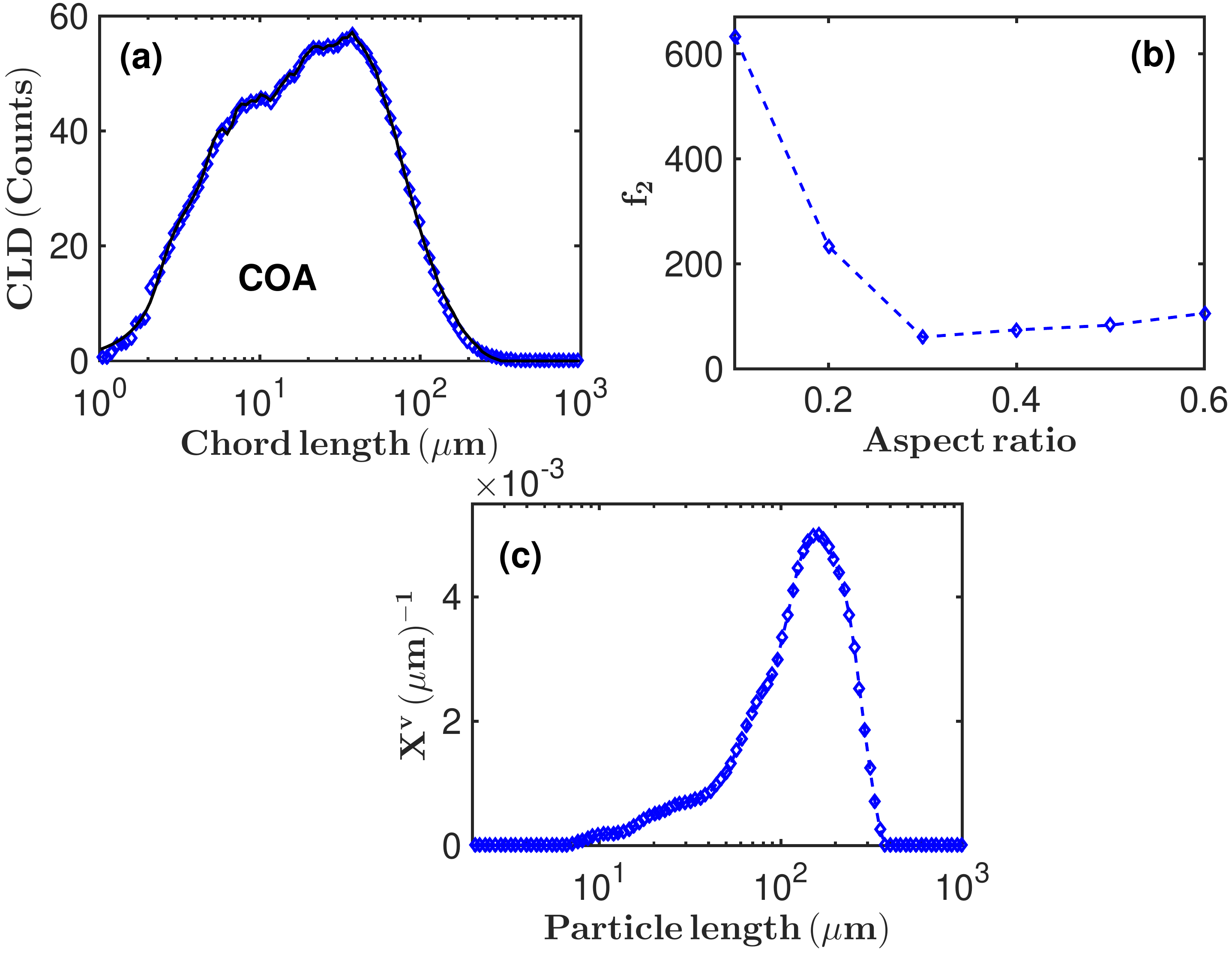}}
        \caption{Similar to Fig. \ref{fig8} but for COA. The calculation was done at the aspect ratio $r=0.3$ where the objective function $f_2$ (using $\lambda_1=0.54$) reaches a minimum in (b). The number based PSD (used for calculating the CLD) was calculated using $\lambda_2= 0.05$ in Eq. \eqref{eq21} while the volume based PSD (shown in  (c)) was calculated using $\lambda_2=10^{-7}$ in Eq. \eqref{eq21}.}
        \label{fig9}
         \end{figure}

\begin{figure}[tbh]
                \centerline{\includegraphics[width=0.8\textwidth]{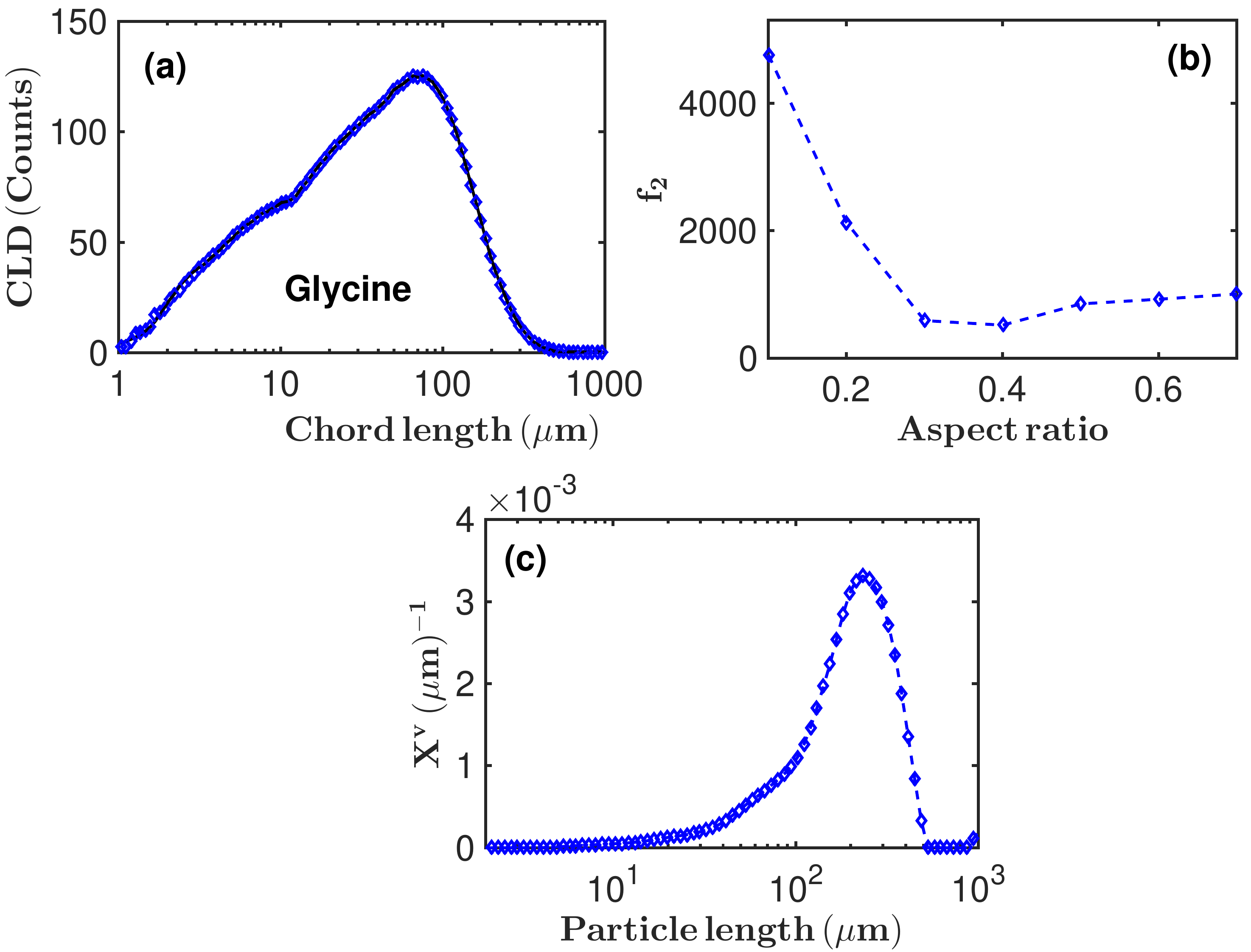}}
                \caption{Similar to Fig. \ref{fig9} but for Glycine with calculations done at $r=0.4$. The values of $\lambda_1 = 0.41$ and $\lambda_2 = 0.01$ were used in Eqs. \eqref{eq20} and \eqref{eq21} respectively for the number and based PSD. The value of $\lambda_2 =10^{-6}$ was used in Eq. \eqref{eq21} for the volume based PSD.}
                \label{fig10}
                 \end{figure}

Figures \ref{fig9} and \ref{fig10} are similar to Fig. \ref{fig8} but for COA and Glycine respectively. Figure \ref{fig9}(b) shows the objective function $f_2$ (in Eq. \eqref{eq20}) with aspect ratio for COA.  The function $f_2$ in Fig. \ref{fig9}(b) predicts an aspect ratio $r=0.3$ for COA. This is reasonable when compared with crystals in Fig. \ref{fig2}(b) and the shape descriptor in Fig. \ref{fig5}(b). Also the mode of the histogram in Fig. \ref{fig6}(e) is close to the aspect ratio $r=0.3$. The predicted aspect ratio of $r=0.3$ in Fig. \ref{fig9}(b) is also close to the estimated aspect ratio of $r\approx0.2$ from the shape descriptor in Fig. \ref{fig5}(b). Furthermore, the calculated (in a manner similar to the case of PS in Fig \ref{fig8}(a)) CLD for COA shown by the solid line in Fig. \ref{fig9}(a) has a near perfect match with the measured CLD for the sample shown by the symbols in Fig. \ref{fig9}(a). The calculations were done with $\mathcal{N}=4$ in Eq. \eqref{eq12}. 

                                 \begin{figure}[tbh]
                                \centerline{\includegraphics[width=0.8\textwidth]{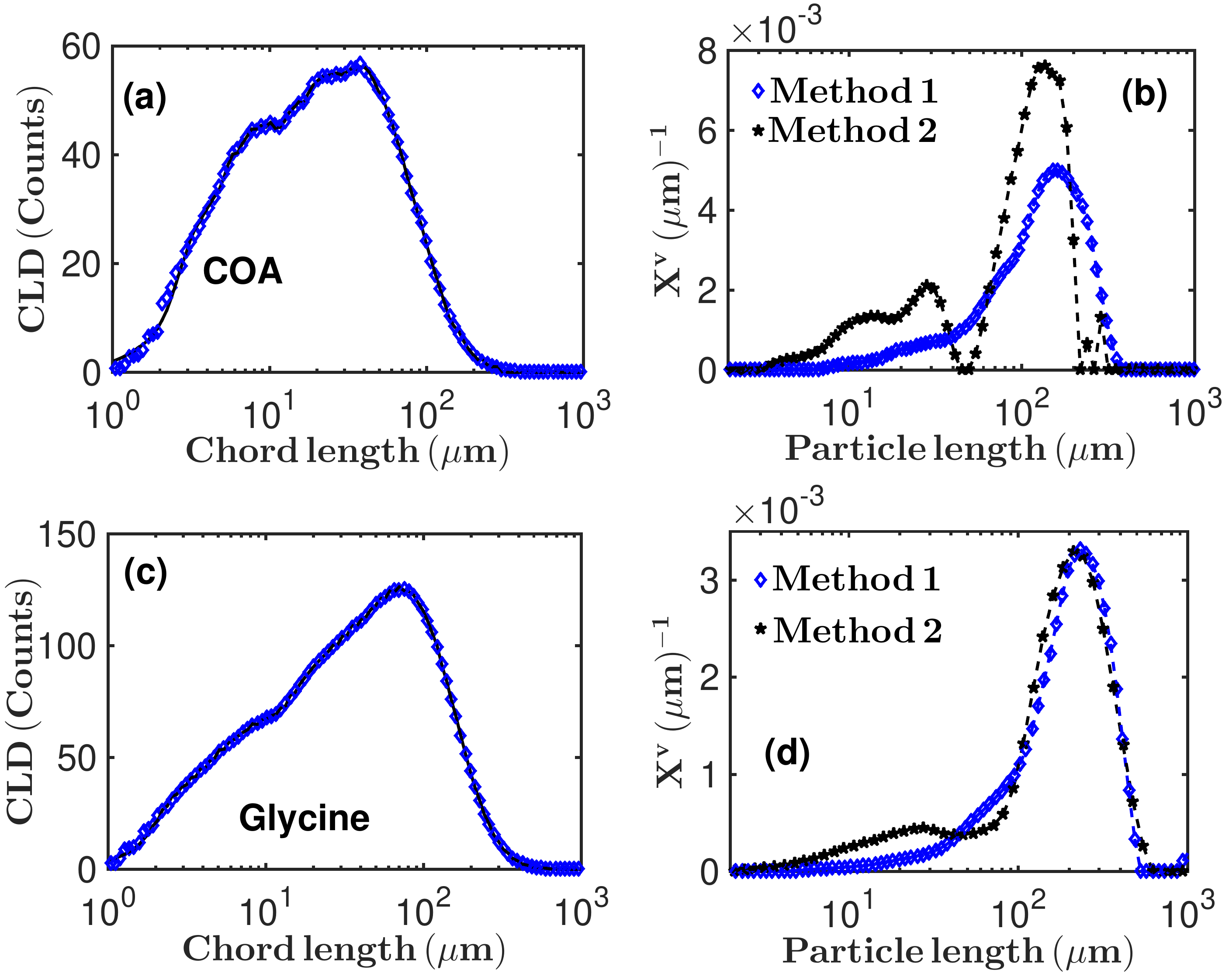}}
                                \caption{(a)\,Experimentally measured (symbols) and calculated (solid line) for COA. The calculated CLD was obtained by solving the forward problem in Eq. \eqref{eq10}, where the matrix $\mathbf{\tilde{A}}$ was calculated by Method 2 as outlined in subsection \ref{ssec6_2}. The number based PSD used in solving the forward problem was obtained from the objective function $f_3$ in Eq. \eqref{eq21} for $\lambda_2=0.1$. (b)\,The blue diamonds are the calculated (by Method 1) volume based PSD for COA shown in Fig. \ref{fig9}(c). The black asterisks are the volume based PSD calculated by Method 2 for COA. The volume based PSD by Method 2 was calculated from Eq. \eqref{eq21} at $\lambda_2=3\times10^{-6}$. (c)\,Similar to (a) but for Glycine. The number based PSD was obtained from Eq. \eqref{eq21} at $\lambda_2=0.01$. (d)\,Similar to (b) but for Glycine, where the value of $\lambda_2=10^{-5}$ has been used for the volume based PSD calculated by Method 2.}
                                \label{fig11}
                                 \end{figure}

The volume based PSD for COA (calculated in a manner similar to the case of PS in Fig. \ref{fig8}(c)) is shown in Fig. \ref{fig9}(c). The calculated PSD in Fig. \ref{fig9}(c) has a left shoulder extending to needle lengths of about 10$\mu$m. This gives a hint of the presence of a significant number of short needles in the COA sample. Some of these short needles can be seen in the image of Fig. \ref{fig2}(b). 

The calculations for Glycine in Fig. \ref{fig10} are similar to the cases of PS and COA in Figs. \ref{fig8} and \ref{fig9} respectively. The predicted particle shape represented by $r=0.4$ in Fig. \ref{fig10}(b) is consistent with the particles in Fig. \ref{fig2}(c) and shape descriptor (which yields $r\approx 0.4$) in Fig. \ref{fig5}(c) as well as the histogram in Fig. \ref{fig6}(f). The calculated CLD (solid line Fig. \ref{fig10}(a)) also matches the experimentally measured CLD for the Glycine sample (symbols in Fig. \ref{fig10}(a)). The calculated volume based PSD for this sample at $r=0.4$ is shown in Fig. \ref{fig10}(c).
The volume based PSD in Fig. \ref{fig10}(c) also has a left shoulder extending to about 10$\mu$m similar to the case of COA in Fig. \ref{fig9}(c). The calculations in Fig. \ref{fig10} for Glycine were done with $\mathcal{N}=2$ which was sufficient to get a good match for the measured CLD.

\subsection{Results from Method 2}
\label{ssec8-2}

The aspect ratios in Figs. \ref{fig6}(b) and \ref{fig6}(c) show a spread over a significant range of particle sizes. Hence the technique referred to as Method 2 in subsections \ref{ssec6_2} and \ref{ssec7_2} was also applied in the analysis of the data from COA and Glycine.

The solid line in Fig. \ref{fig11}(a) shows the calculated CLD for COA using Method 2. The calculation was done by searching for the optimum values of $L_{min}$ and $L_{max}$ while constructing different transformation matrices as outlined in subsection \ref{ssec6_2}. The search for the optimum values of $L_{min}$ and $L_{max}$ (and hence the optimum transformation matrix) is done by minimising the objective function $f_1$ in Eq. \eqref{eq17}. Once the optimum transformation matrix is found, then a number based PSD is calculated by minimising the objective function $f_3$ (with $\lambda_2 = 0.1$) in Eq. \eqref{eq21}. The CLD corresponding to this number based PSD is shown by the solid line in Fig. \ref{fig11}(a). The transformed CLD $\hat{C}_j^{\ast}$ in Eq. \eqref{eq22} is calculated using the optimum transformation matrix and the number based PSD obtained with Eq. \eqref{eq17}.

The calculated CLD in Fig. \ref{fig11}(a) (solid line) has a near perfect match with the experimentally measured CLD (shown by the symbols in Fig. \ref{fig11}(a)) for COA. This is similar to the situation in Fig. \ref{fig9}(a) where the calculation was done with Method 1. The degree of agreement of the calculated CLD in Fig. \ref{fig11}(a) with the experimentally measured CLD demonstrates the level of accuracy that can be achieved with Method 2. Note that the aspect ratios of each of the subgroups of each bin (in Fig. \ref{fig7}) were  assigned equal weights; a simple approach that is sufficient for reasonable results in this case.
The volume based PSD for COA (obtained by Method 2) suggests particle sizes from about 3$\mu$m to about 400$\mu$m (Fig. \ref{fig11}(b)). This is close to the prediction of particle sizes from about 7$\mu$m to about 400$\mu$m by Method 1. Even though the ranges of particle sizes predicted by both methods are close, Method 2 has the advantage that
aspect ratio is not used as a fitting parameter which removes the issue of estimating the optimum aspect ratio from the problem. The aspect ratio is assumed to vary according to imaging data available.
 
Although the particle sizes estimated from these 2D images are not very accurate because of the focusing problem highlighted earlier, a comparison of the estimated PSD from the images with the volume based PSD obtained by both methods can still be made. This comparison shows good agreement of the estimated volume based PSD from the images with the volume based PSD estimated by Methods 1 and 2. Details are given in section 6 of the supplementary information.
 The peak of the volume based PSD obtained by Method 2 is higher than that obtained by Method 1 in this case. This is because the volume based PSD by Method 2 is slightly narrower within the size range of about 50$\mu$m to about 200$\mu$m (Fig. \ref{fig11}(b)) so that the main peak gets higher to satisfy the normalisation constraint in Eq. \eqref{eq26}. The main peak is accompanied by a smaller peak 
at a particle length close to 30$\mu$m (Fig. \ref{fig11}(b)) suggesting a bimodal distribution for the COA particles. 
 However, this feature of a bimodal distribution is not picked up by Method 1 (Fig. \ref{fig11}(b)). This could be because Method 2 is more efficient in picking out bimodal distributions in a population of particles where there is a variation of aspect ratio for particles of different sizes (see section 7 of the supplementary information for details) than Method 1.

The situation for Glycine is similar to that of COA. The solid line in Fig. \ref{fig11}(c) shows the calculated (calculated in a manner similar to the case of COA in Fig. \ref{fig11}(a)) CLD for Glycine. The calculated CLD in Fig. \ref{fig11}(c) also has a near perfect match with the experimentally measured (symbols) CLD in Fig. \ref{fig11}(c). This is similar to the case of COA in Fig. \ref{fig11}(a). The calculated volume based PSD (by Method 2) in Fig. \ref{fig11}(d) for Glycine also covers about the same range of particle sizes as in the case of Method 1 (Fig. \ref{fig11}(d)) with the PSD very similar from both methods. 

The volume based PSD for COA obtained by the two Methods (Fig. \ref{fig11}(b)) cover a size range of 
$\leq 10\mu$m
to about 400$\mu$m, while the CLD data for COA (in Fig. \ref{fig11}(a)) shows a maximum chord length of about 300$\mu$m\footnote{Note that CLD is number based and therefore much less sensitive to presence of a small number of large particles.}. The PSD in Fig. \ref{fig11}(b) agrees with the image data in Fig. \ref{fig6}(b) for COA where the scatter plot is dense in the region between about 30$\mu$m to about 300$\mu$m with a small number of particles of sizes $\gtrsim 300\mu$m. Particles of small sizes below about 30$\mu$m are not picked up by the image processing algorithm because objects smaller than that are rejected by the algorithm to reduce the risk of processing background noise as real objects. This is the reason why particles of sizes $\lesssim 30\mu$m do not contribute to the scatter plot of Fig. \ref{fig6}(b) even though the volume based PSD for COA in Fig. \ref{fig11}(b) suggests the presence of these particles.

The situation with Glycine is similar to that of COA as seen in Fig. \ref{fig11}(d). The volume based PSD obtained by both methods cover a size range from about 10$\mu$m to about 500$\mu$m in agreement with the CLD for Glycine (in Fig. \ref{fig11}(c)) which shows the longest chord to be about 400$\mu$m. The data in Fig. \ref{fig11}(d) also agrees with the image data in Fig. \ref{fig6}(c) which shows particle sizes up to about 500$\mu$m. There may be a larger number of large particles (of sizes close to 500$\mu$m) in the Glycine slurry than in the COA slurry so that their contribution to the CLD is more significant.

\section{Conclusions}
\label{sec9}

Two different methods have been developed to constrain the search space of aspect ratio(s) for particle size estimation using CLD and imaging data. Both methods estimate aspect ratio from images and then use the information in the estimation of aspect ratio and/or PSD from CLD data.

In the first method, the PSD estimation from CLD data is carried out using a single representative aspect ratio for all the particles in the slurry. However, the search space for this representative aspect ratio is reduced by means of data from the images of the particles captured in-line during the process. This reduces the risk of predicting an aspect ratio which is not representative of the particles in the slurry, and hence an unreliable PSD. 

In the second method, a range of aspect ratios (also estimated from the images of the particles captured in-line) is assigned to particles of different sizes. This takes aspect ratio estimation out of the problem, and hence eliminates the risk of estimating a PSD at an aspect ratio which is not representative of the particles in the slurry.

The techniques presented in this work have been developed to be applied in situ, and an in-line imaging tool has been used in this work. The currently available in-line imaging tools are not suitable (when used on their own) for obtaining accurate PSD and aspect ratio due to various issues outlined in the text. The limitations of using images from these in-line tools alone to get aspect ratio and/or PSD estimates also show up in the large error bars in Figs. 14(d) to 14(f) in subsection 4.6 of the supplementary information. Hence the methods presented here combine imaging and CLD data obtained in-line to obtain more robust estimates of PSD and aspect ratio. Note that the methods presented here can be applied to combine CLD with imaging captured with any in situ tools. The images need to be of sufficient quality so that aspect ratio information can be obtained from them using a suitable image processing algorithm.

\section*{Acknowledgement}

This work was performed within the UK EPSRC funded project \\
 (EP/K014250/1) `Intelligent Decision Support and Control Technologies for Continuous Manufacturing and Crystallisation of Pharmaceuticals and Fine Chemicals' (ICT-CMAC). The authors would like to acknowledge financial support from EPSRC, AstraZeneca and GSK. The authors are also grateful for useful discussions with industrial partners from AstraZeneca, GSK, Mettler-Toledo, Perceptive Engineering and Process Systems Enterprise. The authors would also like to acknowledge discussions/suggestions from Alison Nordon and Jaclyn Dunn.
 
\newpage

\setcounter{section}{0}

\setcounter{equation}{0}

\setcounter{figure}{0}

\setcounter{footnote}{0}

\begin{center}
\LARGE{\textbf{Supplementary Information}}
\end{center}

 \section{Probability density function (PDF) for single particle chord length distribution (CLD)}
  \label{Supsec1}
  
  The Li and Wilkinson (LW) model gives a probability density function (PDF) which can be used to calculate the relative likelihood of obtaining a chord of length $s$ from a particle of length $\overline{L}$ and aspect ratio $r$ \cite{Li2005n1}. The LW model was used in this work because it covers the entire range of aspect ratios $r\in[0,1]$. The PDF of the LW model is derived from an ellipsoid of semi major axis length $a$, semi minor axis length $b$ and aspect ratio $r=b/a$ \cite{Li2005n1}. For such an ellipsoid, the probability $p_{\overline{L}_i}(s_{j,\alpha},s_{j+1,\alpha})$ of obtaining a chord whose length lies between $s_i$ and $s_{i+1}$ from a particle of length $\overline{L}_i = 2a_i$ depends on the angle $\alpha$ between the cutting chord and the $x$ axis \cite{Li2005n1}. The angular dependent PDF is given by
        \begin{equation}
        p_{\overline{L}_i}(s_{j,\alpha},s_{j+1,\alpha}) = \begin{cases}
        \sqrt{1 - \left(\frac{s_j}{2a_i} \right)^2} - \sqrt{1 - \left(\frac{s_{j+1}}{2a_i} \right)^2}, & \textrm{for}~ s_j < s_{j+1} \leq 2a_i \\
        \sqrt{1 - \left(\frac{s_j}{2a_i}\right)^2}, & \textrm{for}~ s_j \leq 2a_i < s_{j+1} \\
        0, & \textrm{for}~ 2a_i < s_j < s_{j+1}, \end{cases} 
        \label{eqs1}
        \end{equation}
      for $\alpha = \pi/2$ or $3\pi/2$
         \begin{equation}
         p_{\overline{L}_i}(s_{j,\alpha},s_{j+1,\alpha}) = \begin{cases}
         \sqrt{1 - \left(\frac{s_j}{2ra_i} \right)^2} - \sqrt{1 - \left(\frac{s_{j+1}}{2ra_i} \right)^2}, & \textrm{for}~ s_j < s_{j+1} \leq 2ra_i \\
         \sqrt{1 - \left(\frac{s_j}{2ra_i}\right)^2}, & \textrm{for}~ s_j \leq 2ra_i < s_{j+1} \\
         0, & \textrm{for}~ 2ra_i < s_j < s_{j+1}, \end{cases} 
         \label{eqs2}
         \end{equation}
         for other values of $\alpha$
             \begin{equation}
             p_{\overline{L}_i}(s_{j,\alpha},s_{j+1,\alpha}) = \begin{cases}
             \sqrt{1 - \frac{r^2 + t^2}{1 + t^2}\left(\frac{s_j}{2ra_i} \right)^2} & \\
             - \sqrt{1 - \frac{r^2 + t^2}{1 + t^2}\left(\frac{s_{j+1}}{2ra_i} \right)^2}, & \textrm{for}~ s_j < s_{j+1} \leq 2ra_i\sqrt{\frac{1 + t^2}{r^2 + t^2}} \\
             \sqrt{1 - \frac{r^2 + t^2}{1 + t^2}\left(\frac{s_j}{2ra_i}\right)^2}, & \textrm{for}~ s_j \leq 2ra_i\sqrt{\frac{1+t^2}{r^2+t^2}} < s_{j+1} \\
             0, & \textrm{for}~ 2ra_i\sqrt{\frac{1+s^2}{r^2+s^2}} < s_j < s_{j+1}, \end{cases} 
             \label{eqs3}
             \end{equation}
      where $t=\tan{(\alpha)}$. The angle independent PDF is then given as
       \begin{equation}
              p_{\overline{L}_i}(s_j,s_{j+1}) = \frac{1}{2\pi}\int_0^{2\pi} p_{\overline{L}_i}(s_{j,\alpha},s_{j+1,\alpha})d\alpha.
              \label{eqs4}
              \end{equation}
  
  \section{Determining the number of particles for making aspect ratio estimates}
  \label{Supsec2}
  
   Consider a situation where a total of $N_{obj}$ are detected by the image processing algorithm. A number  $N_k<N_{obj}$ of objects are chosen at random from the $N_{obj}$ objects detected, and then the average aspect ratio $\overline{r}_{kp}$ of this sample is calculated. Then another random sample (with the same number of objects $N_k$) is selected, and again the average aspect ratio $\overline{r}_{kp}$ is calculated. This process is repeated $N_{pick}$ times (at the same sample size) and each time the value of $\overline{r}_{kp}$ is calculated for that sample size $N_k$. This implies that the index $p$ takes values from 1 to $N_{pick}$. Then the sample size is increased up to $N^T$, and each time the overall process is repeated. The standard deviation for different sample sizes can be calculated as
   
                  \begin{figure}[tbh]
                 \centerline{\includegraphics[width=\textwidth]{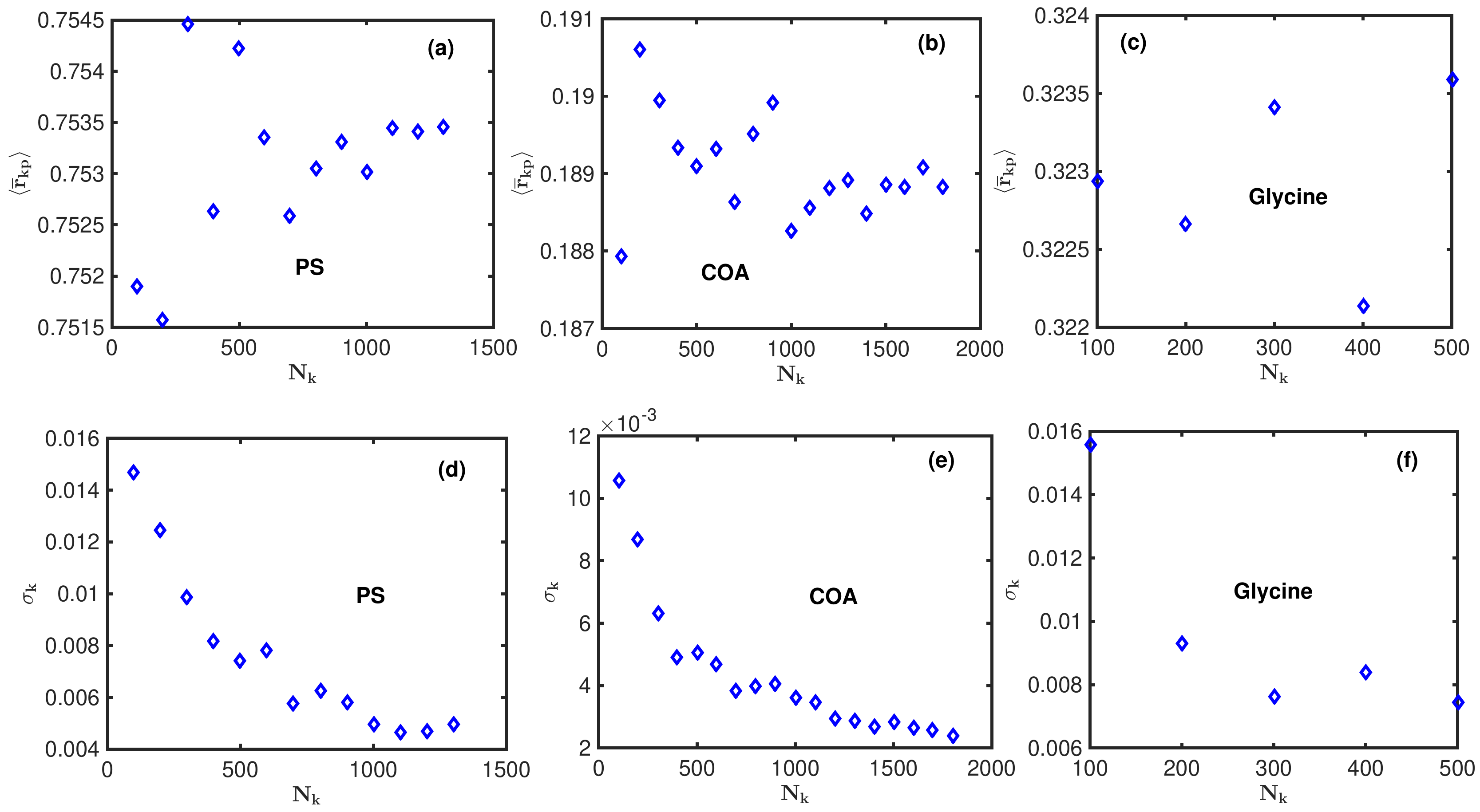}}
                 \caption{Scatter plot of the mean aspect ratio for all random selections of samples of different sizes for (a)\,PS, (b)\,COA and (c)\,Glycine. Variation of the standard deviation of the mean aspect ratio for all random selections of samples of various sizes for (d)\,PS, (d)\,COA and (f)\,Glycine.}
                 \label{figs1}
                  \end{figure}
                 
       \begin{equation}
       \sigma_k = \left[\frac{1}{N_{pick}-1}\sum_{p=1}^{N_{pick}}{\left(\langle\overline{r}_{kp}\rangle - \overline{r}_{kp}\right)^2} \right]^{1/2},
       \label{eqs5}
       \end{equation}
       where
            \begin{equation}
            \langle\overline{r}_{kp}\rangle = \frac{1}{N_{pick}}\sum_{p=1}^{N_{pick}}{\overline{r}_{kp}}.
             \label{eqs6}
            \end{equation}
      
     The average aspect ratio $\langle\overline{r}_{kp}\rangle$ for all random selections $N_{pick}=100$ of samples of different sizes from $N_k=100$ are shown in Figs. \ref{figs1}(a) to \ref{figs1}(c) for the three materials PS (Fig. \ref{figs1}(a)), CoA (Fig. \ref{figs1}(b)) and Glycine (Fig. \ref{figs1}(c)). The figures show small variations of the mean aspect ratio with different sample size. However, the standard deviation $\sigma_k$ consistently decreases with increasing sample size (as seen in Figs. \ref{figs1}(d) to \ref{figs1}(f)) as expected. The results in Fig. \ref{figs1} clearly show that the calculated mean value from the objects detected in the images  become more representative of the particles in the slurry as the number of detected objects increase. However, detecting more objects implies processing more images and the time to do this depends on the acquisition frequency of the image acquisition device and the image processing algorithm. Hence a decision needs to be made on the degree of accuracy that is sufficient for a particular process. Once that decision is made then the average aspect ratio can be retrieved at that sample size. In any case, Fig. \ref{figs1}(d) to \ref{figs1}(f)
  show that the standard deviation $\sigma_k$ begins to level off at sample size $N_k\approx 500$. This implies that the error incurred in under sampling the particles in the slurry become minimal for sample sizes $N_k\gtrsim 500$.
  
      \section{Calculating volume based PSD}
     \label{Supsec3}
              
     The technique for calculating the volume based PSD is the same as that presented in \cite{Agimelen2015}. A generalisation of the technique is necessary for the case of Method 2 (described in subsection 6.2 of the main text) where multiple aspect ratios are assigned to particles of the same characteristic length. The updated technique is described in this section.
        
     Obtain the normalised number based PSD $\hat{X}_i$ as 
             \begin{equation}
           \hat{X}_i = \frac{X_i}{\sum_{i=1}^N{X_i}},
             \label{eqs7}
             \end{equation}
     where $X_i$ is the number based PSD calculated with the inversion algorithm. Then calculate the CLD $\hat{C}^{\ast}_j$ given as
              \begin{equation}
           \hat{C}^{\ast}_j = \sum_{i=1}^N{A_{ji}\hat{X}_i},
              \label{eqs8}
              \end{equation}
     where $A_{ji}$ is the transformation matrix corresponding to the number based PSD $X_i$. The CLD $\hat{C}^{\ast}_j$ could be associated with the number based PSD $\hat{X}_i$ (as in Eq. \eqref{eqs8}) or the volume based PSD $X_i^v$ depending on the weighting applied to the matrix $A_{ji}$. The technique for weighting the matrix $A_{ji}$ in order to associate the transformed CLD $\hat{C}^{\ast}_j$ to the volume based PSD is described below.
     
     The volume based PSD is defined as \cite{Holdich2002}
               \begin{equation}
            X_i^v = \frac{\hat{X}_iV_i^3}{\sum_{i=1}^N{\hat{X}_i}V_i^3},
               \label{eqs9}
               \end{equation}
     where $V_i$ is the volume of the particle with characteristic size $\overline{L}_i$. The shape of the  particles in this work have been approximated with ellipsoids so that the volume of each particle is given as 
                \begin{equation}
            V_i = \frac{4\pi a_ib_ic_i}{3},
                        \label{eqs10}
                \end{equation}
     where $a_i$, $b_i$ and $c_i$ are the semi axes lengths in the $x$, $y$ and $z$ directions respectively of the ellipsoid of characteristic length $\overline{L}_i = 2a_i$. In this case, the origin of coordinates has been placed at the centre of the ellipsoid with the $z$ direction parallel to the major axis of the ellipsoid. Assuming the axes lengths $b_i$ and $c_i$ are equal, then using $b_i = \overline{r}_ia_i$ (where $\overline{r}_i$ is the mean aspect ratio of all particles of the same characteristic length $\overline{L}_i$) in Eq. \eqref{eqs10} and substituting in Eq. \eqref{eqs9} gives
                 \begin{equation}
            X_i^v = \frac{\hat{X}_i\overline{r}_i^2a_i^3}{\sum_{i=1}^N{\hat{X}_i\overline{r}_i^2a_i^3}}.
                         \label{eqs11}
                 \end{equation}
      
      Substituting Eq. \eqref{eqs11} in Eq. \eqref{eqs8} gives
                  \begin{equation}
            \hat{C}_j = \sum_{i=1}^N{\overline{A}_{ji}\overline{X}^v_i},
                          \label{eqs12}
                  \end{equation}
          where
          		\begin{align}
          		\overline{A}_{ji} & = \frac{A_{ji}}{\overline{r}_i^2\overline{L}_i^3} \nonumber \\
          		\overline{X}_i^v & = X^v_i\sum_{i=1}^N{\hat{X}_i\overline{r}_i^2\overline{L}_i^3}.
          		\label{eqs13}
          		\end{align}
          		
     Equation \eqref{eqs12} is the forward problem for the volume based PSD. If the weighted (due to Eq. \eqref{eqs13}) volume based PSD $\overline{X}^v_i$ is known, then the CLD $\hat{C}^{\ast}_j$ in Eq. \eqref{eqs8} can be calculated using Eq. \eqref{eqs12}. In the case of Method 1 (described in subsection 6.1 of the main text) where the same aspect ratio is used for all the particles in the slurry, then the quantities $\overline{A}_{ji}$ and $\overline{X}_i^v$ reduce to 
           		\begin{align}
           		\overline{A}_{ji} & = \frac{A_{ji}}{\overline{L}_i^3} \nonumber \\
           		\overline{X}_i^v & = X^v_i\sum_{i=1}^N{\hat{X}_i\overline{L}_i^3}.
           		\label{eqs14}
           		\end{align}
     
     Since the volume based PSD is usually not known, then an inverse problem can be formulated by searching for fitting parameters $\gamma^v_i$ which minimise an objective function $f$ of the form 
            		\begin{equation}
            		f = \sum_{j=1}^M{\left[\hat{C}^{\ast}_j - \sum_{i=1}^N{\overline{A}_{ji}\overline{X}_i^v} \right]^2},
            		\label{eqs15}
            		\end{equation}
    where   
             		\begin{equation}
             	\overline{X}_i^v = e^{\gamma_i^v}\quad i=1,2,3,\ldots, N.
             		\label{eqs16}
             		\end{equation}
  This approach was demonstrated in \cite{Agimelen2015} to correctly reproduce the volume based PSD. The function $f$ is minimised using the Levenberg-Marquardth (the Matlab implementation) using suitable initial trial solution for $\gamma_i^v$. The initial trial solution for $\gamma_i^v$ is constructed as $\gamma_i^v=\ln(X_ia_i^3)$, where $a_i$ is the semi major axis length (defined in Eq. \eqref{eqs10}) for bin $i$ and $X_i$ is the number based PSD calculated from Eq. (17) of the main text. When a smooth solution vector $\overline{X}_i^v$ is required, then $X_i$ is calculated from Eq. (21) of the main text and the corresponding $\overline{X}_i^v$ is calculated from the same Eq. More details on the procedure for choosing the smoothing parameter $\lambda_2$ in Eq. (21) of the main text will be given in subsection \ref{Supssec4-4}.
 
 \section{Choice of algorithms parameters}
  \label{Supsec4}
              
 The motivation for choosing different parameter values for the algorithms used in this work are presented in the following subsections.
   
  \subsection{Choice of number of particle subgroups in each bin in Method 2}
    \label{Supssec4-1}
   
    The Method 2 presented in subsection 6.2 of the main text outlines a technique for assigning multiple aspect ratios to particles of the same size. In this method, the characteristic particle size $\overline{L}_i$ representing the size of particles in bin $i$ is associated with multiple aspect ratios from $r_i^{min}$ to $r_i^{max}$ (see subsection 6.2 of the main text for details). To achieve this, the particles in bin $i$ of characteristic size $\overline{L}_i$ are subdivided into $N_r$ subgroups. Each subgroup is assigned an aspect ratio 
                   \begin{equation}
                  r_i^k\in \left[r_i^{min},r_i^{max}\right].
                   \label{eqs17}
                   \end{equation}

                     \begin{figure}[tbh]
                    \centerline{\includegraphics[width=0.5\textwidth]{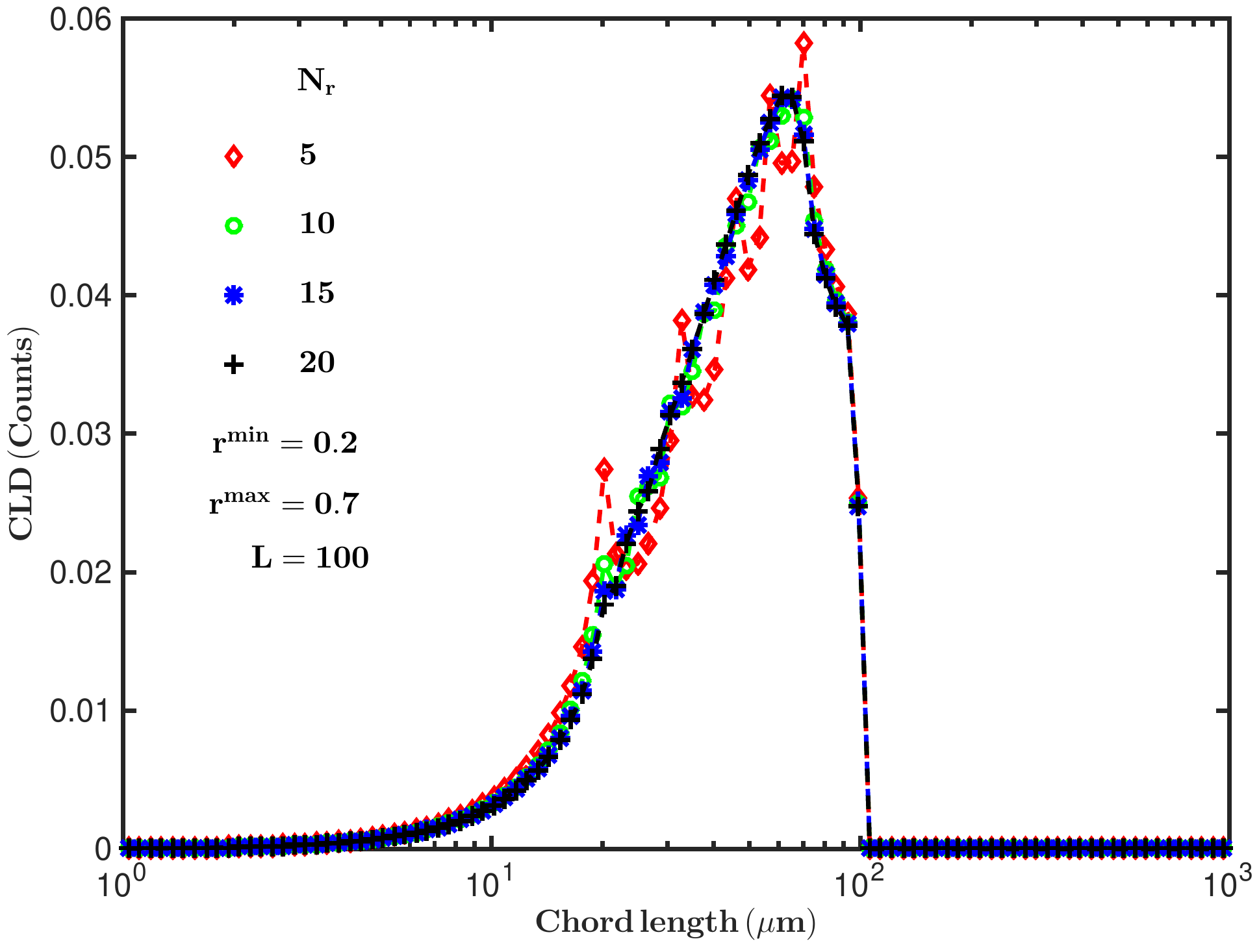}}
                    \caption{The mean CLD of a group of $N_r$ particles of the same size $\overline{L}=100\mu$m but different aspect ratios $r\in[0.2,0.7]$.}
                    \label{figs2}
                     \end{figure}
           
                \begin{figure}[tbh]
               \centerline{\includegraphics[width=\textwidth]{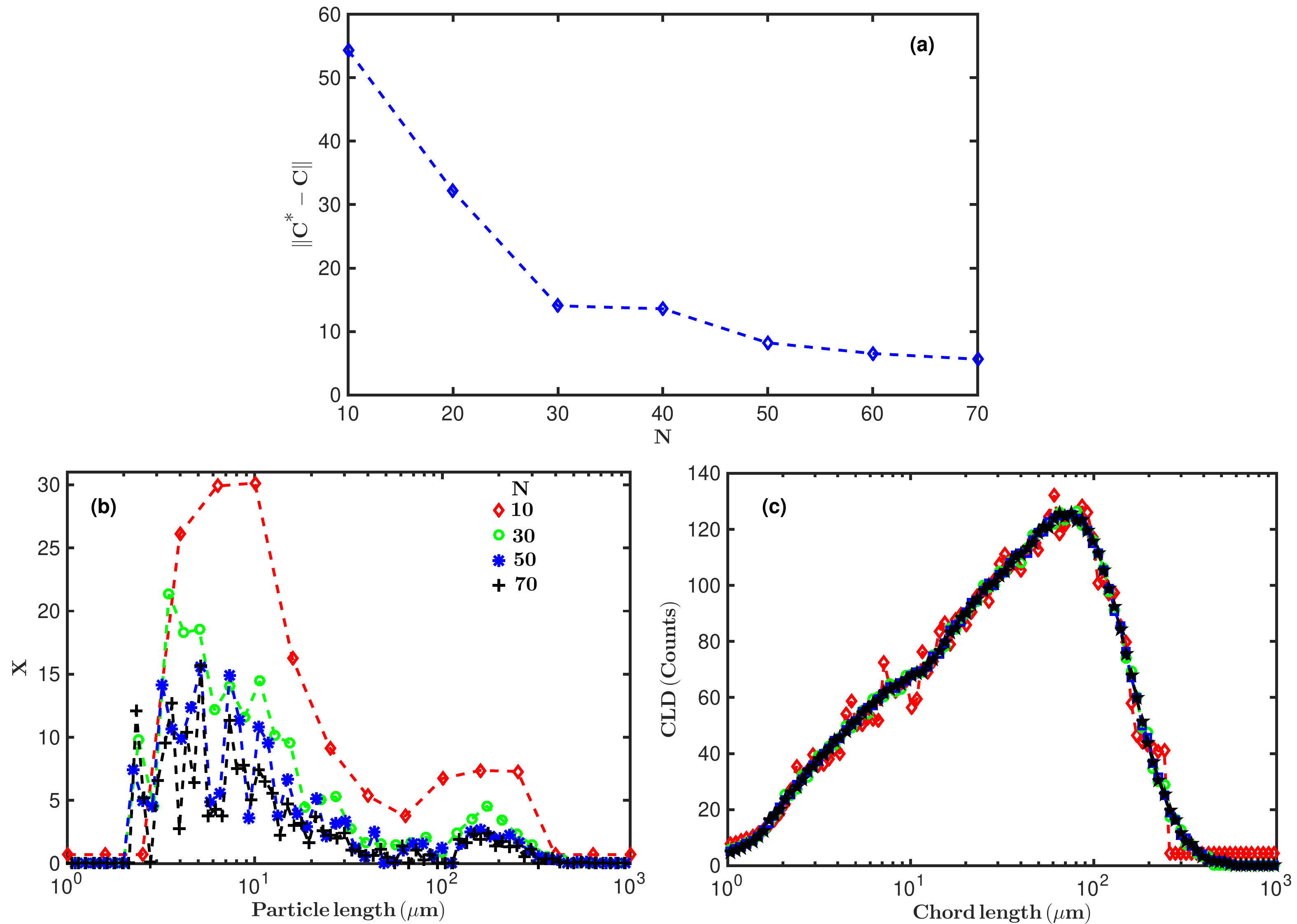}}
               \caption{(a)\,Variation of the $L_2$ norm in Eq. \eqref{eqs18} for different number of particle size bins $N$. (b)\,The number based PSD for different number of particle size bins. (c)\,Corresponding CLDs for different number of particle size bins.}
               \label{figs3}
                \end{figure}
  
   This then allows the construction of the 3D transformation matrix $\mathbf{\mathcal{A}}$ from which the 2D transformation matrix $\mathbf{A}$ is constructed by averaging across slices of the 3D matrix as in Eq. 15 of the main text. Since the CLD for a particle of characteristic size $\overline{L}$ reaches a peak at a size corresponding to the width of the particle \cite{Agimelen2015}, then each column of the average 2D matrix $\mathbf{A}$ (obtained using Eq. 15 of the main text) for a particle of size $\overline{L}$ will contain oscillations. This is due to the variations in the widths of the particles of the same size but different aspect ratios. Hence, for more accurate calculations, then the value of $N_r$ needs to be chosen sufficiently large such that the oscillations are minimised. 
   
   Figure \ref{figs2} shows the average CLD for a group of particle of the same characteristic size $\overline{L}=100\mu$m. The $N_r$ particles in each group are assigned uniformly spaced aspect ratios   
  from $r^{min}=0.2$ to $r^{max}=0.7$. The Fig. shows that the oscillations reduce as the value of $N_r$ increases. The oscillations become negligible at $N_r\gtrsim 20$.  However, a value of $N_r = 50$ was used in the calculations in the main text for more accuracy.
     
  \subsection{Choice of number of particle size bins in Methods 1 and 2}
  \label{Supssec4-2}
  
  It was demonstrated in \cite{Agimelen2015} that the number of particle size bins $N$ needed to get reasonable PSD estimate is of the order of $N\gtrsim70$. Since Method 1 (described in subsection 6.1 of the main text) uses a single representative aspect ratio to describe the shape of the particles (which is similar to the approach in \cite{Agimelen2015}) in a slurry, then the number $N=70$ of particle size bins was used in Method 1.
  
  However, Method 2 (described in subsection 6.2 of the main text) assigns a range of aspect ratios to particles of the same characteristic length $\overline{L}$. This makes it necessary to search for an optimum number of particle size bins which yields accurate solutions and physically realistic PSDs. 
  
  Figure \ref{figs2}(a) shows the behaviour of the $L_2$ norm defined in Eq. (19) of the main text (repeated here for convenience)
            \begin{equation}
           \|\mathbf{C}^{\ast} - \mathbf{C}\| = \sqrt{f_1}, 
            \label{eqs18}
            \end{equation}
 where
     \begin{equation}
    f_1 = \sum_{j=1}^{M}{\left[C_j^{\ast} - \sum_{i=1}^N{\tilde{A}_{ji}X_i}\right]^2}
     \label{eqs19}
     \end{equation}
   and
           \begin{equation}
          X_i = e^{\gamma_i}, i=1,2,3, \ldots, N.
           \label{eqs20}
           \end{equation}
  The number based PSD $X_i$ in Eq. \eqref{eqs20} is the solution vector which minimises the objective function $f_1$ in \eqref{eqs19}, and the CLD $C_j$ is calculated from the forward problem described in Eq. (10) of the main text. The $L_2$ norm in Eq. \eqref{eqs18} measures the degree of agreement between the experimentally measured CLD $C_j^{\ast}$ with the calculated CLD $C_j$.
  
  The $L_2$ norm in Fig. \ref{figs3}(a) decreases as the number of particle size bins $N$ increases, and begins to flatten out at about $N=50$. This indicates that the calculations become more accurate as the number of particle size bins increases. However, the calculated number based PSD $\mathbf{X}$ in Fig. \ref{figs3}(b) becomes increasingly noisy as $N$ increases. The opposite situation holds for the calculated CLD $\mathbf{C}$ in Fig. \ref{figs3}(c) which becomes less noisy as $N$ increases. Hence larger values of $N$ give more accurate CLDs but the corresponding PSDs become more physically unrealistic. This then leads to a choice of $N=50$ in the calculations shown in the main text. This is because the calculations becomes sufficiently accurate as suggested by Figs. \ref{figs3}(a) and \ref{figs3}(c). Furthermore, the calculations are computationally more expensive at larger values of $N$.

  \subsection{Choice of $\lambda_1$ value in Method 1}
  \label{Supssec4-3}
  
  The Method 1 presented in subsection 6.1 of the main text uses a mean aspect ratio to represent the shape of the particles in a slurry. It involves the estimation of the mean aspect ratio $\overline{r}$ from images, after which the best representative aspect ratio $r$ is searched from the range
        \begin{equation}
      r\in[\overline{r} - \mathcal{N}\sigma_r, \overline{r} + \mathcal{N}\sigma_r],
        \label{eqs21}
        \end{equation}
   where $\sigma_r$ is the standard deviation of all aspect ratios estimated from images and $\mathcal{N}$ is the number of standard deviations chosen. Then for each aspect ratio $r$, a solution vector $\mathbf{X}$ (which is the number based PSD) which minimises the objective function $f_1$ (given in Eq. \eqref{eqs19}) 
  is calculated. 
  
  When the value of $\mathcal{N}$ is made sufficiently large, then the calculated CLD $C_j$ from the number based PSD $X_i$ will match the experimentally measured CLD $C^{\ast}_j$ for one or more values of $r$. The degree of agreement is quantified by calculating the $L_2$ norm in Eq. \eqref{eqs18}.

   The desired situation would have been the case where the calculated CLD $C_j$ at the average aspect ratio $\overline{r}$ matches the experimentally measured CLD $C_j^{\ast}$. However, this is usually not the case for a number of reasons. For example, the objects in the images are not always in focus and a number of objects are rejected because they make contact with the image frame. Also the particles do not all have the same shape, so that the estimated average aspect ratio $\overline{r}$ will not necessarily be representative of all the particles in the slurry. Hence it is necessary to search for a representative aspect ratio from a suitable range around $\overline{r}$ as given in Eq. \eqref{eqs21}. The search for a suitable aspect ratio is similar to the situation presented in \cite{Agimelen2015}, however in this work, the search range has been narrowed down by means of the estimated average aspect ratio $\overline{r}$ from the images. 
   
                         \begin{figure}[tbh]
                        \centerline{\includegraphics[width=\textwidth]{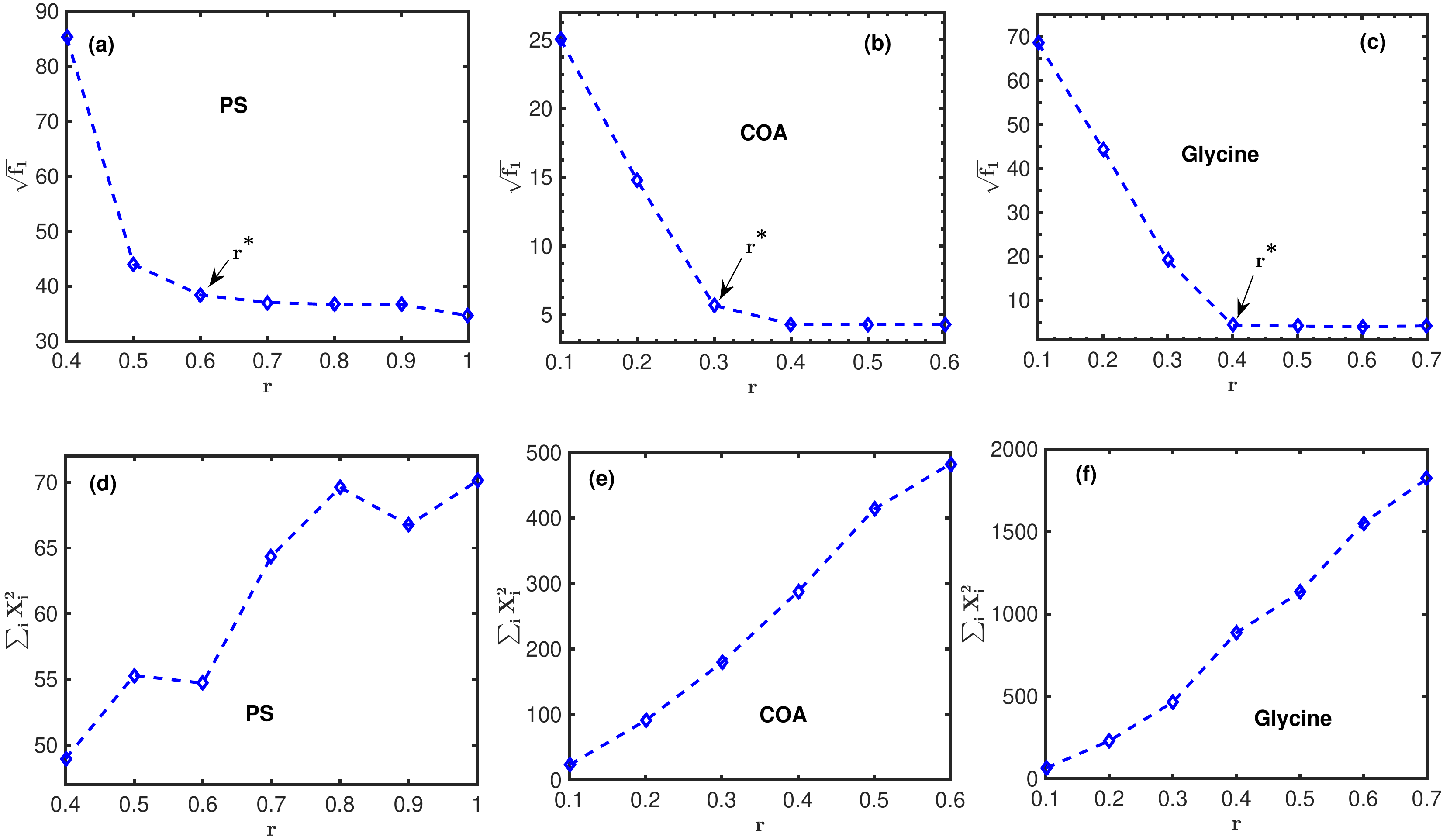}}
                        \caption{The $L_2$ norm in Eq. \eqref{eqs18} for (a)\,PS, (b)\,COA and (c)\,Glycine. The square of the size of the number based PSD retrieved from the objective function $f_1$ in Eq. \eqref{eqs18} for (d)\,PS, (e)\,COA and (f)\,Glycine.}
                        \label{figs4}
                         \end{figure}
              
                                                  \begin{figure}[tbh]
                                                 \centerline{\includegraphics[width=\textwidth]{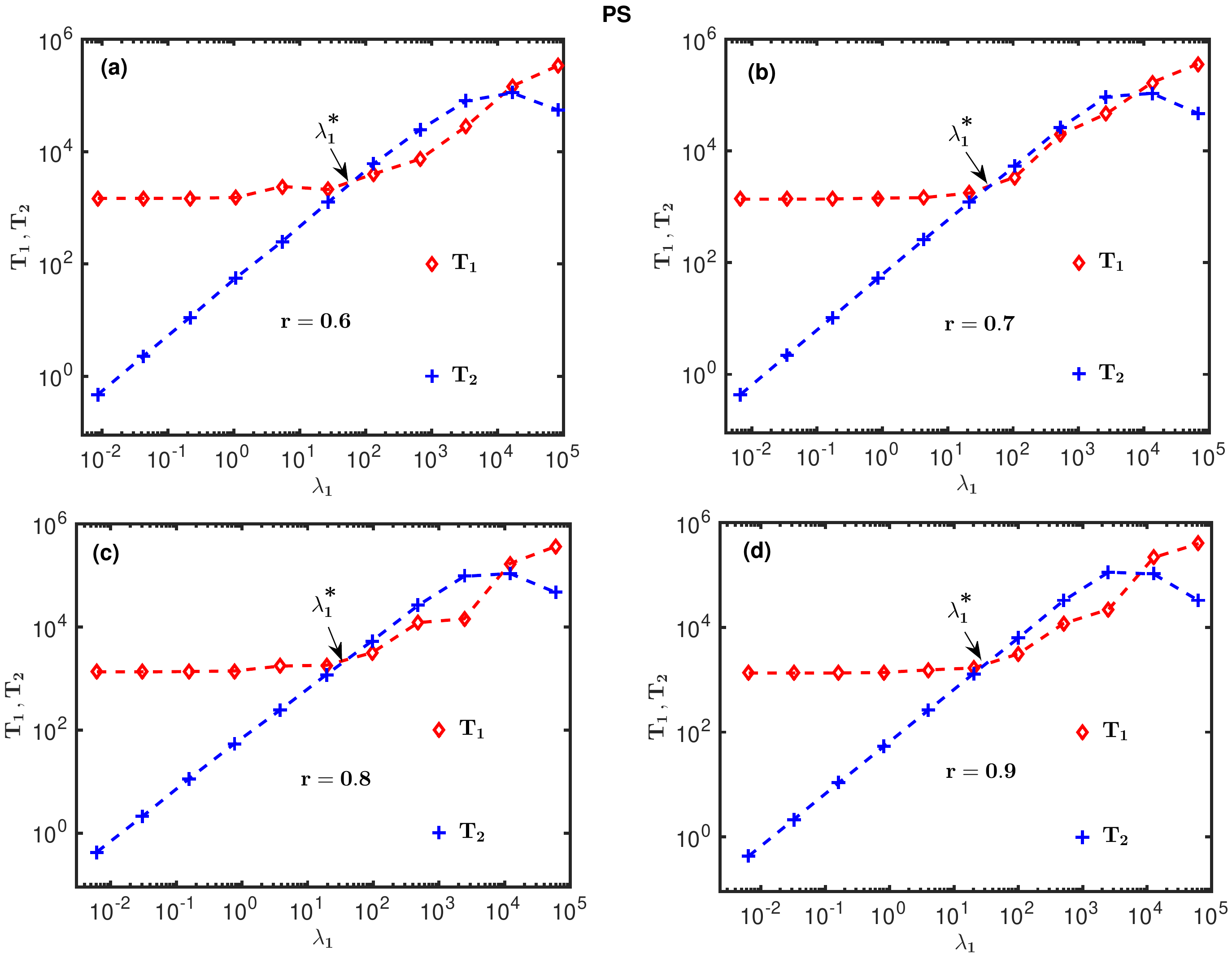}}
                                                 \caption{The behaviour of the terms $T_1$ and $T_2$ in Eq. \eqref{eqs22} for PS at the aspect ratios indicated in (a) to (d).}
                                                 \label{figs5}
                                                  \end{figure}
                                       
                                                          \begin{figure}[tbh]
                                                         \centerline{\includegraphics[width=\textwidth]{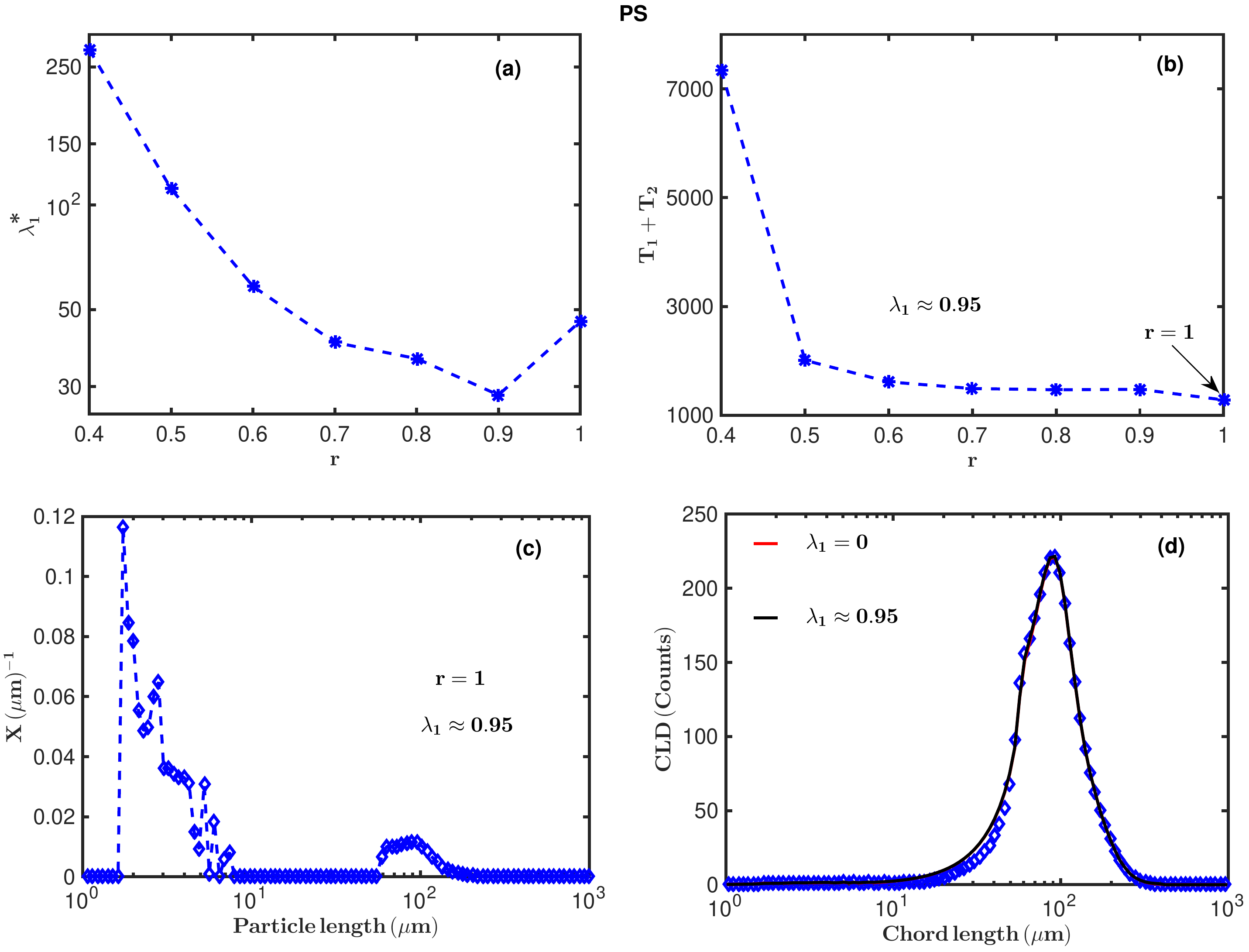}}
                                                         \caption{The behaviour (with aspect ratio) of the critical value of $\lambda_1$ at which the $T_2$ term first crosses the $T_1$ term in Eq. \eqref{eqs22} for PS. (b)\,The objective function $f_2$ in Eq. \eqref{eqs22} with aspect ratios for PS. (c)\,The number based PSD (for PS) calculated at $r=1$ where the objective function in (b) reaches a minimum. (d)\,The corresponding CLD (black line) to the number based PSD in (c). The red line is the calculated CLD at $\lambda_1=0$ and the symbols are the experimentally measured CLD.}
                                                         \label{figs6}
                                                          \end{figure}
                                               
                                                    \begin{figure}[tbh]
                                                   \centerline{\includegraphics[width=\textwidth]{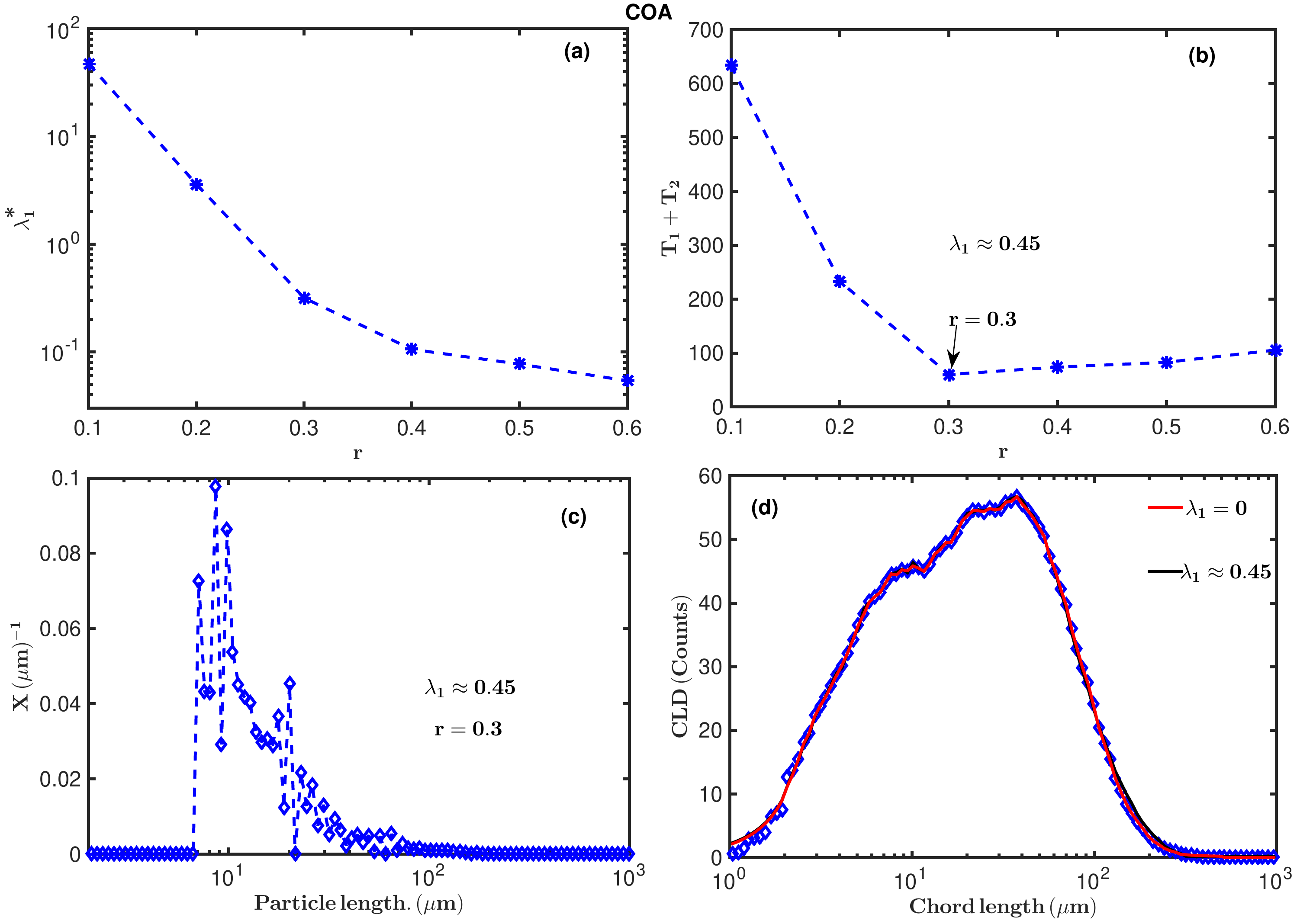}}
                                                   \caption{Same as in Fig. \ref{figs6} but for COA.}
                                                   \label{figs7}
                                                    \end{figure}
                                         
                                                   \begin{figure}[tbh]
                                                  \centerline{\includegraphics[width=\textwidth]{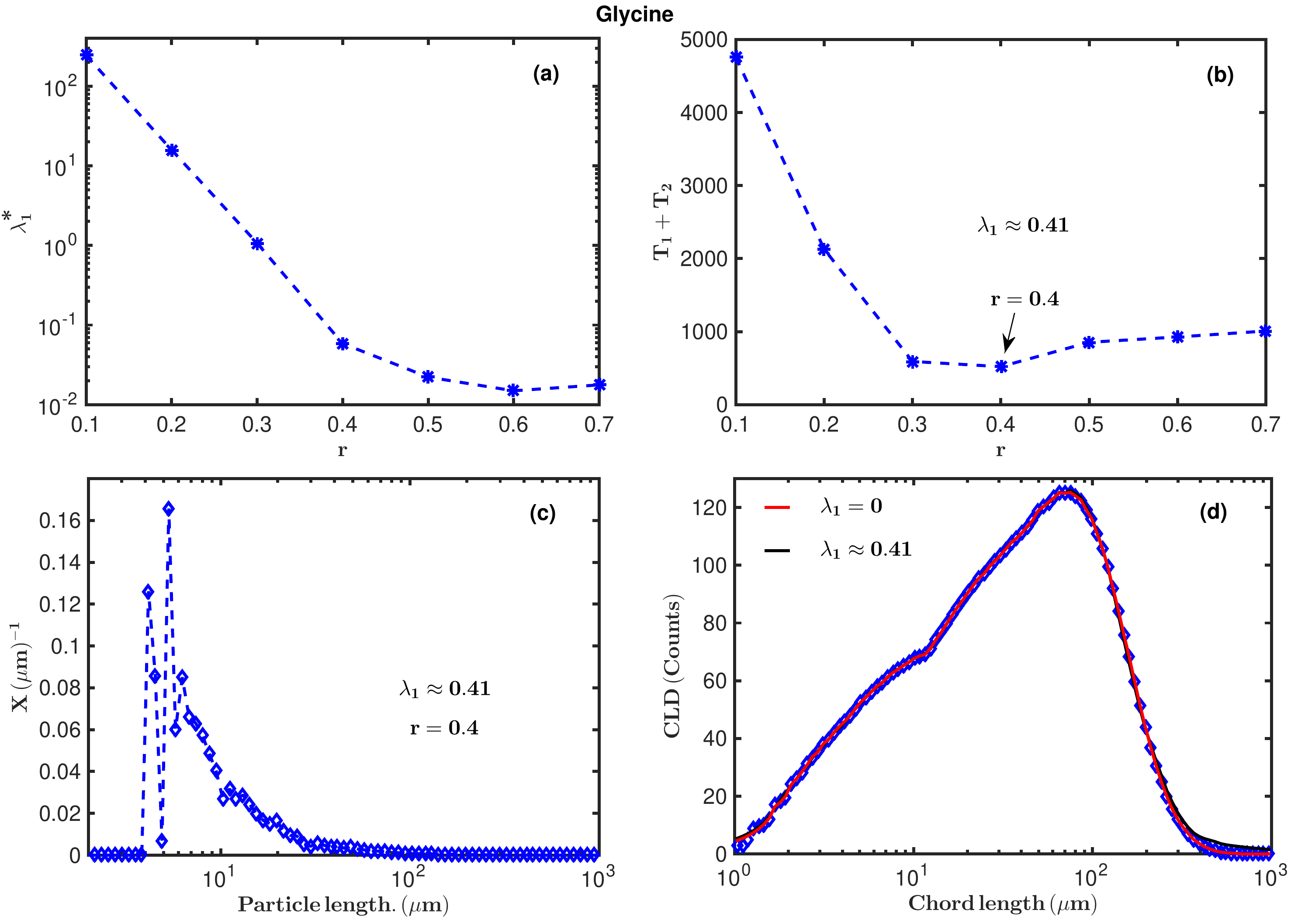}}
                                                  \caption{Same as in Fig. \ref{figs6} but for Glycine.}
                                                  \label{figs8}
                                                   \end{figure}
                                        
                                \begin{figure}[tbh]
                                  \centerline{\includegraphics[width=\textwidth]{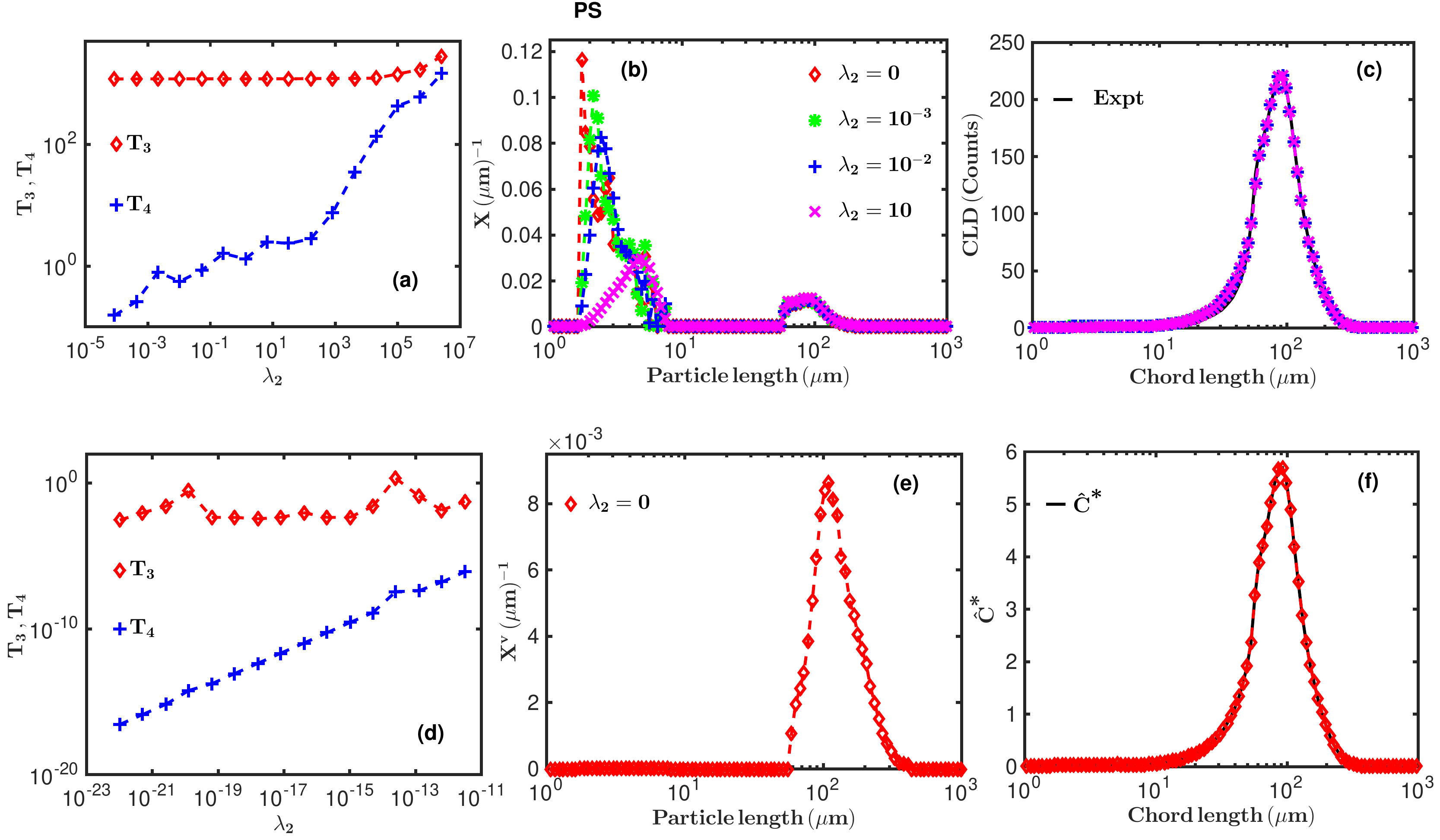}}
                                  \caption{(a)\,The terms $T_3$ and $T_4$ (for the number based PSD) in Eq. \eqref{eqs27} at various values of $\lambda_2$ for PS. (b)\,The number based PSD (for PS) computed from Eq. \eqref{eqs27} at the various values of $\lambda_2$ indicated. (c)\,The calculated CLDs corresponding to the number based PSDs in (b). The black line is the experimentally measured CLD. (d)\,The terms $T_3$ and $T_4$ (for the volume based PSD) at different values of $\lambda_2$ in Eq. \eqref{eqs27}. (e)\,The calculated (from Eq. \eqref{eqs27}) volume based PSD for PS at the value of $\lambda_2$ indicated. All calculations performed by Method 1 described in subsection 6.1 in the main text.}
                                  \label{figs9}
                                  \end{figure}
                         
                                                     \begin{figure}[tbh]
                                                       \centerline{\includegraphics[width=\textwidth]{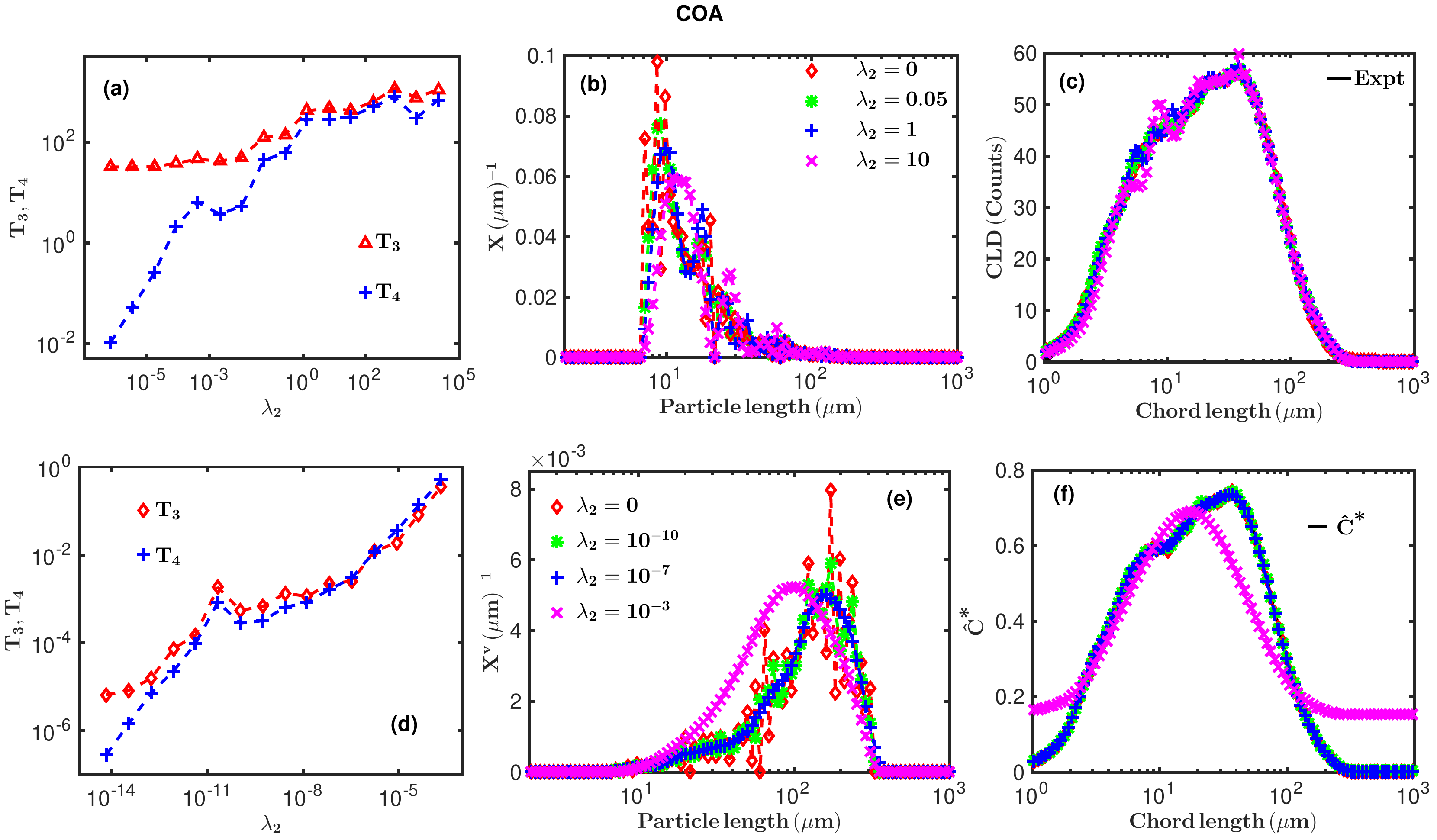}}
                                                       \caption{Same as in Fig. \ref{figs9} but for COA.}
                                                       \label{figs10}
                                                       \end{figure}
                                       
                                                 \begin{figure}[tbh]
                                                   \centerline{\includegraphics[width=\textwidth]{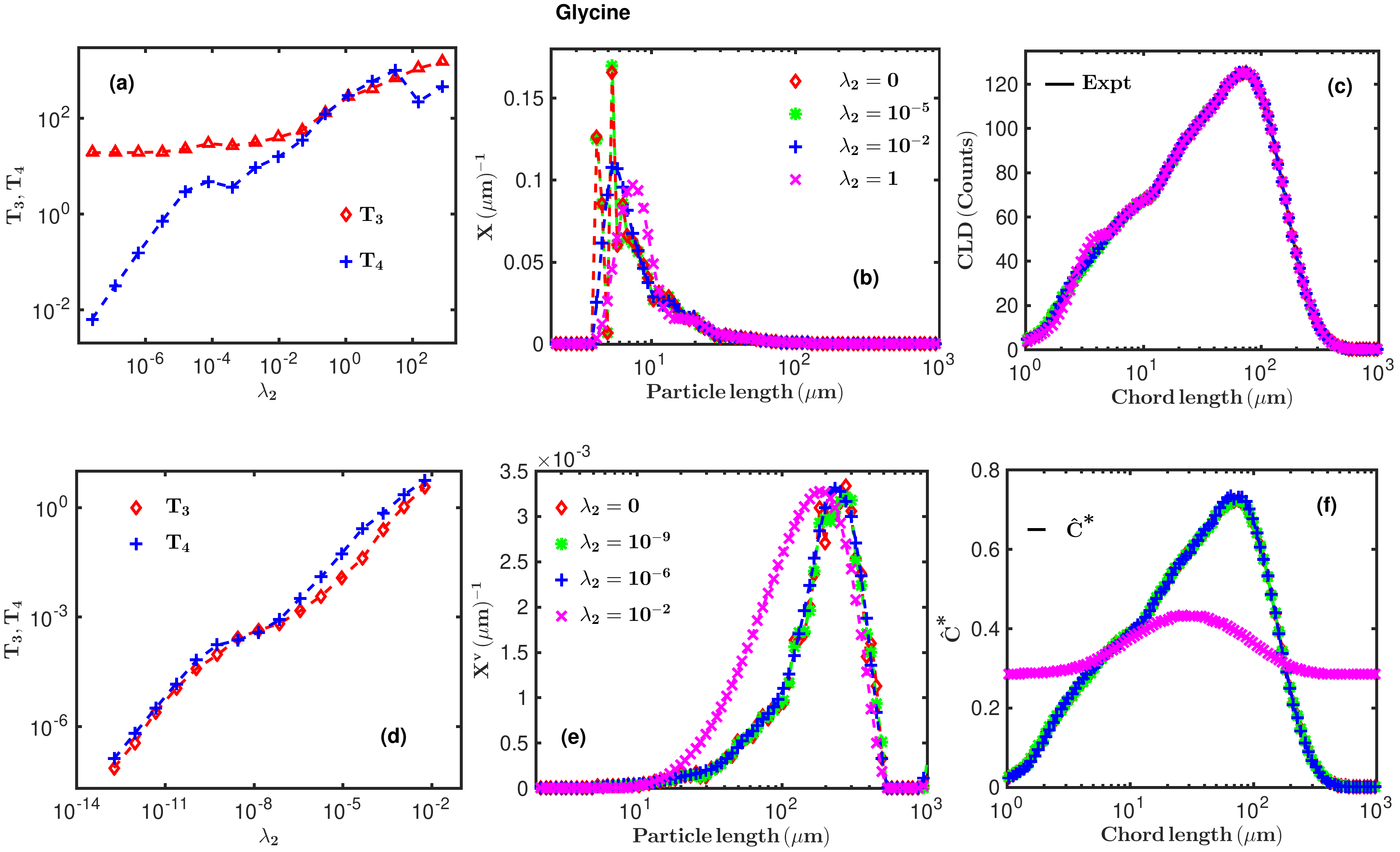}}
                                                   \caption{Same as in Fig. \ref{figs9} but for Glycine.}
                                                   \label{figs11}
                                                   \end{figure}
                                   
                                                        \begin{figure}[tbh]
                                                          \centerline{\includegraphics[width=\textwidth]{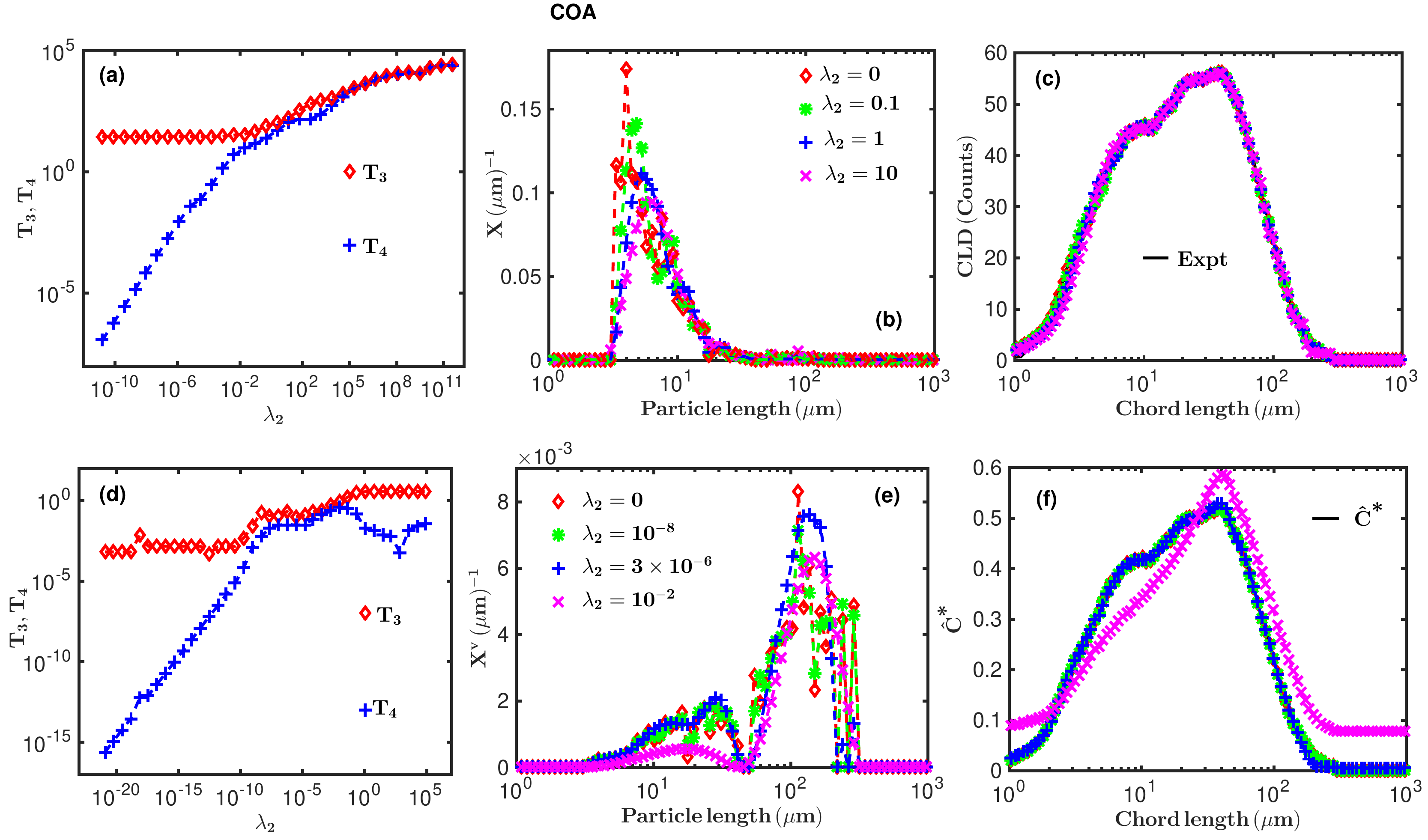}}
                                                          \caption{Similar to Fig. \ref{figs9} but calculations by Method 2 (described in subsection 6.2 of the main text) for COA.}
                                                          \label{figs12}
                                                          \end{figure}
                                                 
                                                       \begin{figure}[tbh]
                                                         \centerline{\includegraphics[width=\textwidth]{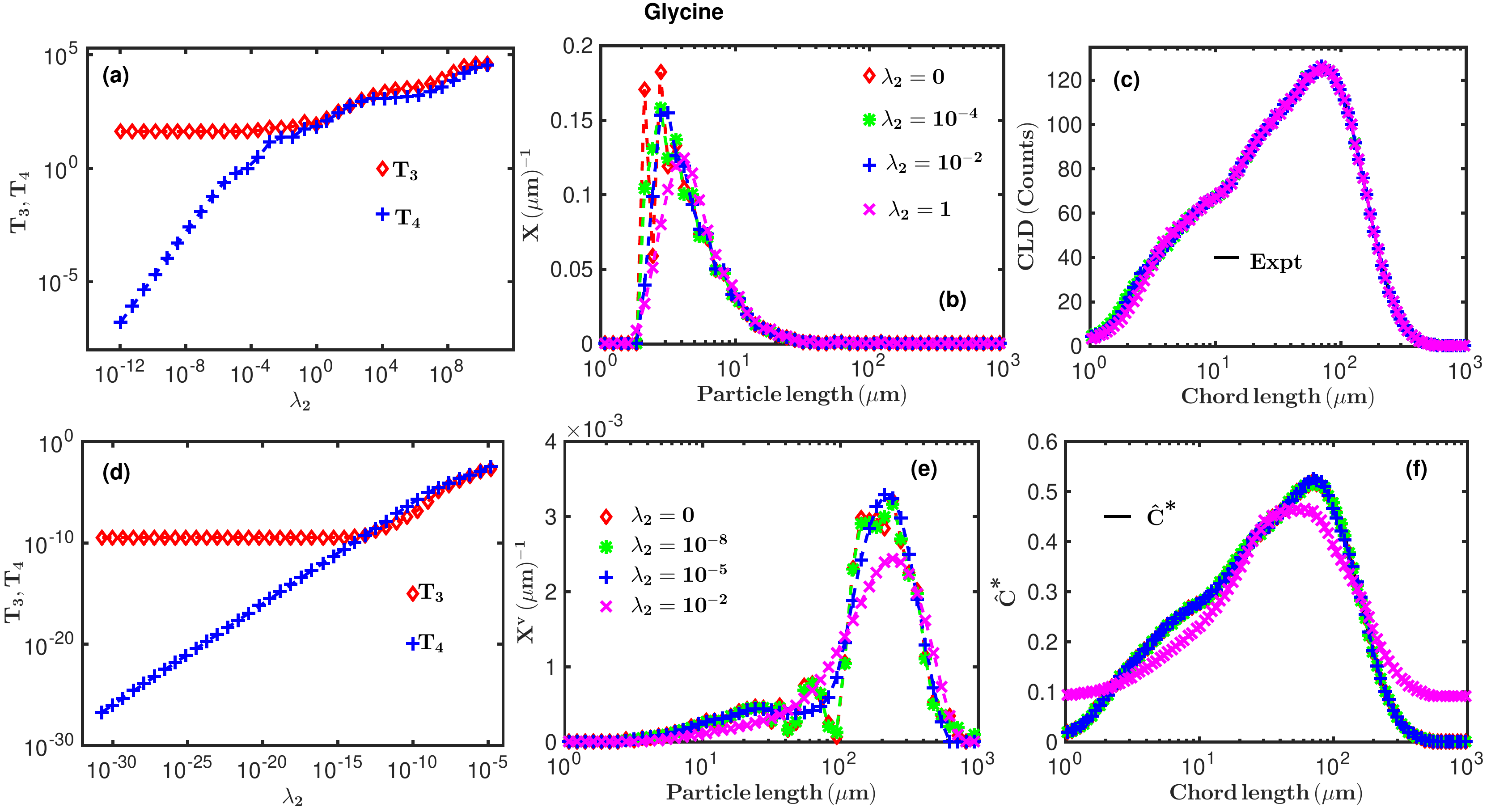}}
                                                         \caption{Same as in Fig. \ref{figs12} but for Glycine.}
                                                         \label{figs13}
                                                         \end{figure}
                                                
                                                              \begin{figure}[tbh]
                                                                \centerline{\includegraphics[width=\textwidth]{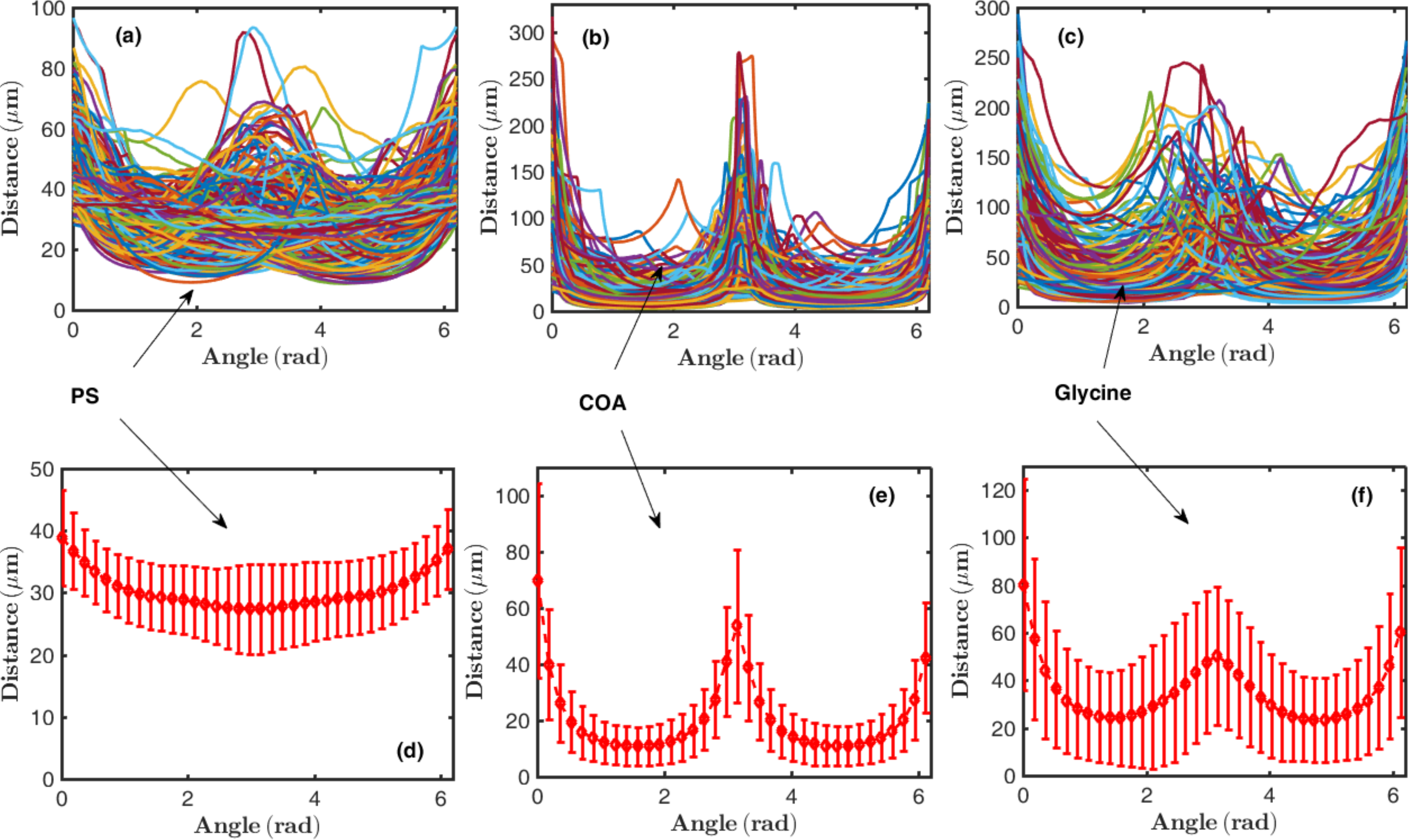}}
                                                                \caption{The shape descriptor (defined in Eq. (4) of the main text) for individual particles of (a)\,PS, (b)\,COA and (c)\,Glycine. The mean shape descriptor with error bars for (d)\,PS, (e)\,COA and (f)\,Glycine. Each error bar is one standard deviation (standard deviation of the distances from the centroid for all objects detected at each angular position) above and below each data point.}
                                                                \label{figs14}
                                                                \end{figure}
                                                       
      \begin{figure}[tbh]
       \centerline{\includegraphics[width=\textwidth]{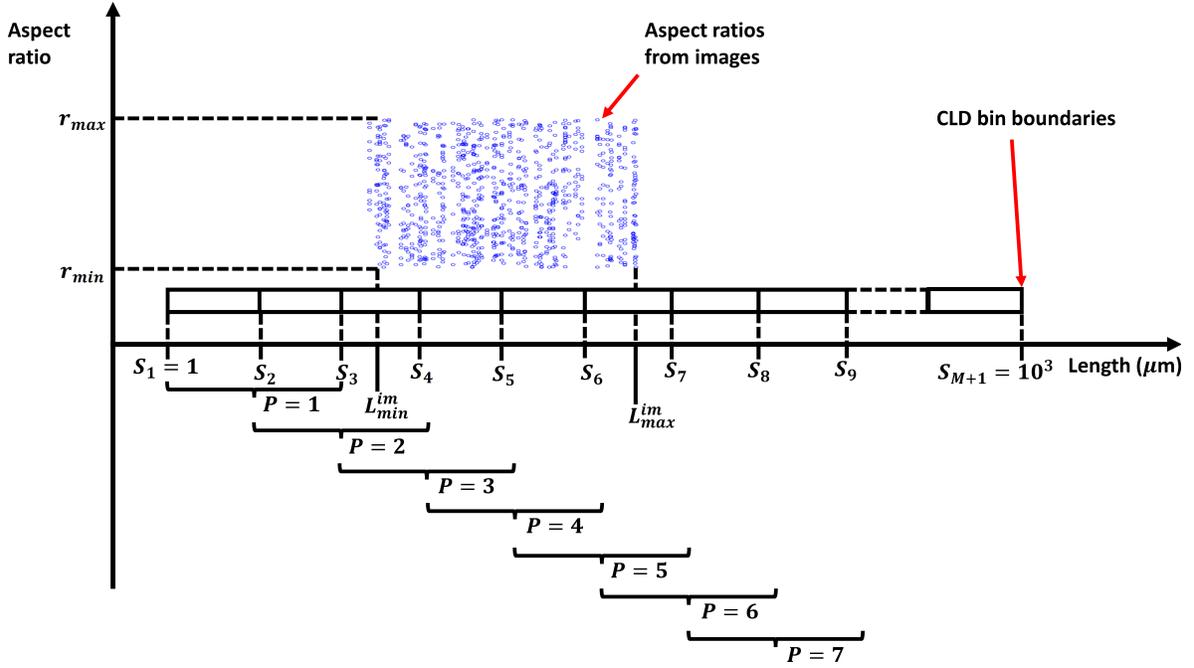}}
        \caption{Illustration of the scheme to combine the windowing technique introduced in \cite{Agimelen2015} with aspect ratio from images.}
         \label{figs15}
        \end{figure}
      
    Since the inversion is not necessarily carried out at the estimated average aspect ratio $\overline{r}$, then a situation similar to that discussed in \cite{Agimelen2015} arises, whereby the $L_2$ norm in Eq. \eqref{eqs18} becomes nearly flat after some critical aspect ratio $r^{\ast}$. This can be seen for PS ($r^{\ast}\approx 0.6$), COA ($r^{\ast}\approx 0.3$) and Glycine ($r^{\ast}\approx 0.4$) in Figs. \ref{figs4}(a) to \ref{figs4}(c). The critical aspect ratio shifts to the left as the aspect ratio of the particles decreases as can be seen by comparing Figs \ref{figs4}(a) to \ref{figs4}(c) to Figs. 2(a) to 2(c) in the main text.
   
   This calls for a technique to retrieve a unique aspect ratio which is reasonable when compared with the shape of the particles of interest. An objective function $f_2$ which imposes a penalty on the size of the calculated number based PSD $X_i$ was introduced in \cite{Agimelen2015} to pick this unique aspect ratio. 
   The objective function $f_2$ is given as (same as Eq. (20) in the main text)
       \begin{equation}
      f_2 = \underbrace{\sum_{j=1}^{M}{\left[C_j^{\ast} - \sum_{i=1}^N{\tilde{A}_{ji}X_i}\right]^2}}_{T_1} + \underbrace{\lambda_1\sum_{i=1}^N{X_i^2}}_{T_2}.
       \label{eqs22}
       \end{equation}
 The term $T_1$ is the original objective function $f_1$ in Eq. \eqref{eqs19}, and the parameter $\lambda_1$ sets the level of penalty on the PSD size which is contained in the term $T_2$. The form of the penalty term was chosen (as discussed in \cite{Agimelen2015}) because the total variation of the solution vector shows a general increase with aspect ratio as seen in Figs. \ref{figs4}(d) to \ref{figs4}(f) for PS, COA and Glycine.  If the value of $\lambda_1$ is chosen carefully, then a unique aspect ratio can be found which is
 reasonable when compared with the shape of the particles. The method for estimating $\lambda_1$ is outlined below.
 
 The idea is to estimate a value of $\lambda_1$ such that the penalty term $T_2$ just balances the term $T_1$ in Eq. \eqref{eqs22}. The procedure is as follows: \\
 For a given aspect ratio $r$, obtain an initial estimate $\lambda_1^0$ from
       \begin{equation}
      \lambda_1^0 = \frac{T_1^0}{T_2^0},
       \label{eqs23}
       \end{equation}
  where $T_1^0$ and $T_2^0$ are computed from Eq. \eqref{eqs10} as
        \begin{align}
       T_1^0 & = \sum_{j=1}^{M}{\left[C_j^{\ast} - \sum_{i=1}^N{\tilde{A}_{ji}X_i}\right]^2} \nonumber \\
       T_2^0 & =  \sum_{i=1}^N{X_i^2}.
        \label{eqs24}
        \end{align}
   Then construct the objective function $f_2$ in Eq. \eqref{eqs22} for different values of $\lambda_1$ from the range
        \begin{equation}
      \lambda_1 \in \ldots, \lambda_1^0(\Delta\lambda_1)^{-2}, \lambda_1^0(\Delta\lambda_1)^{-1}, \lambda_1^0, \lambda_1^0\Delta\lambda_1, \lambda_1^0(\Delta\lambda_1)^{2}, \ldots,
        \label{eqs25}
        \end{equation}
   where the value of $\Delta\lambda_1$ chosen depends on the data set being analysed. The value of $\Delta\lambda_1 = 5$ was used for the PS, COA and Glycine samples in this work. Then the value of $\lambda_1$ at which the term $T_2$ (in Eq. \eqref{eqs22}) just becomes equal to the term $T_1$ (also in Eq. \eqref{eqs22}) is chosen as the optimum value of $\lambda_1$ for that aspect ratio. The behaviour of the terms $T_1$ and $T_2$ for different aspect ratios for PS are shown in Figs. \ref{figs5}(a) to \ref{figs5}(d). The situation is the same for COA and Glycine. The term $T_2$ just matches the term $T_1$ at a critical value of $\lambda_1$ (indicated as $\lambda_1^{\ast}$) in Figs. \ref{figs5}(a) to \ref{figs5}(d). However, as Figs. \ref{figs5}(a) to \ref{figs5}(d) show, there is a wide disparity in the values of $\lambda^{\ast}_1$ as the aspect ratio increases. This is seen more clearly in Fig. \ref{figs6}(a) which shows the variation of $\lambda_1^{\ast}$ with aspect ratio $r$ for PS, and in Fig. \ref{figs7}(a) and \ref{figs8}(a) for COA and Glycine respectively. This suggests that the large variance in the values of $\lambda_1^{\ast}$ be removed before a meaningful average can be made. This is achieved by normalising the values of $\lambda_1^{\ast}$ by the standard deviation of the $\lambda_1^{\ast}$ values for a particular sample and then taking the mean value as given in Eq. \eqref{eqs26}
           \begin{equation}
          \lambda_1 = \frac{1}{\sigma_{\lambda}\mathcal{N}_r}\sum_{q=1}^{\mathcal{N}_r}{\lambda_1^{\ast}(r_q)},
           \label{eqs26}
           \end{equation}
  where $\mathcal{N}_r$ is the number of aspect ratios at which the calculations were performed and  $\sigma_{\lambda}$ is the standard deviation of the $\lambda_1^{\ast}$ values. Normalising the data set (the values of $\lambda_1^{\ast}$) by their standard deviation reduces the large variance in the set so that the variance of the data set becomes unity. The estimated $\lambda_1$ (by Eq. \eqref{eqs26}) is just large enough to produce a shallow minimum in the objective function $f_2$. Then the predicted aspect ratio is consistent with the actual shape of the particles.
  
  The estimated value of $\lambda_1\approx 0.95$ from Eq. \eqref{eqs26} for PS gives an estimated aspect ratio of $r=1$ as seen in Fig. \ref{figs6}(b). This is a reasonable estimate of aspect ratio since the particles of PS are near spherical as seen in Fig. 2(a) of the main text. The estimated number based PSD at $\lambda_1\approx 0.95$ and $r=1$ (using the objective function $f_2$ in Eq. \eqref{eqs22}) is shown in Fig. \ref{figs6}(c). The corresponding calculated CLD at $\lambda_1\approx 0.95$ and $r=1$ is shown by the black solid line in Fig. \ref{figs6}(d). The experimentally measured CLD for PS is shown by the symbols in Fig. \ref{figs6}(d). The calculated CLD at $\lambda_1=0$ (using $f_1$ in Eq. \eqref{eqs19}) and $r=1$ for PS is shown by the red solid line in Fig. \ref{figs6}(d).
  
  The number based PSD for PS in Fig. \ref{figs6}(c) shows large particle counts at small particle sizes close to $10\mu$m. This is because of surface roughness in the PS particles which contribute a significant number of short chords approximately less than $10\mu$m to the CLD. The counts from these short chords are not very obvious in Fig. \ref{figs6}(d) due to the concentration of the PS sample which makes the longer chords  dominate the CLD. However, the counts from short chords are more obvious in a more dilute system.
  
  Both the red and black solid lines in Fig. \ref{figs6}(d) have a near perfect match with the experimentally measured CLD (symbols in Fig. \ref{figs6}(d)), which shows that the estimated value of $\lambda_1$ by Eq. \eqref{eqs26} is reasonable.

  Figure \ref{figs8}(a) shows that the behaviour of $\lambda_1^{\ast}$ as a function of aspect ratio, for COA is similar to the case of PS. The estimated value of $\lambda_1\approx 0.45$ (using Eq. \eqref{eqs26}) gives an aspect ratio of $r=0.3$ (using $f_2$ in Eq. \eqref{eqs22}) as seen in Fig. \ref{figs7}(b). This estimated aspect ratio agrees with the needle-like shape of the particles of COA as seen in Fig. 2(b) of the main text. The estimated number based PSD at $\lambda_1\approx 0.45$ and $r=0.3$ for COA is shown in Fig. \ref{figs7}(c). The corresponding calculated CLD at $\lambda_1\approx 0.45$ and $r=0.3$ for COA (shown by the black solid line in Fig. \ref{figs7}(d)) shows a very good agreement with the measured CLD (symbols in Fig. \ref{figs7}(d)). The calculated CLD at $\lambda_1 = 0$ and $r=0.3$ is very close to the calculated CLD at $\lambda_1\approx0.45$ similar to the case of PS. This indicates that the estimated value of $\lambda_1$ is just at the right level.
  
  The situation for Glycine is similar to those of the PS and COA samples. The $\lambda_1^{\ast}$ values also show a decrease with aspect ratio as seen in Fig. \ref{figs8}(a). The application of Eq. \eqref{eqs26} leads to an estimated $\lambda_1\approx 0.41$ and subsequently $r=0.4$. The estimated value of $r=0.4$ also agrees with the prism-like shape of the particles of Glycine. Similar to the cases of PS and COA, the estimated value of $\lambda_1\approx 0.41$ is just at the right level. This is because the calculated CLD at $\lambda_1\approx 0.41$ and $r=0.4$ (black solid line in Fig. \ref{figs8}(d)) has a near perfect match with the calculated CLD at $\lambda_1 = 0$ and $r=0.4$ (red solid line in Fig. \ref{figs8}(d)) and both show a near perfect match with the measured CLD (symbols in Fig. \ref{figs8}(d)).

  \subsection{Choice of $\lambda_2$ value for both Methods 1 and 2}
  \label{Supssec4-4}
  
  The number based PSDs estimated from Eq. \eqref{eqs22} are not necessarily smooth as in the cases of the PS, COA and Glycine samples seen in Figs. \ref{figs6}(c), \ref{figs7}(c) and \ref{figs8}(c) respectively. They contain different degrees of oscillations. This is because the penalty function $T_2$ in Eq. \eqref{eqs22} is not guaranteed to force the algorithm to search for a smooth solution. Similarly, when the volume based PSDs are estimated from Eq. \eqref{eqs15}, they may not necessarily be smooth.
  
  In order to retrieve smooth solutions a new objective function was introduced in section 7 of the main text. This objective function $f_3$ (which contains a penalty function that enforces smoothness) is defined in Eq. (15) of the main text but repeated here for convenience
    \begin{equation}
   f_3 = \underbrace{\sum_{j=1}^{M}{\left[C_j^+ - \sum_{i=1}^N{A^+_{ji}X^+_i}\right]^2}}_{T_3} + \underbrace{\lambda_2\sum_{i=1}^{N}{\left[\nabla_h^2\left[X^+_i\right]\right]^2}}_{T_4},
    \label{eqs27}
    \end{equation}
    where $\nabla_h^2$ is the central difference approximation to the second derivative of the solution vector $X^+_i$ as defined in Eq. (24) of the main text. In the case of a number based PSD, the CLD $C^+_j=C_j^{\ast}$ (the experimentally measured CLD), and $A^+_{ji}=\tilde{A}_{ji}$ (which is the transformation matrix in the forward problem in Eq. 10 of the main text). For a number based PSD, the solution vector $X_i^+=X_i$. However, in the case of the volume based PSD, the vector $X_i^+=X^v_i$. In this case, the matrix $A^+_{ji}$ will be scaled as described in Eq. \eqref{eqs13} and $C_j^+=\hat{C}^{\ast}$ in Eq. \eqref{eqs8}.
    
    It is desirable to have some systematic way of estimating the $\lambda_2$ parameter at which the term $T_4$ (which is the penalty term in Eq. \eqref{eqs27}) just balances the term $T_3$ (which is the same as the function $f$ in Eq. \eqref{eqs15}). However (as will be demonstrated soon), the term $T_4$ does not always equal the term $T_3$ at reasonable values of $\lambda_2$. In most instances, the term $T_4$ just approaches  $T_3$. This then leads to a situation whereby the value of $\lambda_2$ is chosen in such a way that the smooth solution does not deviate too much from the unsmoothed solution. In addition to this, the behaviour of the CLD $\hat{C}_j$ in Eq. \eqref{eqs12} in comparison to the CLD $\hat{C}^{\ast}_j$ in Eq. \eqref{eqs8} is also used in the selection of $\lambda_2$. The $\hat{C}^{\ast}_j$ in Eq. \eqref{eqs8} is obtained directly from the number based PSD. If the calculation of the volume based PSD from Eq. \eqref{eqs27} is correct, then the CLD $\hat{C}_j$ obtained from this volume based PSD using Eq. \eqref{eqs12} should match the CLD $\hat{C}^{\ast}_j$ from Eq. \eqref{eqs8}. Hence a value of $\lambda_2$ is accepted if the CLD $\hat{C}_j$ calculated from the volume based PSD (obtained at that value of $\lambda_2$) does not deviate significantly from the CLD $\hat{C}^{\ast}_j$, and the volume based PSD (calculated at the specified value of $\lambda_2$) does not deviate significantly from the volume based PSD calculated at $\lambda_2=0$. Similarly, the calculated number based PSD from Eq. \eqref{eqs27} is rejected it its corresponding CLD (calculated from the forward problem in Eq. (10) of the main text) deviates significantly from the measured CLD.
    
    Figure \ref{figs9}(a) shows the behaviour of the terms $T_3$ and $T_4$ for different values of $\lambda_2$ for PS (for the case of the number based PSD) at aspect ratio\footnote{This is the case where the particles are assigned a single representative aspect ratio estimated using $f_2$ in Eq. \eqref{eqs22}. This approach was referred to as Method 1 in subsection 6.1 of the main text.} $r=1$. The term $T_4$  approaches the term $T_4$ for $\lambda_2\gtrsim 10^5$. The number based PSD obtained at $\lambda_2 = 10^{-3}$ (shown by the green asterisks in Fig. \ref{figs9}(b)) has about the same degree of oscillations as the number based PSD obtained at $\lambda_2=0$ (red diamonds in Fig. \ref{figs9}(b)). At $\lambda_2 = 10^{-2}$ the number based PSD (blue crosses in Fig. \ref{figs9}(b)) becomes smoother and has a peak which is close to the unsmoothed PSD obtained at $\lambda_2=0$. However, as the value of $\lambda_2$ is increased further, the number based PSD drifts significantly from the unsmoothed PSD as seen by the position of the peak of the number based PSD obtained at $\lambda_2 = 10$ (shown by the magenta crosses in Fig. \ref{figs9}(b)). 
    
    Even though the calculated CLD at $\lambda_2=10$ (in Fig. \ref{figs9}(c)) still matches the experimentally measured CLD, the significant drift of the peak of the number based PSD at $\lambda_2=10$ (seen in Fig. \ref{figs9}(b)) suggests that a smaller value of $\lambda_2$ is appropriate to obtain a smooth number based PSD for this PS sample. Hence the value of $\lambda_2= 10^{-2}$ was chosen for this sample.
    
    In the case of the volume based PSD (for PS), the term $T_4$ (in Fig. \ref{figs9}(d)) approaches the term $T_3$ at sufficiently large values of $\lambda_2$ similar to the case of the number based PSD in Fig. \ref{figs9}(a). The unsmoothed ($\lambda_2=0$) volume based PSD estimated for PS is shown in Fig. \ref{figs9}(e). Since the volume based PSD at $\lambda_2=0$ does not contain oscillations (as seen in Fig. \ref{figs9}(e)), then the computed solution at $\lambda_2=0$ was accepted. Furthermore, the CLD $\hat{C}_j$ (for PS) calculated at $\lambda_2=0$ (red diamonds in Fig. \ref{figs9}(f)) matches the CLD $\hat{C}^{\ast}_j$ (solid black line in Fig. \ref{figs9}(f)).
    
    The case of COA is similar to that of PS. The term $T_3$ never crosses the term $T_4$ within reasonable values of $\lambda_2$ as seen in Fig. \ref{figs10}(a). The $\lambda_2$ values are said to be unreasonable when the smoothed number based PSD begins to drift from the unsmoothed solution (obtained at $\lambda_2=0$) as seen in Fig. \ref{figs10}(b). This is also reflected in the calculated  CLDs which begin to deviate significantly from the experimentally measured CLD when the $\lambda_2$ values become too large as in Fig. \ref{figs10}(c). 
    
    The oscillations at the peak of the unsmoothed ($\lambda_2=0$) number based PSD had been smoothed out in the number based PSD obtained at $\lambda_2=0.05$ as in Fig. \ref{figs10}(b). The corresponding CLD at $\lambda_2=0.05$ has a near perfect match with the experimentally measured CLD for COA as in Fig. \ref{figs10}(c). Since the peaks of the PSDs obtained at the higher values of $\lambda_2$ ($\lambda_2\gtrsim 0.05$) drift from the unsmoothed number based PSD (Fig. \ref{figs10}(b)) and their corresponding CLDs deviate from the experimentally measured CLD, then the CLD corresponding to the number based PSD obtained at $\lambda_2=0.05$ was presented in Fig. 9(b) of the main text.
    
    Similar to the $T_4$ term for the number based PSD, the $T_4$ term for the volume based PSD does not cross the $T_3$ term at reasonable values of $\lambda_2$ (see Fig. \ref{figs10}(d)). The value of $\lambda_2 = 10^{-7}$ was chosen for smoothing the volume based PSD for COA. This is because Fig. \ref{figs10}(e) shows that the peaks of the calculated volume based PSDs drift significantly from the unsmoothed ($\lambda_2=0$) volume based PSD for large values of $\lambda_2$ ($\lambda_2\gtrsim 10^{-7}$). Similarly, the CLD $\hat{C}_j$ deviates significantly from the CLD $\hat{C}^{\ast}_j$ for $\lambda_2\gtrsim 10^{-7}$ as seen in Fig. \ref{figs10}(f).
    
    Unlike the cases of the PS and COA samples, the $T_4$ term for the number based PSD crosses the $T_3$ term (at $\lambda_2\approx 0.4$) for Glycine as seen\footnote{This is one of the few instances where the $T_4$ term crosses the $T_3$ term in Eq. \eqref{eqs27}.} in Fig. \ref{figs11}(a). However, as Fig. \ref{figs11}(b) shows, the oscillations in the unsmoothed solution (at $\lambda_2=0$) are already smoothed out for a smaller value of $\lambda_2=0.01<0.4$. Similarly to the cases of the PS and COA samples, the peak of the smoothed number based PSD drifts from the peak of the unsmoothed PSD for large values of $\lambda_2$ (as in the case of $\lambda_2=1$ in Fig. \ref{figs11}(b)). Similarly, the calculated CLD deviates from the experimentally measured CLD at large values of $\lambda_2$ as seen in the case of $\lambda_2=1$ in Fig. \ref{figs11}(c). Hence the value of $\lambda_2=0.01$ was chosen for the number based PSD for Glycine.
    
    Also in contrast to the cases of PS and COA, the $T_4$ term of the volume based PSD is about equal to the $T_3$ term (for $\lambda_2\lesssim 10^{-6}$) of the in the case of Glycine as seen in Fig. \ref{figs11}(d). However, the selection of $\lambda_2$ for the volume based PSD is done in a manner similar to those of PS and COA. The value of $\lambda_2=10^{-6}$ was chosen for the volume based PSD for Glycine since the left shoulder seen in the volume based PSD at $\lambda_2 = 0$ (Fig. \ref{figs11}(e)) becomes less obvious for $\lambda_2\gtrsim 10^{-6}$. Also, the CLD $\hat{C}_j$ deviates significantly from the CLD $\hat{C}^{\ast}_j$ for $\lambda_2\gtrsim 10^{-6}$ as seen in Fig. \ref{figs11}(f).
      
    In the case\footnote{The case where multiple aspect ratios are assigned to different particles of the same length as described in subsection 6.2 of the main text} of Method 2, the behaviour of the terms $T_3$ and $T_4$ in Eq. \eqref{eqs27} is similar to the case of Method 1. Figure \ref{figs12}(a) shows the behaviour of the terms $T_3$ and $T_4$ (in Eq. \eqref{eqs27}) with $\lambda_2$ for COA. The term $T_4$ does not cross the term $T_3$ (within reasonable values of $\lambda_2$). The term $T_4$ becomes nearly equal to the term $T_3$ for $\lambda_2\gtrsim 0.1$. The number based PSD estimated for COA at $\lambda_2=0.1$ is shown in Fig. \ref{figs12}(b). It has about the same degree of fluctuation (at the peak) as the estimated number based PSD at $\lambda_2=0$ (in Fig. \ref{figs12}(b)). However, as $\lambda_2$ is increased to 1, the fluctuations at the peak are penalised as seen in Fig. \ref{figs12}(b). As the value of $\lambda_2$ is increased further to $\lambda_2=10$, the peak of the number based PSD begins to drift from the unsmoothed number based PSD as seen in Fig. \ref{figs12}(b). At this value of $\lambda_1=10$, the calculated CLD begins to deviate from the experimentally measured CLD and calculated CLDs at smaller values of $\lambda_2$. This situation is shown in Fig. \ref{figs12}(c).
    
    The behaviour of the $T_3$ and $T_4$ terms for the volume based PSD (for COA) is similar to the case of the number based PSD. The $T_4$ term shows a near linear increase for $\lambda_2\lesssim 10^{-8}$ as seen in Fig. \ref{figs12}(d). This suggests that the smoothing effect of the $T_4$ term is ineffective at these small values of $\lambda_2$. However, the smoothing effect kicks in for $\lambda_2\gtrsim 10^{-8}$. At this value of $\lambda_2=10^{-8}$, the fluctuations in the estimated volume based PSD only reduce slightly as seen in Fig. \ref{figs12}(e). However, at $\lambda_2\approx 3\times 10^{-6}$, the fluctuations smoothen out as seen in Fig. \ref{figs12}(e). As the value of $\lambda_2$ is increased to $10^{-2}$, the peak of the estimated volume based PSD begins to drift from the unsmoothed volume based PSD as seen in Fig. \ref{figs12}(e). Also, the minor peak of the estimated volume based PSD at $\lambda_2=10^{-2}$ decreases significantly. Furthermore, the calculated CLD $\hat{C}_j$ at $\lambda_2=10^{-2}$ deviates significantly from the CLD $\hat{C}^{\ast}_j$ as seen in Fig. \ref{figs12}(f). This indicates that the value of $\lambda_2=3\times 10^{-6}$ is more suitable for the COA sample.
       
    The situation for Glycine is similar to that of COA where the $T_4$ term does not cross the $T_3$ term as seen in Fig. \ref{figs13}(a). Similarly, Fig. \ref{figs13}(b) suggests a suitable value of $\lambda_2$ for the number based PSD to be $\lambda_2\approx 10^{-2}$. Figure \ref{figs13}(c) also show that the calculated CLD for the higher value of $\lambda_2 = 1$ deviates slightly from the experimentally measured CLD and the calculated CLDs at smaller values of $\lambda_2$. The situation for the volume based PSD is similar to that of COA. The smoothing effect of the $T_4$ term kicks in at $\lambda_2\gtrsim 10^{-12}$ (Fig. \ref{figs13}(d)), and Fig. \ref{figs13}(e) suggests a suitable value of $\lambda_2$ to be about $10^{-5}$. This is also supported by Fig. \ref{figs13}(f) which shows that the CLD $\hat{C}_j$ deviates significantly from the CLD $\hat{C}^{\ast}_j$ for $\lambda_2>10^{-5}$.

   \subsection{Image processing algorithm parameters}
   \label{Supssec4-5}
   
   The tunable parameters of the image processing algorithm described in section 4 of the main text are summarised in Table \ref{Tabs1}. The parameter names in Table \ref{Tabs1} are the same as they were used in the code. The description of each parameter and its effect is given in Table \ref{Tabs1}.

   \begin{table}
   \caption{Summary of tunable parameters of the image processing algorithm.}
   \label{Tabs1}
   \begin{center}
   \begin{tabular}{p{4cm}|p{2.5cm}|p{2cm}|p{7cm}}
   \hline
   \hline
   \textbf{Parameter name} & \textbf{Description} & \textbf{Default value} & \textbf{Effect} \\
  \hline
  \\[-0.45cm]
  \multicolumn{4}{l}{\parbox{15cm}{\textbf{Step 1: Traditional spatial median filtering is used to reduce noise in the image.}}} \\
  \hline
  prefilter.median\_size & Size of the filter window & 5 pixels & The bigger the value the stronger the effect. Effect is similar to a low pass filter but preserves sharp (high contrast) edges in the image. \\
  \hline
   \multicolumn{4}{l}{\parbox{15cm}{\textbf{Step 2: A classical method of high pass filter (Laplacian of Gaussian and threshold) is used to detect objects edges.}}} \\
   \hline
  object\_detector.Log \_size &  Size of the filter window & 89 pixels & Large window sizes remove noise at the expense of blurred objects.\\
  \hline
  object\_detector.Log \_power & Gaussian curve standard deviation & 0.23327 & The bigger the value, the lower the contrast needs to be at the edges for the edge be detected; this does increase the risk of misclassifying noise as edge. \\
   \hline
   object\_detector.edge \_threshold & Threshold of detection & 251 & The higher the value, the sharper the edge has to be in order to be classified as edge. This reduces noise at the expense of not detecting an edge if the value is too high. \\
    \hline
     \multicolumn{4}{l}{\parbox{15cm}{\textbf{Step 3: Object processing - detected objects are filtered to discard objects that do not possess desirable properties.}}} \\
     \hline
   object \_processing.imclose \_size & Size of disk for the operation of morphological closing & 8 pixels & The bigger the value, the bigger the gaps in the object that will be filled. This reduces noise in the result. The trade-off is that the shape of the original object might be lost. \\
   \hline
   object \_processing.filter \_area\_minimum & Objects area less than specified value are discarded & 900 pixels & If an object is too small, the chance is that it is noise. The bigger the value, then the more that small objects are discarded. The trade-off is that potentially useful objects can be discarded. \\
   \hline
   \multicolumn{4}{l}{\parbox{15cm}{\textbf{In the final stage, solid objects are reduced to their boundary only, and the coordinates of pixels on the boundaries are reported.}}} \\
      \hline
   border \_processing.border \_thickness & Desired thickness of the border & 1 pixel & Optionally, setting this to 2 pixels will produce higher angular resolution output. However, this is often not needed. \\
   \hline
   \hline
   \end{tabular}
   \end{center}
   \end{table}
   
  \subsection{Shape descriptors for individual objects}
   \label{Supssec4-6}
   
  The mean aspect ratio $\overline{r}$ (used in Method 1 to construct a range of aspect ratios within which  to search for a representative value) is obtained from the mean shape descriptor of the objects from images as outlined in section 4 of the main text. Although this mean shape descriptor is reasonably smooth as seen in Figs. 5(d), 5(e) and 5(f) of the main text, the shape descriptors for individual objects contain some degree of oscillations. Figures \ref{figs14}(a), \ref{figs14}(b) and \ref{figs14}(c) show the shape descriptors for the individual particles in the PS, COA and Glycine samples respectively. They contain different degrees of oscillations due to variations in the shapes of the particles. There is also a contribution from errors in the detection of particles' boundaries because some particles are only partly in focus. There is also a contribution from impurities in the samples. This leads to the large error bars on the mean shape descriptors for the three samples as seen in Figs. \ref{figs14}(d), \ref{figs14}(e) and \ref{figs14}(f). The length of each error bar in Figs. \ref{figs14}(d), \ref{figs14}(e) and \ref{figs14}(f) is two times the standard deviation of the distance from the centroid. The size of the error bars in Figs. \ref{figs14}(d), \ref{figs14}(e) and \ref{figs14}(f) strongly indicates that a reliable estimate of the PSD cannot be made from current in-line imaging sensors (the PVM in this case) alone. Hence the need to combine the imaging data with CLD data as carried out in this work. However, the use of image data along with CLD data leads to significant improvements in the PSD estimates. The contribution to the size of the error bars by the variation in particle shape justifies the method which assigns multiple aspect ratios to particles of the same length rather than using an average aspect ratio for all particles in the slurry.

  \section{Integrating windowing technique with aspect ratios from images in Method 2}
  \label{Supsec5}
  
  The windowing technique developed in \cite{Agimelen2015} allows the values of $L_{min}$ and $L_{max}$ to be estimated in-line using the bin boundaries of the CLD histogram. The quantities $L_{min}$ and $L_{max}$ (defined in Eqs. (8) and (9) of the main text) are the minimum and maximum sizes of the particles in a slurry. The size of a window is the number of CLD bins covered by that window, and the position of a window is set by the index $P$. For example, in the case shown in Fig. \ref{figs15}, a window of size 2 is placed at different positions (indicated by $P=1,2,\ldots,7$ in the Fig.) along the CLD bin boundaries. At the first window position (indicated by $P=1$ in Fig. \ref{figs15}), the values of $s_1$ and $s_3$ (where $s_1$ and $s_3$ are the chord lengths corresponding to the boundaries covered by the window at $P=1$) are assigned to $L_{min}$ and $L_{max}$ respectively. Similarly, at $P=2$, $L_{min}=s_2$ and $L_{max}=s_4$ and so on \cite{Agimelen2015}.
  
  Consider a simple distribution of aspect ratios shown by the blue circles in Fig. \ref{figs15}. The aspect ratios in Fig. \ref{figs15} were simulated using uniformly distributed random numbers in the interval (0,1). In a real experiment, the aspect ratios will be estimated from images. The aspect ratios $r\in[r_{min},r_{max}]$ correspond to particle sizes $L\in [L_{min}^{im}, L_{max}^{im}]$ as indicated in Fig. \ref{figs15}, where $L_{min}^{im}$ and $L_{max}^{im}$ are the minimum and maximum particle sizes obtained  from all the images analysed.
  
  The chord lengths corresponding to the bin boundaries of the CLD histogram cover the range $s_1=1\mu$m to $s_{M+1}=10^3\mu$m as indicated in Fig. \ref{figs15}, where $M$ is the number of CLD bins. However, the particle sizes estimated from images do not cover this range ($L_{min}^{im} > 1\mu$m and $L_{max}^{im}<10^3\mu$m as illustrated in Fig. \ref{figs15}). This calls for appropriate approximations in the situations where the $L_{min}$ and $L_{max}$ values returned by the window do not lie entirely within the range set by $L_{min}^{im}$ and $L_{max}^{im}$. The following procedure was implemented in this work:
  
  \begin{enumerate}
  \item{At position $P=1$ where $L_{max}<L_{min}^{im}$, assign $r_{max}$ to all $N_r$ particle subgroups (the particle subgroups were discussed in subsection 6.2 of the main text). This approach was adopted because Figs. 6(b) and 6(c) of the main text suggest that the small particles are more rounded than the larger particles.}
  
  \item{At $P=2$ where $L_{min}<L_{min}^{im}$ and $L_{min}^{im}<L_{max}<L_{max}^{im}$}, assign $r_{max}$ to the first 50\% of the $N_r$ subgroups and for the remaining 50\%, assign aspect ratios (from images) corresponding to particle sizes between $L_{min}^{im}$ and $L_{max}$.
  
  \item{At $P=4$ where $L_{min}^{im}<L_{min}<L_{max}^{im}$ and $L_{min}^{im}<L_{max}<L_{max}^{im}$, assign aspect ratios (from images) corresponding to particle sizes between $L_{min}$ and $L_{max}$ to all $N_r$ subgroups.}
  
  \item{At $P=6$ where $L_{min}^{im}<L_{min}<L_{max}^{im}$ and $L_{max}>L_{max}^{im}$}, assign aspect ratios (from images) corresponding to particle sizes between $L_{min}$ and $L_{max}^{im}$ to the first 50\% of the $N_r$ subgroups, and assign aspect ratio $\overline{r}$ to the remaining 50\% of the $N_r$ subgroups, where $\overline{r}$ is the mean aspect ratio estimated from images. This approach was adopted because Figs. 6(b) and 6(c) suggest that the larger particles have aspect ratio closer to the mean value than to $r_{max}$.
  
  \item{At $P=7$, where $L_{min}>L_{max}^{im}$, assign aspect ratio $\overline{r}$ to all $N_r$ subgroups.}
  \end{enumerate}
  This approach allows the computation of the 3 D matrix defined in Eq. (15) of the main text.
  
                 \begin{figure}[tbh]
                  \centerline{\includegraphics[width=\textwidth]{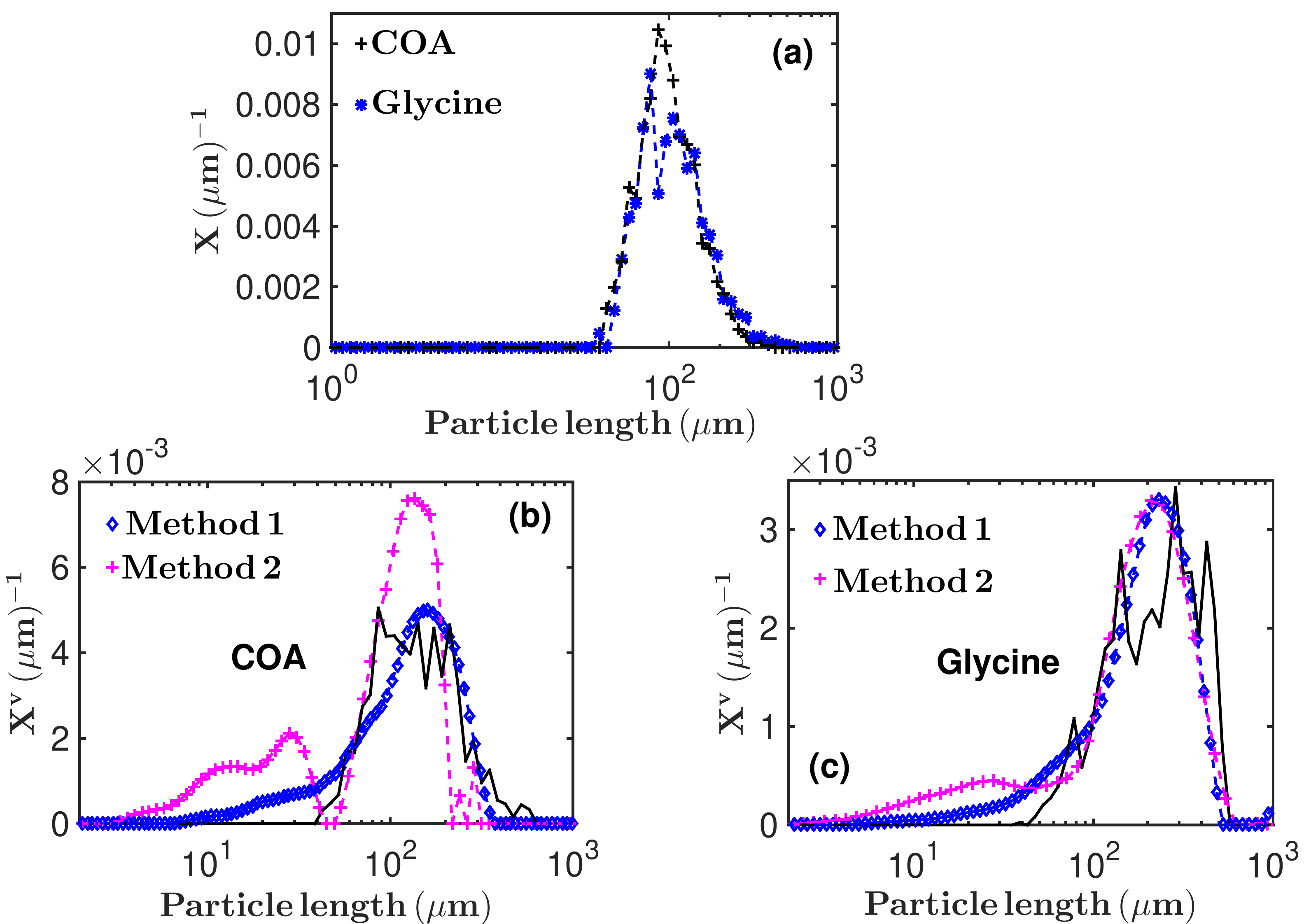}}
                   \caption{(a)\,Estimated (from imaging data) number based PSD for the COA and Glycine samples (as indicated in the Fig.). This estimate does not include counts of particles whose lengths are less than about $30\mu$m as the image processing algorithm used here is not able to distinguish between these small particles and background noise in the images. (b)\,Estimated volume based PSD for COA by Methods 1 and 2 (as indicated). The black solid line is the corresponding estimate from imaging data for the same COA sample. (c)\,Similar to (b) but for Glycine.}
                    \label{figs16}
                   \end{figure}
                  
                                  \begin{figure}[tbh]
                                   \centerline{\includegraphics[width=0.6\textwidth]{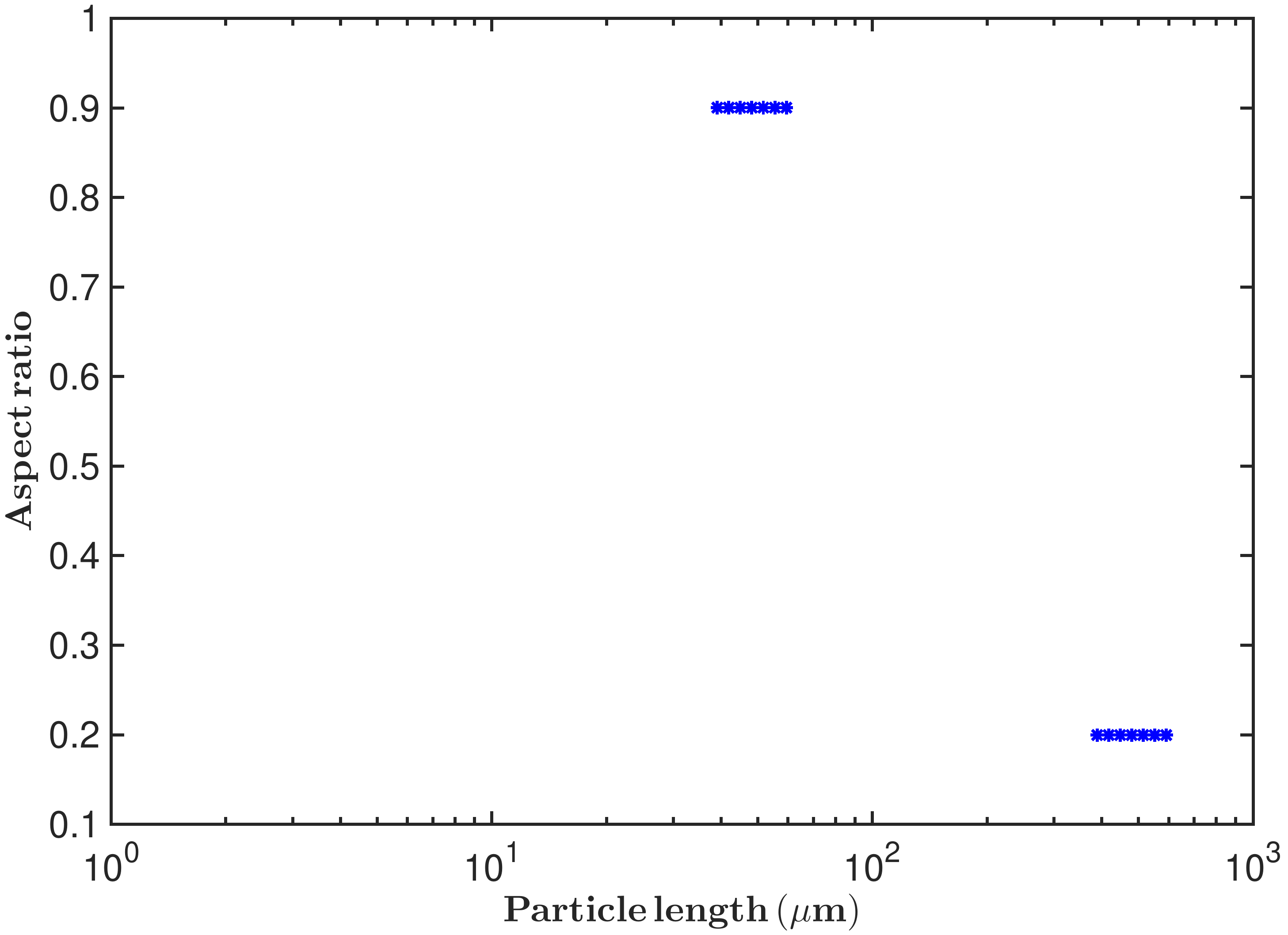}}
                                    \caption{Distribution of aspect ratios for the two subpopulations of particles used to test the relative performance of Methods 1 and 2.}
                                     \label{figs17}
                                    \end{figure}
                                  
      \begin{figure}[tbh]
       \centerline{\includegraphics[width=0.9\textwidth]{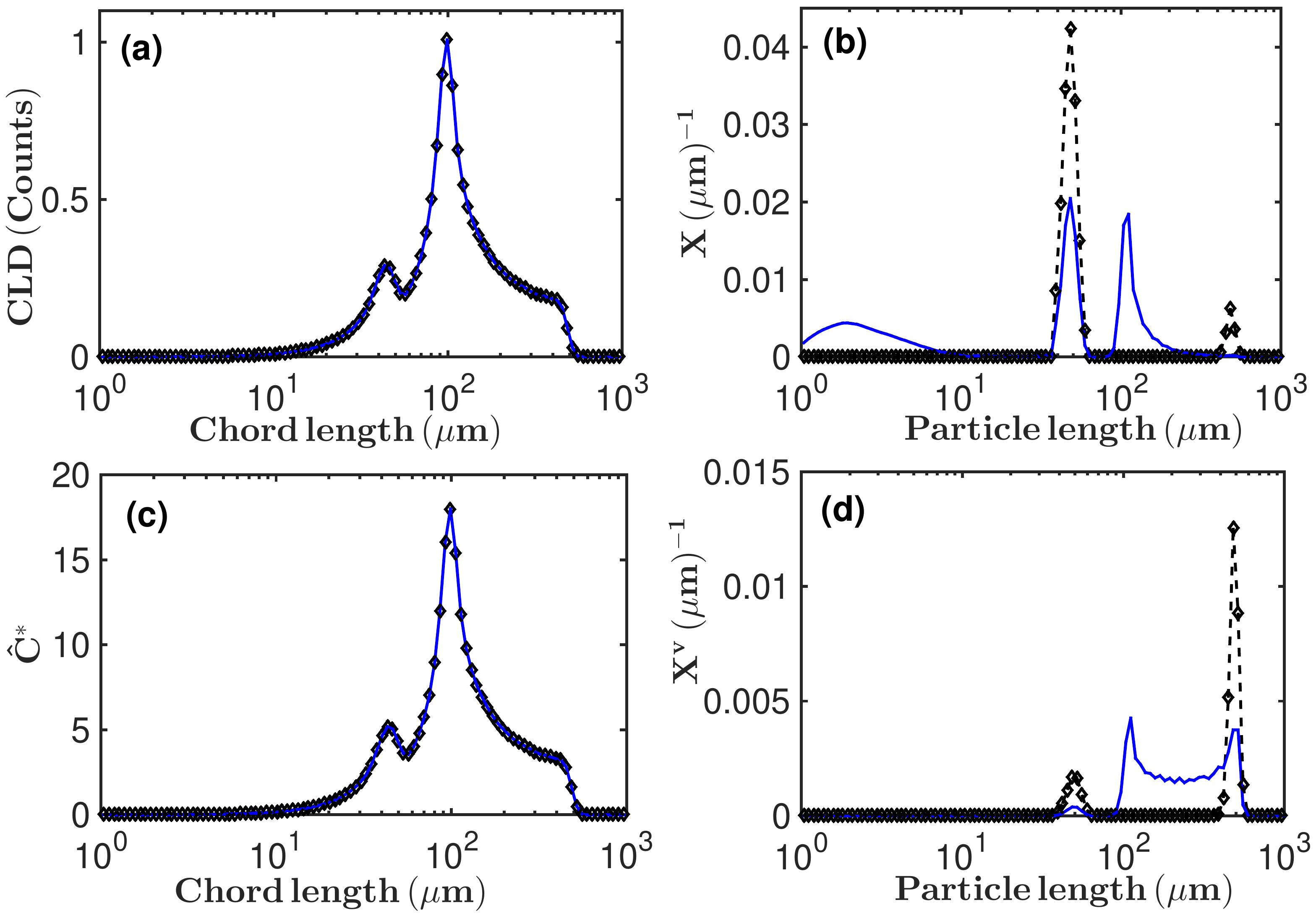}}
       \caption{(a)\,The black diamonds are the simulated CLD for the two subpopulations of particles whose aspect ratio distribution is shown in Fig. \ref{figs17}. The blue solid line is the calculated CLD for the two subpopulations by Method 1 at an aspect ratio of 0.9. (b)\,The black diamonds represent the bimodal number based PSD for the two subpopulations of particles whose aspect ratio distribution is shown in Fig. \ref{figs17}. The blue solid line is the calculated number based PSD (by Method 1) for this population of particles at an aspect ratio of 0.9. (c)\,The black diamonds are the transformed CLD (defined in Eq. (22) of the main text) for the two subpopulations of particles referred to in (a). The blue solid line is the equivalent transformed CLD calculated by Method 1 at an aspect ratio of 0.9. (d)\,The black diamonds are the volume based PSD corresponding to the number based PSD shown by the black diamonds in (b). The blue solid line is the calculated volume based PSD by Method 1 (at an aspect ratio of 0.9) for the population of particles whose volume based PSD is shown by the black diamonds.}
        \label{figs18}
         \end{figure}

  \section{Comparison of estimated PSD from images and PSDs from Methods 1 and 2}
  \label{Supsec6}
  
 Even though the objects detected from the images suffer from focusing problems as discussed in the main text, an estimate of the PSD from the images can be made. For this purpose the particles are treated as ellipsoids whose two minor axes lengths are equal (as described in section \ref{sec3}) and their semi major axes lengths are given as $a$. The number based PSD for each sample (COA and Glycine) is obtained from a histogram of the counts of particle lengths $L = 2a$ using a geometrically spaced grid (with $N=70$ bins) running from particle lengths of 1$\mu$m to 1000$\mu$m.  The number based PSD is then normalised as described in Eq. (26) of the main text.
  
  The number based PSDs $\mathbf{X}$ estimated in this way for COA and Glycine are shown in Fig. \ref{figs16}(a). The PSDs are a bit noisy due to the focusing issues in the images. The left tails of the PSDs terminate close to 30$\mu$m since objects smaller than about this size are discarded from the image analysis as explained in subsection 4.1 of the main text. Hence small particles of lengths $\lesssim 30\mu$m are not counted in this estimate of the number based PSD. As the small particles are not counted, then the peak of the number based PSD (for COA) in Fig. \ref{figs16}(a) is shifted to the right of the number based PSDs estimated for COA in Fig. \ref{figs10}(b) (by Method 1) and Fig. \ref{figs12}(b) (by Method 2). A similar situation holds for the number based PSD estimated for Glycine in Fig. \ref{figs11}(b) (by Method 1) and in Fig. \ref{figs13}(b) by Method 2.
  
  The volume based PSD $\mathbf{X}^v$ is obtained from the number based PSD (before the normalisation in Eq. 26 of the main text) as described in Eq. \eqref{eqs11}, and then the normalisation in Eq. (26) of the main text is applied. The volume based PSDs for COA and Glycine obtained this way are shown by the solid black lines in Figs. \ref{figs16}(b) and \ref{figs16}(c) respectively. Similar to the case of the number based PSD in Fig. \ref{figs16}(a), the left tails of the volume based PSDs for COA (black solid line in Fig. \ref{figs16}(b)) and Glycine (black solid line in Fig. \ref{figs16}(c)) estimated from images are truncated at particle length $\approx 30\mu$m. The small fluctuations in the number based PSDs in Fig. \ref{figs16}(a) become exaggerated in the volume based PSDs in Figs. \ref{figs16}(b) and \ref{figs16}(c) due to the volume weighting in Eq. \eqref{eqs11}. However, the peaks of the volume based PSD estimated from images agree well with the peaks of the volume based PSDs estimated by both Methods 1 and 2 for COA and Glycine as seen in Figs. \ref{figs16}(b) and \ref{figs16}(c).
  
  \section{Performance of Methods 1 and 2 for a bimodal distribution}
  \label{Supsec7}
  
  The results in Fig. 11(b) of the main text suggests that Method 2 may be more effective for the estimation of the PSD of a population of particles with a bimodal distribution of sizes and a distribution of aspect ratio for a given size. This hypothesis is tested here with an idealised population of particles made up of two separate subpopulations of different sizes and aspect ratios. The first subpopulation consists of particles of aspect ratio $r_1=0.9$ as shown in Fig. \ref{figs17} and a number based distribution of particle lengths $\mathbf{X}$ tightly packed around the mean length $\overline{L}_1\approx 50\mu$m as shown by the black diamonds in Fig. \ref{figs18}(b). The second subpopulation consists of particles of aspect ratio $r_2=0.2$ shown in Fig. \ref{figs17}. The number based distribution of particle lengths of the second subpopulation is tightly packed around the mean length $\overline{L}_2\approx 500\mu$m as shown by the black diamonds in Fig. \ref{figs18}(b). The normalised (according to Eq. (26) of the main text) number based PSD of the population of particles is weighted towards the smaller particles as shown by the black diamonds in Fig. \ref{figs18}(b). 
  
  Using the known number based PSD shown by the black diamonds in Fig. \ref{figs18}(b), a CLD is constructed by solving the forward problem in Eq. 10 of the main text. The transformation matrix $\mathbf{\tilde{A}}$ is constructed using particle sizes running from $L_{min}=1\mu$m to $L_{max}=1000\mu$m. The particles whose size fall within the range of sizes bounding the mean length $\overline{L}_1\approx 50\mu$m (shown by the black diamonds in Fig. \ref{figs18}(b)) are assigned the aspect ratio $r_1 = 0.9$, while particles whose sizes fall within the range of sizes bounding the mean length $\overline{L}_2\approx 500\mu$m (also shown by the black diamonds in Fig. \ref{figs18}(b)) are assigned the aspect ratio $r_2=0.2$. 
  
  The CLD $\mathbf{\overline{C}}$ constructed by the above procedure is then perturbed by adding a small perturbation $\mathbf{\tilde{C}}$ as 
     \begin{equation}
    \mathbf{C} = \mathbf{\overline{C}} + \mathbf{\tilde{C}},
     \label{eqs28}
     \end{equation}
  where
      \begin{equation}
     \mathbf{\tilde{C}} = \epsilon\mathbf{\mathcal{N}}(\mu,\sigma^2).
      \label{eqs29}
      \end{equation}
  The parameter $\epsilon = 10^{-4}$ controls the scale of the perturbation while the Gaussian noise $\mathbf{\mathcal{N}}$ is drawn from the normal distribution with mean $\sigma = 0$ and variance $\sigma^2=1$.
  
  The smallest singular value of the transformation matrix $\mathbf{\tilde{A}}$ is of order $10^{-3}$ in this case. This implies that the noise added to the CLD will be amplified by a factor of $10^3$ by the smaller singular values of the matrix $\mathbf{\tilde{A}}$. Hence the noise scale set by $\epsilon=10^{-4}$ will be amplified to about $10^{-1}$ in the solution vector $\mathbf{X}$ (by the smaller singular values of $\mathbf{\tilde{A}}$) which is larger than the scale of the number based PSD shown by the black diamonds in Fig. \ref{figs18}(b). Hence the noise level set by $\epsilon=10^{-4}$ is quite significant.
  
  Using the simulated CLD $\mathbf{C}$ in Eq. \eqref{eqs28} (black diamonds in Fig. \ref{figs18}(a)) and the known particle size range of $L_{min}=1\mu$m to $L_{max}=1000\mu$m, then the inverse problem is solved by both Methods 1 and 2 described in the main text.
  
  \subsection{Solution by Method 1}
   \label{Supssec7-1}
   
  In this case, the transformation matrix $\mathbf{\tilde{A}}$ is constructed using the known particle size range. The same aspect ratio is assigned to all particles of different lengths. Then the inverse problem is solved using the objective function $f_1$ defined in Eq. (17) of the main text. The calculation is then repeated at different aspect ratios. The use of the objective function in this case is to allow an estimate of the parameter $\lambda_1$ to be made for the objective function $f_2$ defined in Eq. (20) of the main text. Once the value of $\lambda_1$ has been estimated, then the objective function $f_2$ defined in Eq. (20) of the main text is then used to obtain a unique aspect ratio for the population of particles, and this procedure results in an aspect ratio $\overline{r}=0.9$ for the bimodal population of particles. The inverse problem is then solved with the objective function $f_3$ (defined in Eq. (21) of the main text) using the aspect ratio of $\overline{r}=0.9$ for all particles of different lengths. The number based PSD obtained by this procedure is shown by the solid line in Fig. \ref{figs18}(b). The corresponding CLD is shown by the solid line in Fig. \ref{figs18}(a). 
  
  Even though the calculated CLD by Method 1 matches the originally simulated CLD (Fig. \ref{figs18}(a)), the corresponding number based PSD does not match the originally simulated number based PSD in Fig. \ref{figs18}(b). The calculated number based PSD by Method 1 in Fig. \ref{figs18}(b) under estimates the length of the larger subpopulation of particles as seen in Fig. \ref{figs18}(b). The calculated number based PSD by Method 1 also underestimates the peak of the smaller subpopulation of particles. The calculated number based PSD by Method 1 contains a small bulge close to particle length of $2\mu$m (in Fig. \ref{figs18}(b)) due to the noise in the calculation. 
  
  The black diamonds in Fig. \ref{figs18}(c) represent the transformed CLD defined in Eq. (22) of the main text which corresponds to the original volume based PSD (shown by the black diamonds in Fig. \ref{figs18}(d)). This volume based PSD is calculated from the number based PSD by treating the particles as ellipsoids whose two minor dimensions are equal as described in section \ref{sec3}. This then allows the volume based PSD to be calculated using Eq. \eqref{eqs11}. The volume based PSD (shown by the solid line in Fig. \ref{figs18}(d)) calculated by Method 1 using the aspect ratio $\overline{r}=0.9$ and the objective function $f_3$ in Eq. (21) of the main text does not match the original volume based PSD shown by the black diamonds in Fig. \ref{figs18}(d). This is not withstanding the fact that the corresponding transformed CLD (shown by the solid line in Fig. \ref{figs18}(c)) matches the original transformed CLD shown by the black diamonds in Fig. \ref{figs18}(c). 
  
  The mismatch of the calculated PSDs by Method 1 to the original PSDs is due to the fact that Method 1 uses a single aspect ratio of $\overline{r}=0.9$ to match the CLDs at the expense of predicting the wrong particle sizes. This is not surprising as there are no particles with intermediate aspect ratios between $r_1=0.9$ and $r_2=0.2$ in the population to enable Method 1 make a compromise.
  
    \begin{figure}[tbh]
     \centerline{\includegraphics[width=0.9\textwidth]{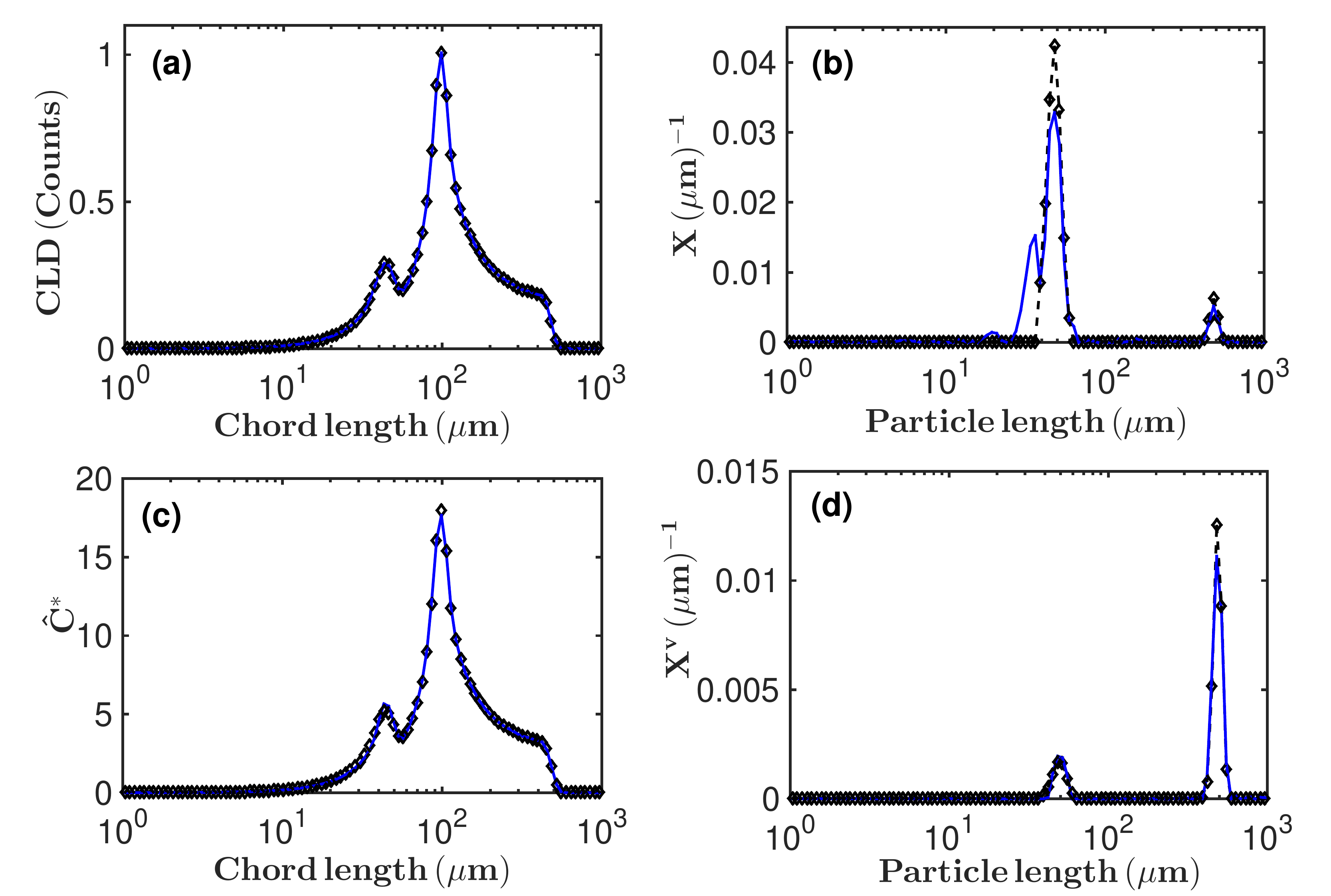}}
    \caption{(a)\,The black diamonds are the simulated CLD shown in Fig. \ref{figs18}(a), while the blue solid line is the calculated CLD by Method 2 using the aspect ratio distribution shown in Fig. \ref{figs17}. (b)\,The black diamonds are the number based PSD shown in Fig. \ref{figs18}(b), while the blue solid line is the number based PSD calculated by Method 2 using the aspect ratio distribution shown in Fig. \ref{figs17}. (c)\,The black diamonds are the transformed CLD shown in Fig. \ref{figs18}(c), while the blue solid line is the corresponding transformed CLD calculated by Method 2 using the aspect ratio distribution in Fig. \ref{figs17}. (d)\,The black diamonds are the volume based PSD shown in Fig. \ref{figs18}(d), while the blue solid line is the predicted volume based PSD by Method 2 using the aspect ratio distribution in Fig. \ref{figs17}.}
    \label{figs19}
    \end{figure}
  
  \subsection{Solution by Method 2}
   \label{Supssec7-2}
   
  Since all the particles in each of the subpopulations have the same aspect ratio, then it implies that there is no distribution of aspect ratios for particles of the same size. Then the calculations by Method 2 can be simplified by collapsing the aspect ratio dimension of the 3D transformation matrix defined in Eq. (15) of the main text. Then it becomes an issue of assigning the aspect ratios $r_1=0.9$ and $r_2=0.2$ to the particles in each of the subpopulations. Although the particle size range is known in this case, however, the objective function $f_1$ (in Eq. (17) of the main text) is still used so that an estimate of the value of the parameter $\lambda_2$ (for the objective function $f_3$ in Eq. (21) of the main text\footnote{Recall that the objective function $f_2$ in Eq. (20) of the main text is not needed in Method 2.}) can be made. Then the objective function $f_3$ in Eq. (21) of the main text is used to estimate the number based PSD and subsequently the CLD. This number based PSD estimated by Method 2 is shown by the solid line in Fig. \ref{figs19}(b) and the corresponding CLD is shown by the solid line in Fig. \ref{figs19}(a). 
  
  The calculated CLD by Method 2 matches the originally simulated CLD (black diamonds in Fig. \ref{figs19}(a)), while the calculated number based PSD by Method 2 correctly predicts the sizes of the two subpopulation as seen in Fig. \ref{figs19}(b). The originally simulated bimodal number based PSD is shown by the black diamonds in Fig. \ref{figs19}(b). The height of the main mode of the calculated number based PSD by Method 2 (solid line in Fig. \ref{figs19}(b)) is slightly lower than the height of the main mode of the originally simulated number based PSD (black diamonds in Fig. \ref{figs19}(b)) due to the effect of noise on the left shoulder of the calculated number based PSD by Method 2. 
  
  Similarly, the calculated volume based PSD by Method 2 correctly predicts the sizes of the two subpopulations of particles as seen by the solid line in Fig. \ref{figs19}(d). The black diamonds represent the originally simulated bimodal volume based PSD. However, the calculated volume based PSD by Method 2 slightly under estimates the height of the main mode of the originally simulated volume based PSD as seen in Fig. \ref{figs19}(d). The transformed CLD (in Eq. (22) of the main text) corresponding to the originally simulated volume based PSD is shown by the black diamonds in Fig. \ref{figs19}(c). The corresponding transformed CLD calculated by Method 2 matches this transformed CLD as seen by the solid line in Fig. \ref{figs19}(c).
  
       \begin{figure}[tbh]
          \centerline{\includegraphics[width=0.2\textwidth]{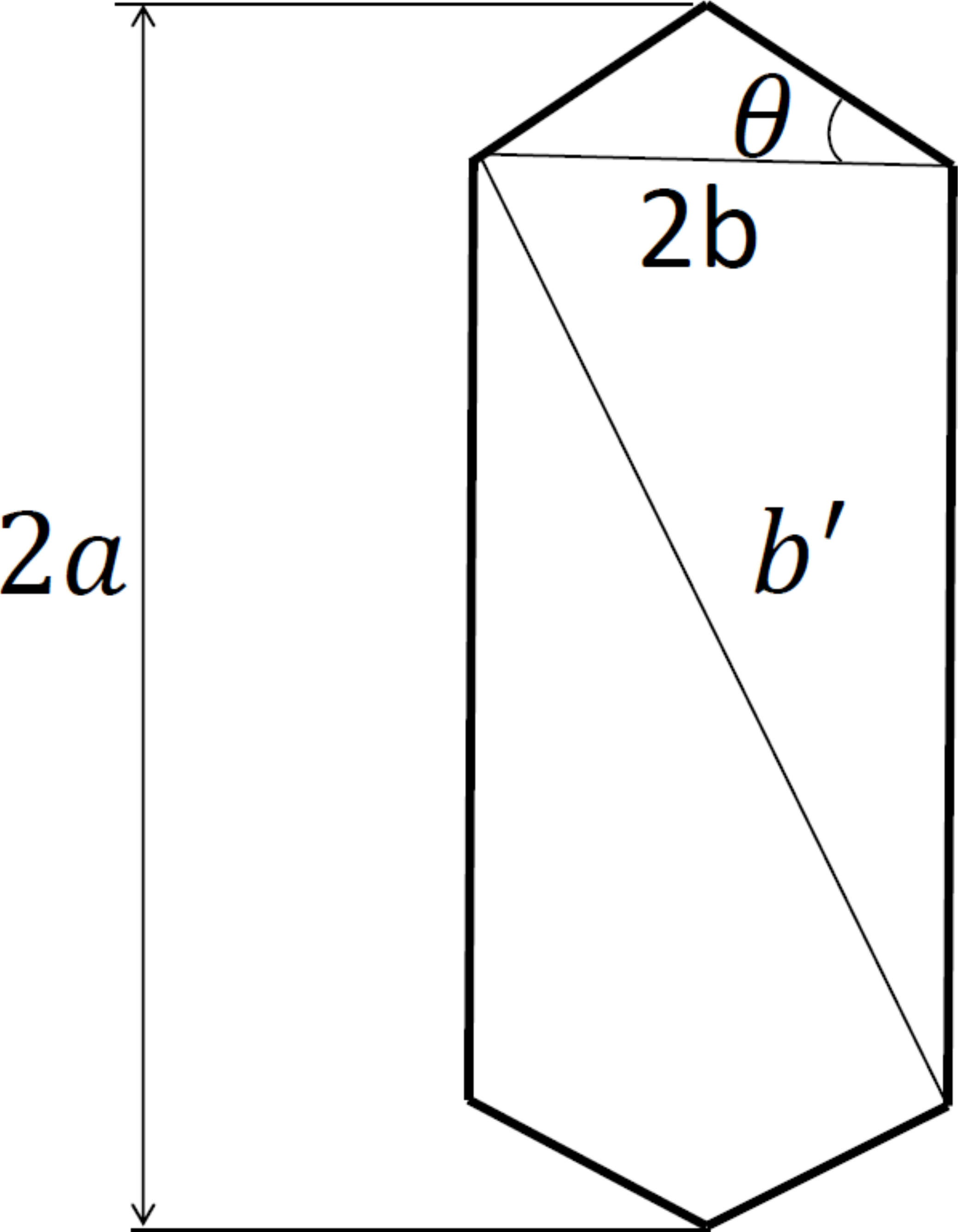}
   			\includegraphics[width=0.2\textwidth]{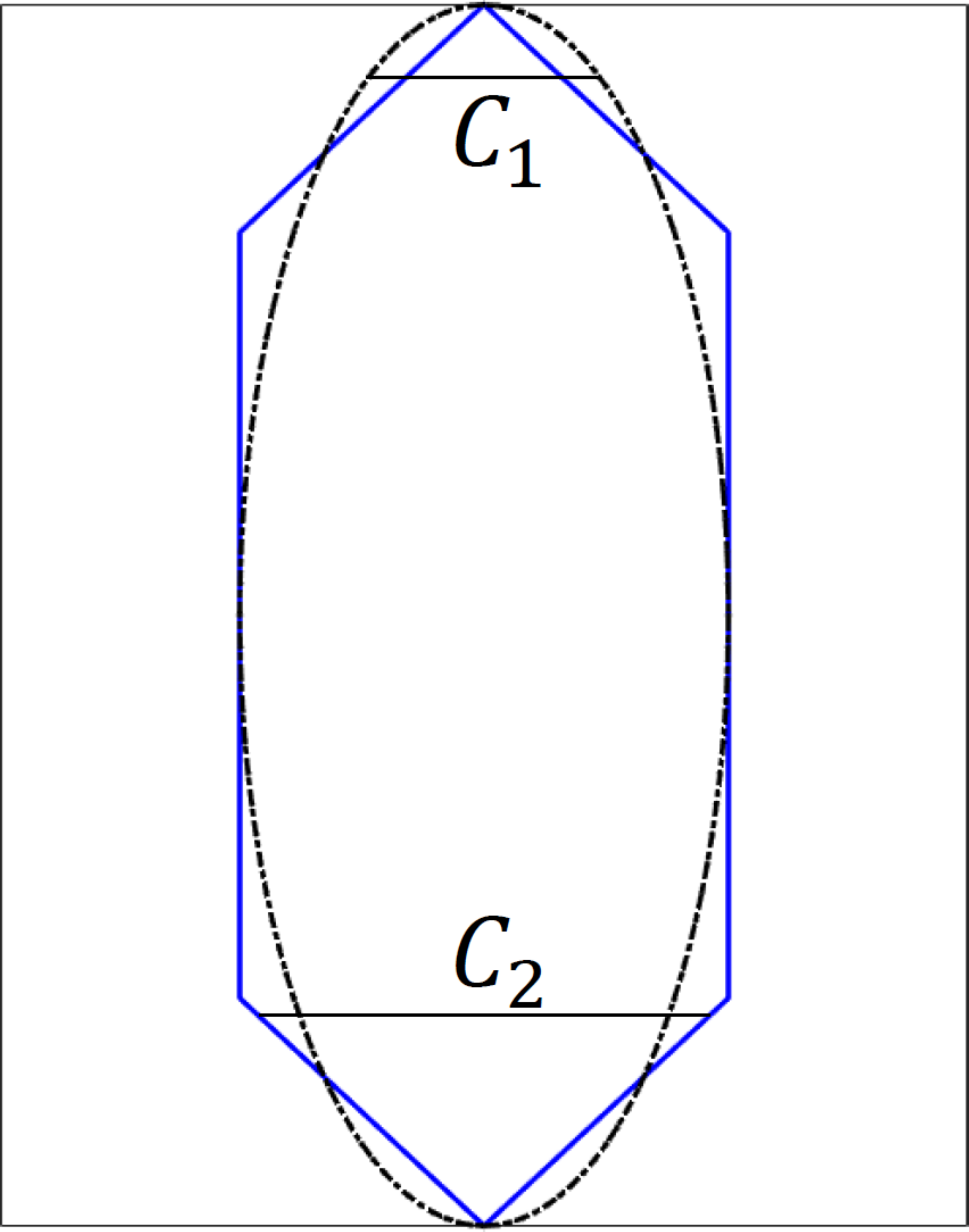}}
       \caption{Left:\, 2D silhouette of a prismatic object (similar to the Glycine particles in Fig. 2(c) of the main text) in which the major (length $2a$) and minor (length $2b$) dimensions lie in the plane of the paper. The object also contains another characteristic dimension shown by the line $b^{\prime}$. The pyramidal cap of the object makes an angle of $\theta$ with the horizontal. Right:\,The prismatic object shown on the left is represented by the blue solid line, while an ellipse of equivalent major and minor dimensions is shown by the dashed line. The chord $C_1$ is over estimated by the ellipse while the chord $C_2$ is under estimated by the ellipse for this value of $\theta = 43^{\textrm{o}}$.}
            \label{figs20}
         \end{figure}

                    \begin{figure}[tbh]
                       \centerline{\includegraphics[width=\textwidth]{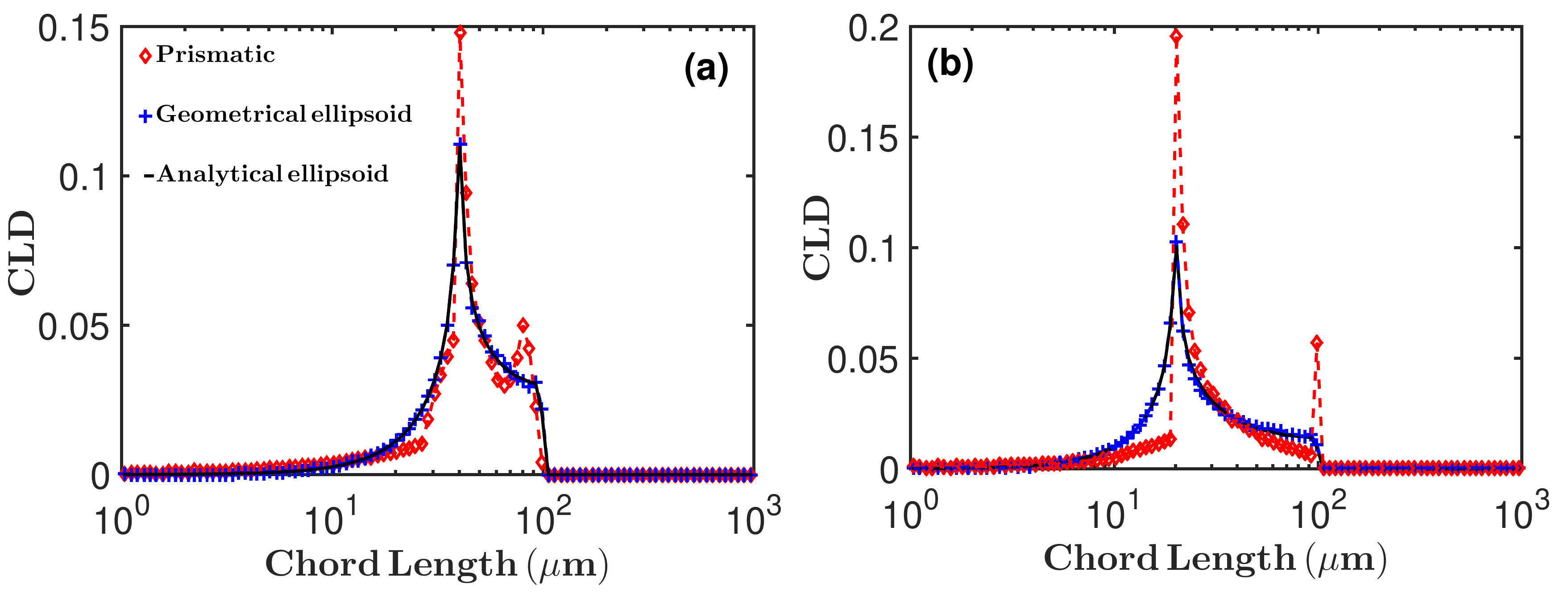}}
                    \caption{(a)\,The red diamonds represent the single particle CLD for a prismatic object (shown on the left of Fig. \ref{figs20}) with dimensions $a=50\mu$m, $b=20\mu$m inclination angle $\theta = 43^{\textrm{o}}$. The blue crosses represent the numerically computed single particle CLD of an ellipse of equivalent dimensions as the prismatic object. The black solid line is the calculated single particle CLD of the same ellipse using the analytic LW model defined in Eqs. \eqref{eq1} to \eqref{eq4}. (b)\,Similar to (a) but with dimensions $a=50\mu$m, $b=10\mu$m inclination angle $\theta = 10^{\textrm{o}}$ for the prismatic object.}
                   \label{figs21}
                    \end{figure}
  
  \section{Error estimate from shape approximation}
   \label{Supsec8}
   
   Crystalline particles are faceted such as the case of Glycine in Fig. 2(c) of the main text. However, all the particles in both the COA and Glycine samples  have been represented as ellipsoids in this work. This approximation will introduce some errors in the CLD calculated with the ellipsoidal model. A robust estimate of this error will involve a sensitivity analysis using models for faceted objects, and then investigating the variation of the difference between the CLDs of the faceted and ellipsoidal models as the shapes change. As this analysis is beyond the scope of this work, then a simple estimate of this error using the faceted object on the left of Fig. \ref{figs20} shall be made here.
   
   The object shown on the left of Fig. \ref{figs20} is the 2D silhouette of a 3D object with the shape of a prism and pyramidal caps as in the case of potassium dihydrogen phosphate (KDP) \cite{Majumder2013}. In this 2D representation, the major (of length $2a$ on the left of Fig. \ref{figs20}) and minor (length $2b$ on the left of Fig. \ref{figs20}) dimensions lie in the plane of the paper. However, the prismatic object also has a third characteristic dimension indicated as $b^{\prime}$ on the left of Fig. \ref{figs20}. This third characteristic dimension is missing from the 2D silhouette of an ellipsoid (where its major and minor axes lie in the plane of the paper) of the same length ($2a$) and width ($2b$) as the prismatic object as shown by the dashed line on the right of Fig. \ref{figs20}. This will introduce an error in the CLD estimated with the ellipsoidal model. The faces of the pyramidal caps are inclined at an angle $\theta$ to the horizontal as seen on the left of Fig. \ref{figs20}. This will cause some chords such as $C_1$ on the right of Fig. \ref{figs20} to the over estimated while some other chord such as $C_2$ on the right of Fig. \ref{figs20} will be under estimated when $\theta$ is large enough. However, when $\theta$ is small, the chord $C_1$ will be under estimated as the ellipsoid will be completely contained within the prismatic object. These effects will also introduce errors in the CLD estimated with the ellipsoidal model. However, the size of the error will depend on the length and aspect ratio ($b/a$) of the particle. It will also depend on the angle of inclination $\theta$. 
   
   The error estimation is carried out here by constructing the CLD of both objects (prismatic and ellipsoid) for given values of the dimensions of the objects and inclination angle of the prismatic object. For a given value of $a$, $b$ and $\theta$, a prismatic object is constructed and the corresponding ellipse is constructed as well. Then each object is rotated through a random angle $\in [0, 2\pi]$, and then horizontal chords are made on each object at each orientation. The CLD is accumulated for 5000 rotations and then the CLD is normalised such that the total probability is unity. The result for the prismatic object for $a = 50\mu$m, $b = 20\mu$m, and $\theta = 43^{\textrm{o}}$ is shown\footnote{The value of $\theta = 43^{\textrm{o}}$ corresponds to KDP particles \cite{Majumder2013}. However, this value is used here to represent the Glycine particles in Fig. 2(c) of the main text since the Glycine and KDP particles have similar shapes. A smaller value of $\theta = 10^{\textrm{o}}$ is used to represent the COA particles in Fig. 2(b) of the main text as the pyramidal caps of the COA particles are not so pronounced as those of Glycine. The parameter values used in this illustration may not necessarily represent the Glycine and COA samples accurately, however, they show the discrepancy that could be introduced by shape differences.} by the red diamonds in Fig. \ref{figs21}(a). The CLD for the corresponding ellipse is shown by the blue crosses in Fig. \ref{figs21}(a). The black solid line in Fig. \ref{figs21}(a) is the calculated CLD (for an ellipse with the same dimensions as the prismatic object) from the analytical LW model defined in Eqs. \eqref{eqs1} to \eqref{eqs4}. 
   
   Of course the analytically calculated CLD for the ellipse matches that of the numerically computed CLD for the same ellipse as seen in Fig. \ref{figs21}(a). However, the ellipsoidal model shows some discrepancies with the prismatic model as seen in Fig. \ref{figs21}(a) due to differences in shape. The level of discrepancy is estimated by taking the $L_2$ norm of the difference between the CLDs from the ellipsoid and the prismatic object. This procedure gives an estimated level of discrepancy of about 6.5\%. However, the level of discrepancy increases when a thinner prismatic object with less inclined pyramidal caps is used. This time the parameter values are given as $a = 50\mu$m, $b = 10\mu$m, and $\theta = 10^{\textrm{o}}$. The CLD for this prismatic object is shown by the red diamonds in Fig. \ref{figs21}(b). The numerically computed CLD for an equivalent ellipse is shown by the blue crosses in Fig. \ref{figs21}(b), while the analytically computed CLD (using the LW model) for the same ellipse is shown by the black solid line in Fig. \ref{figs21}(b). Figure \ref{figs21}(b) shows that the level of discrepancy between the CLDs form the prismatic and ellipsoidal objects increase as the aspect ratio and angle of inclination $\theta$ are reduced. The level of discrepancy this time is about 14\%.

\clearpage

\end{document}